\documentclass[11pt,a4paper,reqno]{amsart}
\usepackage{amsmath,amssymb,amsbsy,amsthm,dsfont,ulem}
\usepackage{graphicx,a4wide,hyperref}
\usepackage{multirow,ulem}
\usepackage[utf8]{inputenc}
\usepackage[T1]{fontenc}
\usepackage{xcolor}
\usepackage{float}
\usepackage{natbib}
\usepackage[ruled,vlined]{algorithm2e}
\usepackage{setspace}
\usepackage{enumitem}
\usepackage{amsfonts}

\usepackage{url}

\DeclareSymbolFont{myletters}{OML}{ztmcm}{m}{it}
\DeclareMathSymbol{\uplambda}{\mathord}{myletters}{"15}

\newcounter{hypA}

\DeclareMathOperator{\gammaM}{\boldsymbol{\gamma}}
\DeclareMathOperator{\betaM}{\boldsymbol{\beta}}
\DeclareMathOperator{\tauM}{\boldsymbol{\tau}}

\DeclareMathOperator*{\argmin}{\arg\!\min}

\newcommand{\diag}{\mathop{\mathrm{diag}}}

\def \1{\mathbf{1}}

\newcommand{\mbf}[1]{\mathbf{#1}}

\newcommand{\R}{\mathbb{R}}

\renewcommand{\a}{\mbf{a}}

\newcommand{\B}{\mbf{B}}

\renewcommand{\u}{\mbf{u}}
\renewcommand{\v}{\mbf{v}}

\newcommand{\x}{\mbf{x}}

\newcommand{\Y}{\mbf{Y}}
\newcommand{\W}{\mbf{W}}

\renewcommand{\t}{\mbf{t}}

\renewcommand{\P}{\mathcal{P}}
\renewcommand{\S}{\mathcal{S}}

\begin{document}

\title[A novel approach for estimating functions in the multivariate setting]{A novel approach for estimating functions in the multivariate setting based on an adaptive knot selection for B-splines with an application to a chemical system used in geoscience}
\author{Mary E. Savino}
\address{Andra, 1/7 Rue Jean Monnet, 92290 Châtenay-Malabry, France and Université Paris-Saclay, AgroParisTech, INRAE, UMR MIA Paris-Saclay,
  91120, Palaiseau, France}
\author{Céline Lévy-Leduc}
\address{Université Paris-Saclay, AgroParisTech, INRAE, UMR MIA Paris-Saclay,
  91120, Palaiseau, France}
  
\begin{abstract}

In this paper, we will outline a novel data-driven method for estimating functions in a multivariate nonparametric regression model based on an adaptive knot selection for B-splines. 
The underlying idea of our approach for selecting knots is to apply the generalized lasso, since the knots of the B-spline basis can be seen as changes in the derivatives of the function to be estimated. This method was then extended to functions depending on several variables by processing each dimension independently, thus reducing the problem to a univariate setting. The regularization parameters were chosen by means of a criterion based on EBIC. The nonparametric estimator was obtained using a multivariate B-spline regression with the corresponding selected knots. Our procedure was validated through numerical experiments by varying the number of observations and the level of noise to investigate its robustness. The influence of observation sampling was also assessed and our method was applied to a chemical system commonly used in geoscience. For each different framework considered in this paper, our approach performed better than state-of-the-art methods. 
Our completely data-driven method is implemented in the \texttt{glober} R package which is available on the Comprehensive R Archive Network (CRAN). 


\end{abstract}


\maketitle

\section{Introduction}

In geochemical models, computing the concentrations of reactive species at equilibrium is well-known to be a
challenging task especially when the number of species is large and/or when the reactions involve the
dissolution or the precipitation of minerals, see \cite{white58}, \cite{smith80} and \cite{decapitani87} for further details.
The numerical resolution of these non-linear problems can become so time consuming that coupling them with other physical
processes may require to be simplified. For instance in the case of reactive transport, the
size of the geometric model has to be drastically limited. To overcome this issue, researchers have been focusing their work on improving the numerical scheme to speed up computations.

However, despite the significant improvements of the numerical solvers and preconditioners
over the past few decades, solving three dimensional large scale modelling of complex reactive transport over many time steps
is still nearly impossible using standard computers.  Consequently, geoscientists are more and more interested in devising approaches
which can provide an estimation of the solution of the full simulation model (sometimes also called surrogate model) from a limited set of observations obtained with the full simulation model from specific input values that can thus replace it.
Hence, the problem can be reformulated as
the estimation of an unknown function $f$ in the following regression model:
\begin{equation} \label{eq:model}
Y_i = f(x_i) + \varepsilon_i, \quad 1 \leq i \leq n,
\end{equation} 
where the $\varepsilon_i$ are i.i.d centered random variables of variance $\sigma^2$ and the $x_i$ are observation points which belong to a compact set $\S$ of $\R^d$, $d\geq 1$.
In the reactive transport modelling field (RTM), several surrogate models have been proposed, we refer the reader to \cite{asher2015review} and \cite{jatnieks2016data} for a comparison of the different approaches.
Artificial neural networks have recently gained a huge interest in RTM (\cite{guerillot2020geochemical}), more especially through Deep Neural Networks (DNNs), since their approximations have a high accuracy compared to other estimators (\cite{laloy2019emulation}).
Nevertheless, despite all the effort for improving the efficiency of DNNs via the conjunction of computational advancements for training ever-larger networks and improvements of backpropagation algorithms, DNNs still remain difficult to exploit when the quantity of training data is not sufficient (\cite{karpatne2018machine}) especially when a high number of parameters needs to be calibrated.
Recently, \cite{savino2022active} proposed an active learning approach to drastically decrease the number of training observations to use by modeling the function to estimate as a sample of Gaussian Processes. This method has given promising results but is not necessarily the most suitable approach
for noisy observation sets. 
In order to circumvent this limitation, nonparametric estimation approaches based on splines are known to be an efficient tool, see \cite{wahba1990spline}
for further details on this kind of methods.

Nonparametric estimation approaches based on splines consist in approximating the function to estimate by a linear combination of splines
  which are functions defined by pre-selecting a well chosen set of knots. In this framework, \cite{friedman1991mars} proposed an efficient approach called Multivariate Adaptive Regression Splines (MARS) which can be used when
  the function to estimate has several input variables.
However, MARS has not shown better performance than other state-of-the-art methods on a concrete RTM application displayed in \cite{jatnieks2016data}. A theorical and experimental comparison has been undertaken by \cite{eckle2019comparison} and demonstrated that DNNs can outperform MARS but with a specific number of parameters and they do not necessarily give better results for every numerical application. This conclusion was also drawn by \cite{ZHANG201382} and \cite{ZHANG201645} in which the authors demonstrated an equivalent accuracy and performance between a back-propagation neural network architecture and MARS on geotechnical applications but a better interpretability and a higher computational efficiency was demonstrated for the latter.

Other articles proposed approximating the function to estimate by a linear combination of B-splines defined in \cite{de1978practical} since they display an attractive stability and a computationnal efficiency. Their ability to approximate complicated functions while being unsensitive to noisy observation sets have made them very interesting in the past few decades. Since their definition depends on a pre-defined sequence of knot locations, many strategies have been developed to optimize the selection of these points in order to avoid overfitting and so to ensure the best approximation of the underlying function.
\cite{osullivan1986statistical} described an innovative method, introduced as O-splines by \cite{wand08comparison}, to estimate a function $f$ by selecting simultaneously the number and the locations of knots from an arbitrary set of values. Its main goal was to penalize an integrated square of the second order derivative also called roughness in order to determine the coefficients of the linear combination of B-splines. However, the computation of this method was tedious with higher order derivatives. To circumvent this issue, \cite{eilers1996flexible} proposed a discrete version of this method called P-splines which uses a discrete penalty matrix and a $\ell_2$-norm penalized least-square criterion (ridge approach) to determine the coefficients of the B-splines defined from evenly-spaced knots (see \cite{wand08comparison} for a detailed comparison between O-splines and P-splines). These P-splines have been used in an impressive list of work of curve fitting (see \cite{eilers2015twenty} for a review) and have been extended to the multivariate setting, see \cite{eilers2003multivariate} for an application to smoothing two-dimensional signals. \cite{li2022general} have adapted these P-splines to apply them to unevenly-spaced knots by defining a general weighted difference penalty matrix adapted to regular and irregular knot spacing.
To drastically limit the number of knots, \cite{goepp2018spline} have proposed a weigthed adaptive ridge method called A-splines which aims at discarding the less relevant knots and by defining new B-splines from the selected knots. This method appeared to be more interpretable than the P-splines method but their statistical performance is equivalent.

Another approach was introduced for B-spline curve fitting by \cite{yuan2013adaptive}. Their idea is to first select the most pertinent B-splines from a multi-resolution basis by applying the Lasso criterion to get the locations of the knots. Then, after a pruning step to reduce once again the number of knots,
the final B-splines are built from these small sets of knots. The estimation of the function is obtained by fitting a linear combination of these B-splines to
the observations by a least-square approach. This method seems to have promising results but is not available for two-dimensional functions yet.

In this paper, we propose a novel data-driven approach for estimating the function $f$ in the multivariate nonparametric regression model (\ref{eq:model}) based on an adaptive knot selection for B-splines. 
Since the knots of a B-spline basis can be seen as changes in the derivatives of $f$, we propose finding the most relevant ones, based on the work of \cite{denis2020novel}, by using the generalized lasso described in \cite{tibshirani2011solution} and further studied in \cite{tibshirani2014adaptive}. 
A B-spline basis is then defined from these selected knots and a least-square approach is undertaken to determine the coefficients of the linear combination of B-splines. \cite{sadhanala2021multivariate} have proposed a multivariate version of trend filtering (a specific generalized lasso form) called Kronecker trend filtering (KTF) to extend it to smoothing functions with multiple input variables. It implies the use of a huge penalty matrix defined as the Kronecker product of univariate trend filtering penalty operators and of a unique regularization parameter common to every input variables. In order to drastically reduce the dimensions of the difference penalty matrix and to allow a better flexibility in the regularization step, the extension of our method to functions with two input variables is presented by simply considering each dimension seperately to reduce the problem to the one-dimensional setting. We also propose a way to extend our approach to higher dimensional settings where the observation points do not necessarily come from a cartesian product of the sets in each dimension.

This paper is organized as follows. Section \ref{sec:method} describes the methodology that we propose for our adaptive knot selection method for the one and two-dimensional settings. Section \ref{sec:numexp} investigates the performance of our approach through numerical experiments. In Section \ref{sec:real}, we apply our method to the data that motivated this study. Finally, in Section \ref{sec:extension}, we extend our approach to more general observation point settings.


\section{Methodology}\label{sec:method}

In this section, we describe our innovative nonparametric method to estimate the function $f$ defined in \eqref{eq:model}. We will introduce our method first for one-dimensional functions ($d=1$), then in a second section we will extend it to the two-dimensional case ($d=2$).

\subsection{Description of our method in the one-dimensional case}\label{sec:one_dim_case}

We propose estimating the function $f$ appearing in (\ref{eq:model}) by approximating it with a linear combination of B-splines of order $M$ ($M \geq 1$) introduced by \cite{de1978practical} in Chapter 9. 

Let $\t = (t_1, \ldots, t_K)$ be a set of $K$ points called knots which are crucial in the definition of the B-spline basis. We define the augmented knot sequence $\tauM$ such that:
$$\tau_{1} = \ldots = \tau_{M} = x_{min},$$
$$\tau_{j+M} = t_j, \quad j = 1, \ldots, K, $$
$$x_{max} = \tau_{K+M+1} =  \ldots = \tau_{K+2M}, $$
$$\tauM = \left(\tau_1, \ldots, \tau_{K+2M}\right) = \bigl(\underbrace{x_{min}, \ldots, x_{min}}_\text{M times}, \underbrace{t_1, \ldots, t_{K}}_{\t}, \underbrace{x_{max}, \ldots, x_{max}}_\text{M times} \bigr), $$ 
where $x_{min}$ and $x_{max}$ are the lower and upper bounds of $\S$, respectively. 

B-splines are defined by \cite[p. 89-90]{de1978practical} and \cite[p. 160]{hastie01statisticallearning} as follows.
Denoting by $B_{i,m}(x)$ the $i$th B-spline basis function of order $m$ for the knot sequence $\tauM$ with $m\leq M$, they are defined by the following recursion:
\begin{equation}\label{eq:Bi1}
B_{i,1}(x) =
\begin{cases}
        1 & \text{if } \tau_i \leq x < \tau_{i+1} \\
        0  & \text{otherwise } 
\end{cases} \quad \text{for } \; i =1, \ldots, K+2M-1,
\end{equation}
and for $m\leq M$,
\begin{equation}\label{eq:Bspline_recurrence}
B_{i,m}(x) = \frac{x- \tau_i}{\tau_{i+m-1} - \tau_i}B_{i,m-1}(x) + \frac{\tau_{i+m} - x}{\tau_{i+m} - \tau_{i+1}}B_{i+1,m-1}(x),
\end{equation}
for $i =1, \ldots, (K+2M-m)$.

In the next section we will describe how to choose the set of knots $\t$ to estimate $f$.

\subsubsection{Creation of a candidate set of knots.}


Let $\Y = (Y_1, \ldots, Y_n)$ and $\x = (x_1, \ldots, x_n)$ where $Y_i$ and $x_i$ are defined in \eqref{eq:model}. In the following, we shall assume that $x_1 < \ldots < x_n$ and $M = q +1$, with $q \geq 0$. Hence, when $q=0$ (resp. $q=1$, $q=2$) $f$ is approximated with piecewise constant (resp. linear, quadratic) functions.  

Since the knots of a B-spline basis can be seen as changes in the $(q+1)$th derivative of $f$, we propose finding them by using the generalized Lasso described in \cite{tibshirani2011solution} and further studied in \cite{tibshirani2014adaptive}. 
In the latter, they define the polynomial trend filtering which consists in approximating $f$ by $\widehat{\betaM}(\uplambda)$ defined as follows:  
\begin{equation}\label{eq:genlasso}
\widehat{\betaM}(\uplambda) = \argmin_{\betaM \in \R^n}\{||\Y - \betaM||^2_2 + \uplambda||D\betaM||_1\},
\end{equation}
where $||y||^2_2= \sum_{i=1}^ny_i^2$ for $y = (y_1, \ldots, y_n)$ and $||u||_1= \sum_{i=1}^m|u_i|$ for $u = (u_1, \ldots, u_m)$, $\uplambda$ is a positive constant which has to be tuned and $D \in \R^{m\times n}$ is a specified penalty matrix, defined recursively as follows:
\begin{equation}\label{eq:D}
D = D_{tf, q+1} =  D_{0} \cdot  D_{tf, q} \quad q \geq 0,
\end{equation}
where ``$tf$'' is the abbreviation of ``trend filtering'', $(q+1)$ is the order of differentiation, $D_{tf, 0} = \textrm{Id}_{\R^n}$, the identity matrix
of $\mathbb{R}^n$, and $D_{0}$ is the penalty matrix for the one-dimensional fused Lasso: 
\begin{equation*}\label{eq:D0}
D_{0}  = 
\begin{bmatrix} 
-1 & 1  & 0 & \ldots & 0 \\
 0 & -1 & 1 & \ldots & 0 \\
\vdots  &   & \ddots & \ddots  & \vdots \\
0 &   0   & \ldots & -1 & 1  
\end{bmatrix}.
\end{equation*}

The penalty matrix $D$ is the discrete difference operator of order $(q+1)$ and thus, $D\widehat{\betaM}$ estimates the $(q+1)$st order derivative of $f$. Hence, observing the locations where $D\widehat{\betaM} \ne 0$ provides a way of finding the B-spline knots.

The matrix $D$ is well-adapted when  the observation points are evenly spaced. When it is not the case, it should be replaced by the following matrix $\Delta^{(q+1)}$ defined recursively as follows:
\begin{equation*}\label{eq:Delta}
\Delta^{(q+1)} = \W_{(q+1)} \cdot D_0\cdot \Delta^{(q)}, \quad q \geq 0,
\end{equation*}
where $\Delta^{(0)} = \textrm{Id}_{\R^n}$ and $\W_{(q+1)}$ is the diagonal weight matrix defined by:
\begin{equation*}\label{eq:W}
\W_{(q+1)} = \diag\left(
\frac{1}{(x_{(q+1)+1} - x_{(q+1)})}, \frac{1}{(x_{(q+1)+2} - x_{(q+1)+1})}, \ldots, \frac{1}{(x_{n} - x_{n-1})}\right).
\end{equation*} 

In both cases (evenly or unevenly-spaced observations), the number of rows of $D$ and $\Delta^{(q+1)}$ equals $m = n - q - 1$.

Let us now more precisely explain how to choose the B-spline knots.
Let $\Lambda= (\uplambda_1, \ldots, \uplambda_k)$ be a grid of penalization parameters $\uplambda_i$. We define the resulting differentiated column vector $\a(\uplambda)$ by: 
\begin{equation}\label{eq:Alambda}
\a(\uplambda) = \Delta^{(q+1)} \cdot \widehat{\betaM}(\uplambda),
\end{equation}
where $\widehat{\betaM}(\uplambda)$ is the solution of problem \eqref{eq:genlasso} when $D =  \Delta^{(q+1)}$ and $\uplambda$ belongs to $\Lambda$.

The ordered vector of selected knots associated to $\uplambda$ is defined as follows:
\begin{equation}
\widehat{\t}_\uplambda = \left(\widehat{t}_j \right)_{j = 1, \ldots, K_\uplambda} = \left(x_{p_j}\right)_{j = 1, \ldots, K_\uplambda}, \quad \text{with } p_j \in \P_\uplambda,
\label{eq:knot_sequence}
\end{equation}  
where
\begin{equation} \label{eq:setP}
\P_\uplambda = \left\{\ell+1, \; a_\ell(\uplambda) \ne 0\ \right\} \quad \textrm{and} \quad K_\uplambda = \sum_{\ell = 1}^{m} \mathds{1}\{a_\ell(\uplambda) \ne 0\},
\end{equation}
$a_\ell(\uplambda)$ denoting the $\ell$th component of $\a(\uplambda)$ and $\mathds{1}\{A\}=1$ if the event $A$ holds and 0 if not.



%
The corresponding B-spline basis $B_{i,M}$ is defined by replacing the $t_j$ in the augmented knot sequence $\tauM$ appearing in \eqref{eq:Bi1} and \eqref{eq:Bspline_recurrence} by $\widehat{t}_j$ found in \eqref{eq:knot_sequence}. Thus, we obtain the following estimator of $f$ for each $\uplambda$ of $\Lambda$:  
\begin{equation}
\widehat{f}_\uplambda(x) = \sum_{i=1}^{q + K_{\uplambda} + 1}\widehat{\gamma_i}B_{i,M}(x),
\label{eq:estimated_model}
\end{equation}
where $\widehat{\gammaM} = (\widehat{\gamma_i})_{1\leq i \leq q + K_\uplambda+1}$ is obtained using the following least-square criterion:
\begin{equation}
\widehat{\gammaM}= \argmin_{\gammaM \in \R^{q+K_\uplambda+1}}{\|\Y - \B(\uplambda)\gammaM\|_2^2} \, ,
\label{eq:least_square}
\end{equation}
where $\B(\uplambda)$ is a $n\times (q+K_\uplambda+1)$ matrix having as $i$th column $\left(B_{i,M}(x_k)\right)_{1\leq k \leq n}$, $i$ belonging to $\{1, \ldots, q+K_\uplambda+1 \}$. 

\subsubsection{Choice of the penalization parameter of the regularized method.}\label{sec:choice_lambda}

In order to choose the penalization parameter $\uplambda$ which leads to the best selection of knots, we use a criterion defined by \cite{chen2008extended} and recommended in \cite{goepp2018spline}, namely the extended Bayesian information criterion also called EBIC:
\begin{equation}
\text{EBIC}(\uplambda)= \text{SS}(\uplambda) + (q + K_\uplambda + 1)\log{n} + 2\log{{q + K_\text{max} + 1 \choose q + K_\uplambda + 1}},
\label{eq:EBIC}
\end{equation}
where $K_\text{max}$ is the maximum number of knots that we can select (here $K_\text{max} = n$) and $\text{SS}(\uplambda)$ is the sum of squares defined by:
\begin{equation}
\text{SS}(\uplambda)= \|\Y - \widehat{\Y}(\uplambda)\|_2^2 \, ,
\label{eq:SS_lambda}
\end{equation}
where 
\begin{equation*}
\widehat{\Y}(\uplambda) = \B(\uplambda) \widehat{\gammaM},
\label{eq:Bspline_regression}
\end{equation*}
with $\widehat{\gammaM}$ and $\B(\uplambda)$ being defined in \eqref{eq:least_square}.
This criterion allows us to get a trade-off between a good approximation of the underlying function without using too many parameters.
The final estimator of $f$ is defined as follows:
\begin{equation}
\widehat{f}(x) = \widehat{f}_{\uplambda_{\text{EBIC}}}(x),
\label{eq:estimated_model_ebic}
\end{equation}
where $\widehat{f}_{\uplambda}(x)$ is defined in \eqref{eq:estimated_model}
and
\begin{equation}
\uplambda_{\text{EBIC}} = \argmin_{\uplambda \in \Lambda}\{\text{EBIC}(\uplambda)\}.
\label{eq:lambda_EBIC}
\end{equation}

\subsubsection{Illustration of our method on a simple case.}\label{sec:illustration_1d}
\begin{figure}
\begin{center}
\includegraphics[width = 6cm]{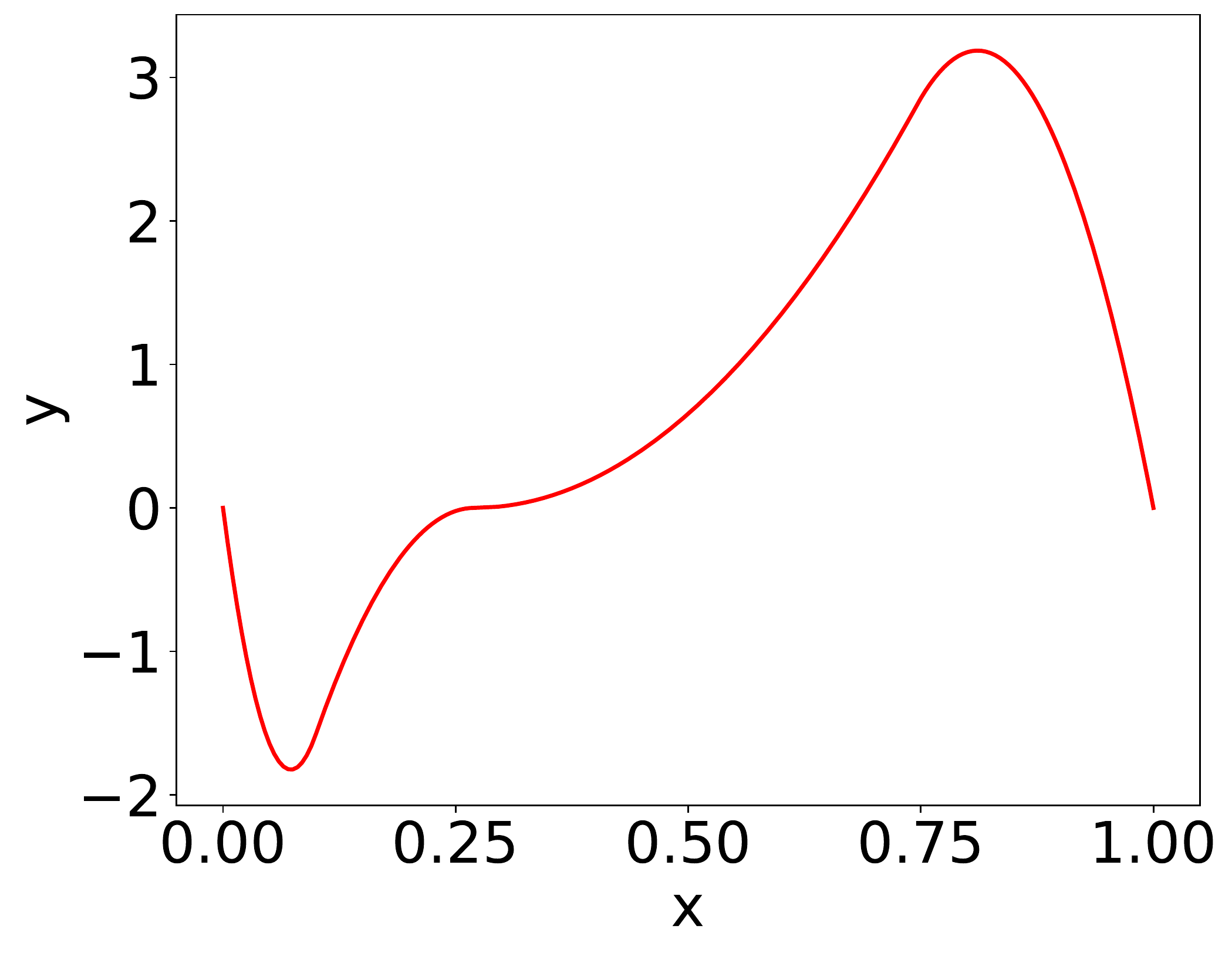}
\includegraphics[width = 6cm]{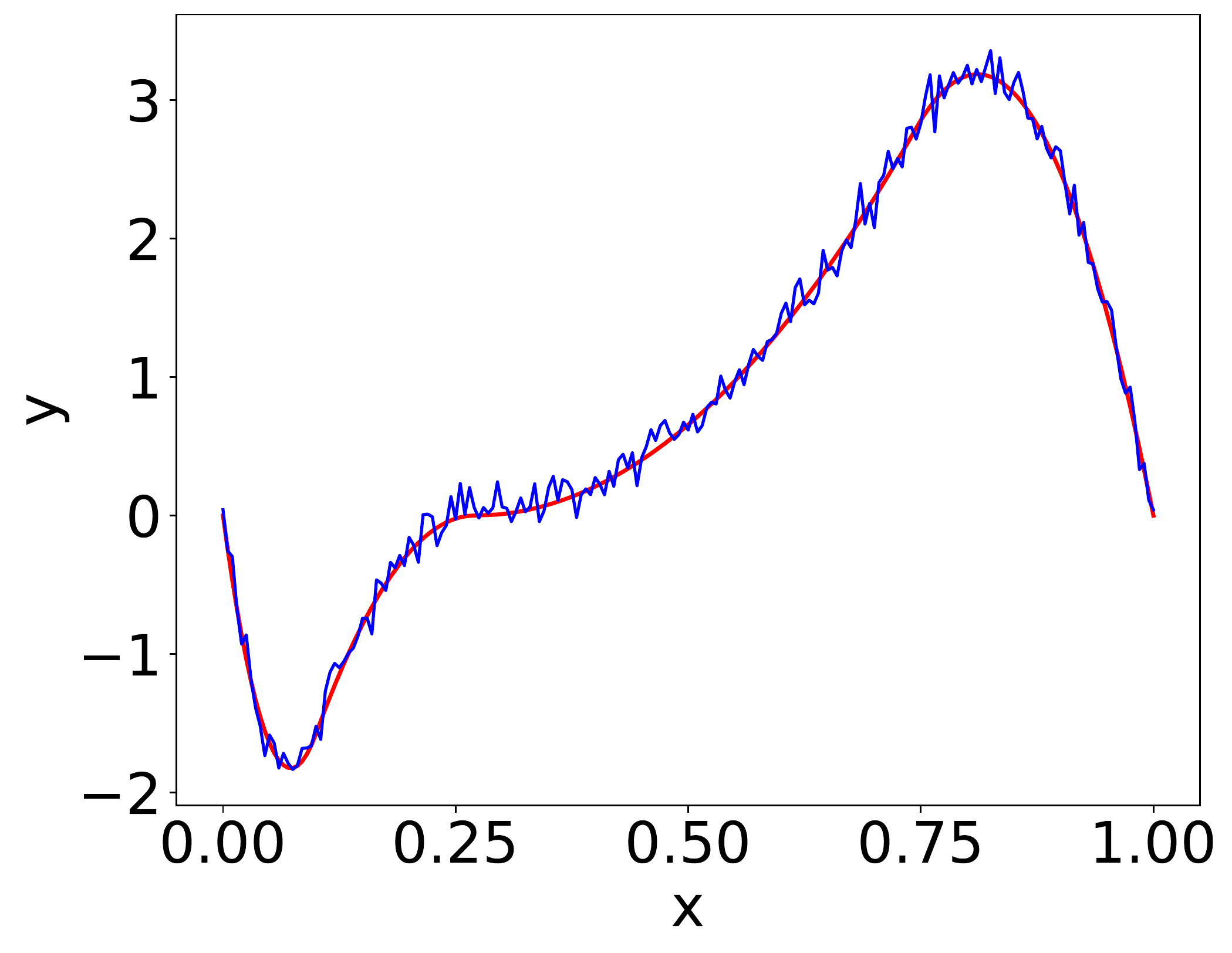}
\caption{Function $f_1$ to estimate (left) and a noisy set of observations $Y_1,\dots,Y_{201}$ with $\sigma = 0.1$ (right). }
\label{fig:functions_test_1D}
\end{center}
\end{figure}

In order to illustrate our method we apply it to a noisy set of observations $\Y = (Y_1, \ldots, Y_n)$ where the $Y_i$ are defined in \eqref{eq:model} and $f = f_1$ is a linear combination of quadratic B-splines ($M=3$) with $\t = (0.1, 0.27,0.745)$ defined as follows:
\begin{equation}\label{eq:f1}
f_1(x) =  -2.5B_{2,3}(x) + 4.3B_{5,3}(x), \quad x \in [0,1].
\end{equation}
In \eqref{eq:model}, the $\varepsilon_i$ are i.i.d Gaussian centered random variables with $\sigma = 0.1$. The set of knots $\t$ belongs to the observation set $\{x_1, \ldots, x_n\}$.
The corresponding  $(f_1(x_i))_{1\leq i \leq n}$
and $(Y_i)_{1\leq i \leq n}$ are displayed in Figure \ref{fig:functions_test_1D} for $n=201$. Since we want to approximate quadratic B-splines, we must choose $q$ such that the method can detect the changes in the third derivative so here $q+1=3$.
In order to assess the performance of our knot selection procedure, we compute the Hausdorff distance defined as follows:
\begin{equation}\label{eq:hausdorff_dist}
d(\t, \widehat{\t}_{\uplambda}) = \max{\left(d_1(\t, \widehat{\t}_{\uplambda}), \; d_2( \t, \widehat{\t}_{\uplambda})\right)},
\end{equation}
where 
\begin{equation*}
d_1(\u, \v) = \sup_{v \in \v}\inf_{u\in\u}\left|u-v\right|, 
\end{equation*}
\begin{equation*}
d_2(\u, \v) = d_1(\v, \u).
\end{equation*}

\begin{figure}[h!]
\begin{center}
\includegraphics[width = 16cm]{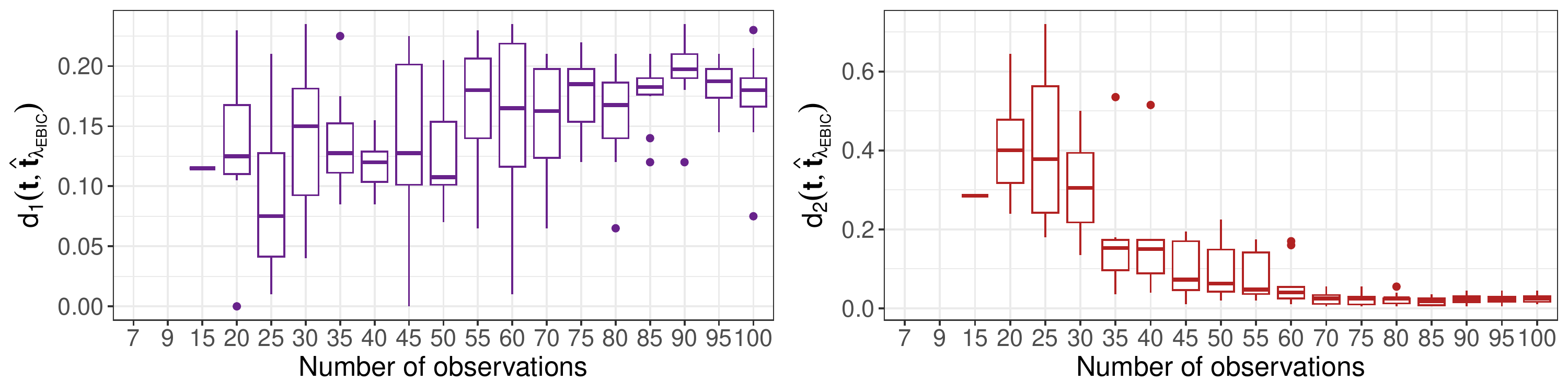}
\includegraphics[width = 8cm]{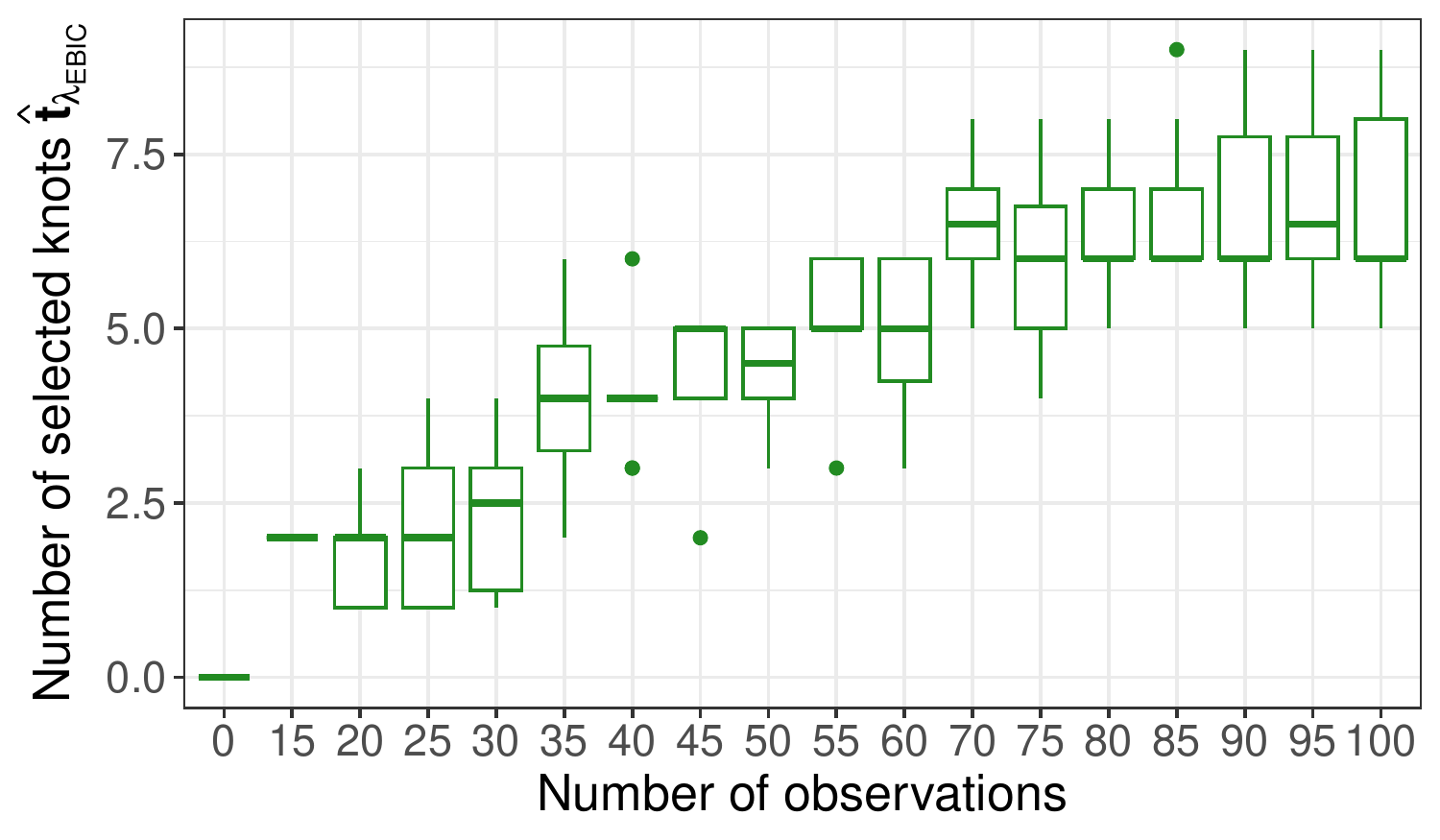}
\caption{Top left: Boxplots for the first part of the Hausdorff distance as a function of $n$. Top right: boxplots for the second part of the Hausdorff distance as a function of $n$. Bottom: number of estimated knots as a function of $n$ with $\uplambda = \uplambda_{\text{EBIC}}$ for estimating $f_1$.}
\label{fig:boxplot_lambda_ebic_f1}
\end{center}
\end{figure}
\begin{figure}[ht]
\begin{center}
\includegraphics[width = 16cm]{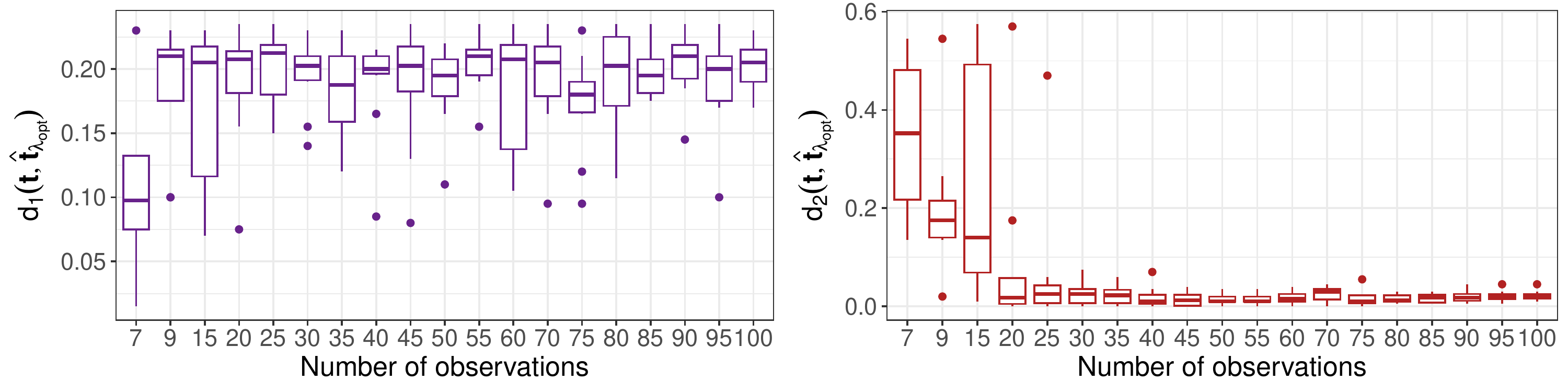}
\includegraphics[width = 8cm]{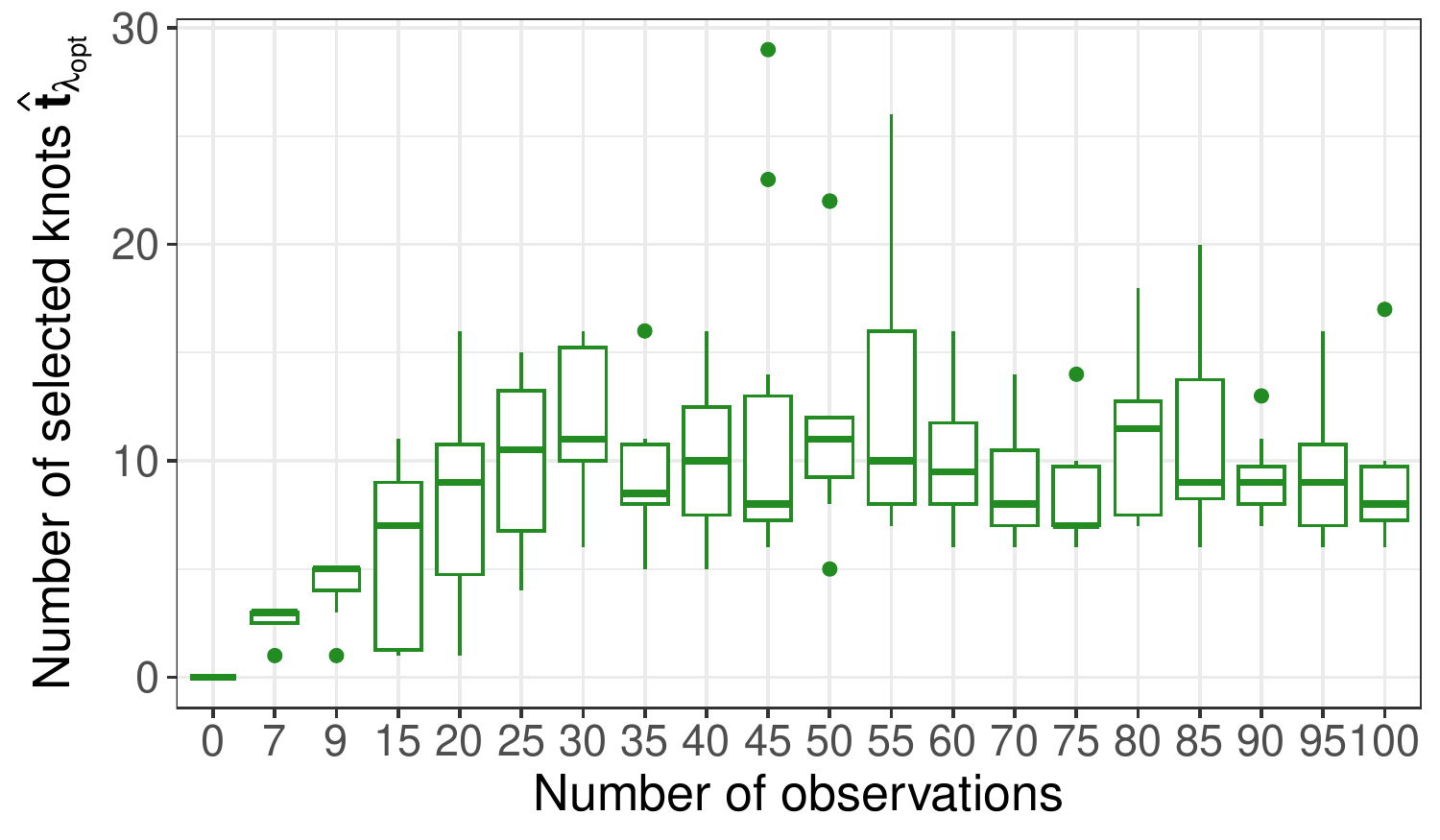}
\caption{Similar to Figure \ref{fig:boxplot_lambda_ebic_f1} by choosing $\uplambda = \uplambda_{\text{opt}}$ for estimating $f_1$.}
\label{fig:boxplot_lambda_opt_f1}
\end{center}
\end{figure}
Figure \ref{fig:boxplot_lambda_ebic_f1} displays the boxplots of the first and second part of the Hausdorff distance and of the number of selected knots $K_\uplambda$ for $\uplambda = \uplambda_\textrm{EBIC}$ obtained from 10 different samplings of $x_1, \ldots, x_n$. The first boxplots are obtained for $n=7$ then new observation points are randomly added to the current observation sets in order to have an increasing number of observations such that $n \leq 100$. We can see from this figure that from $n = 70$ the second part of the Hausdorff distance is close to 0 which means that the estimated knots are near from the real ones. These results are obtained with an almost constant number of selected knots $K_{\uplambda_\textrm{EBIC}} = 6$.

For a comparison purpose, we displayed in Figure \ref{fig:boxplot_lambda_opt_f1} the results obtained for $\uplambda = \uplambda_\textrm{opt}$, where $\uplambda_\textrm{opt}$ is defined by:
\begin{equation*}
\uplambda_\text{opt} = \argmin_{\uplambda \in \Lambda}\left({ \textrm{Normalized sup norm}(\uplambda)}\right),
\label{eq:lambda_opt}
\end{equation*}
with
\begin{equation}\label{eq:norm_sup_norm}
  \textrm{Normalized sup norm}(\uplambda)=\max_{1\leq k\leq N} {\frac{\left|f(x_k)-\widehat{f}_\uplambda(x_k)\right|}{f_{max} - f_{min}}},
\end{equation}
where $\widehat{f}_\uplambda$ is defined in \eqref{eq:estimated_model}.
In \eqref{eq:norm_sup_norm}, $N$ ($N>n$) is the cardinality of the set of evenly-spaced points $\{x_1, \ldots, x_N\}$ of $[0,1]$ which contains the observation points $x_1, \ldots, x_n$ as well as additional points where $f$ has not been observed. Moreover, $f_{min}$ and $f_{max}$ denote the minimum and maximum values of $f$ evaluated on $\{x_1, \ldots, x_N\}$, respectively. We can see from this figure that the performance obtained when $\uplambda$ is optimally chosen is on a par with that of $\uplambda_\textrm{EBIC}$ which means that our procedure for choosing $\uplambda$ is almost optimal. 

\begin{figure}
\begin{center}

\includegraphics[width = 7.5cm]{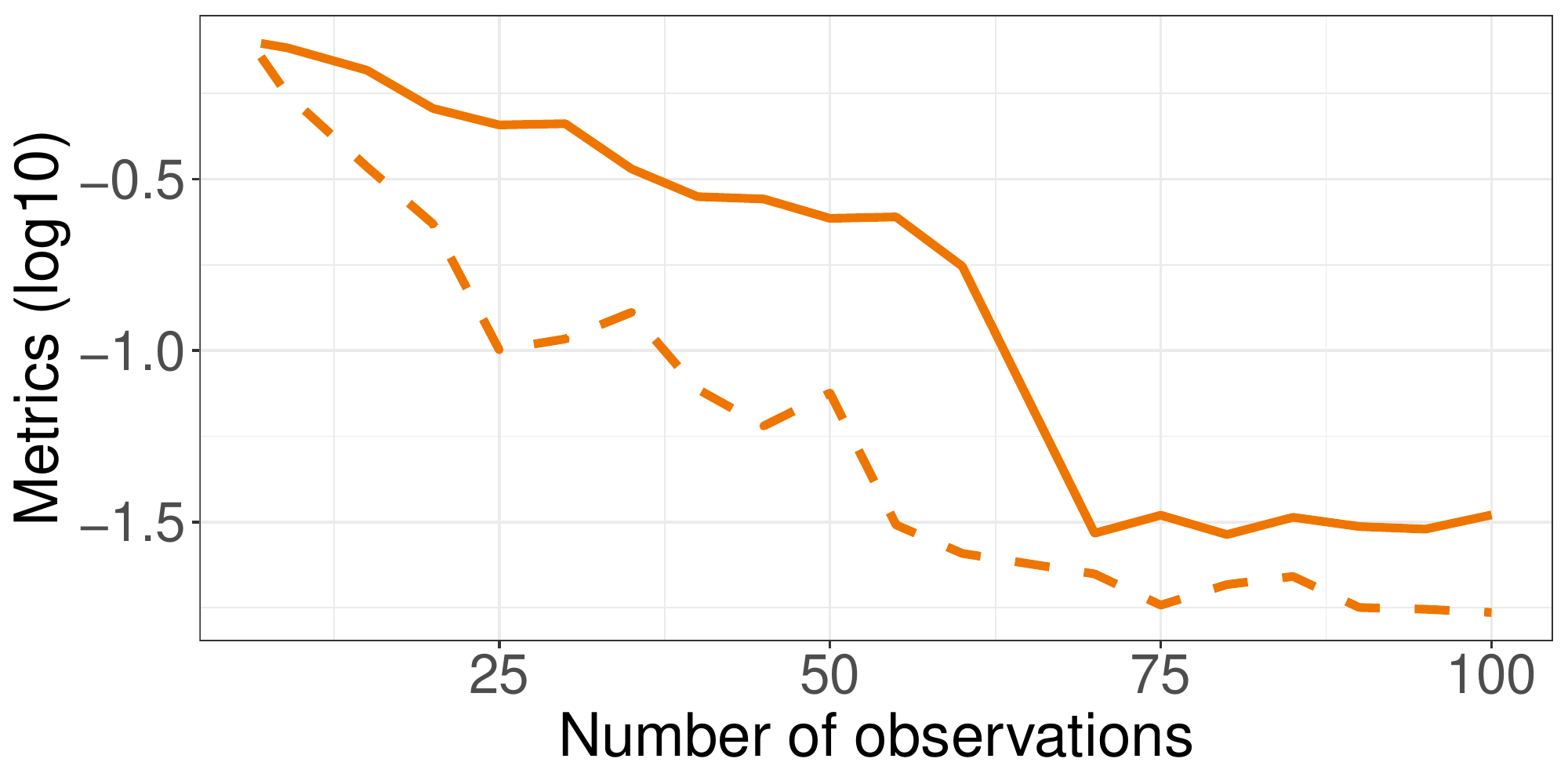}
\includegraphics[width = 7.5cm]{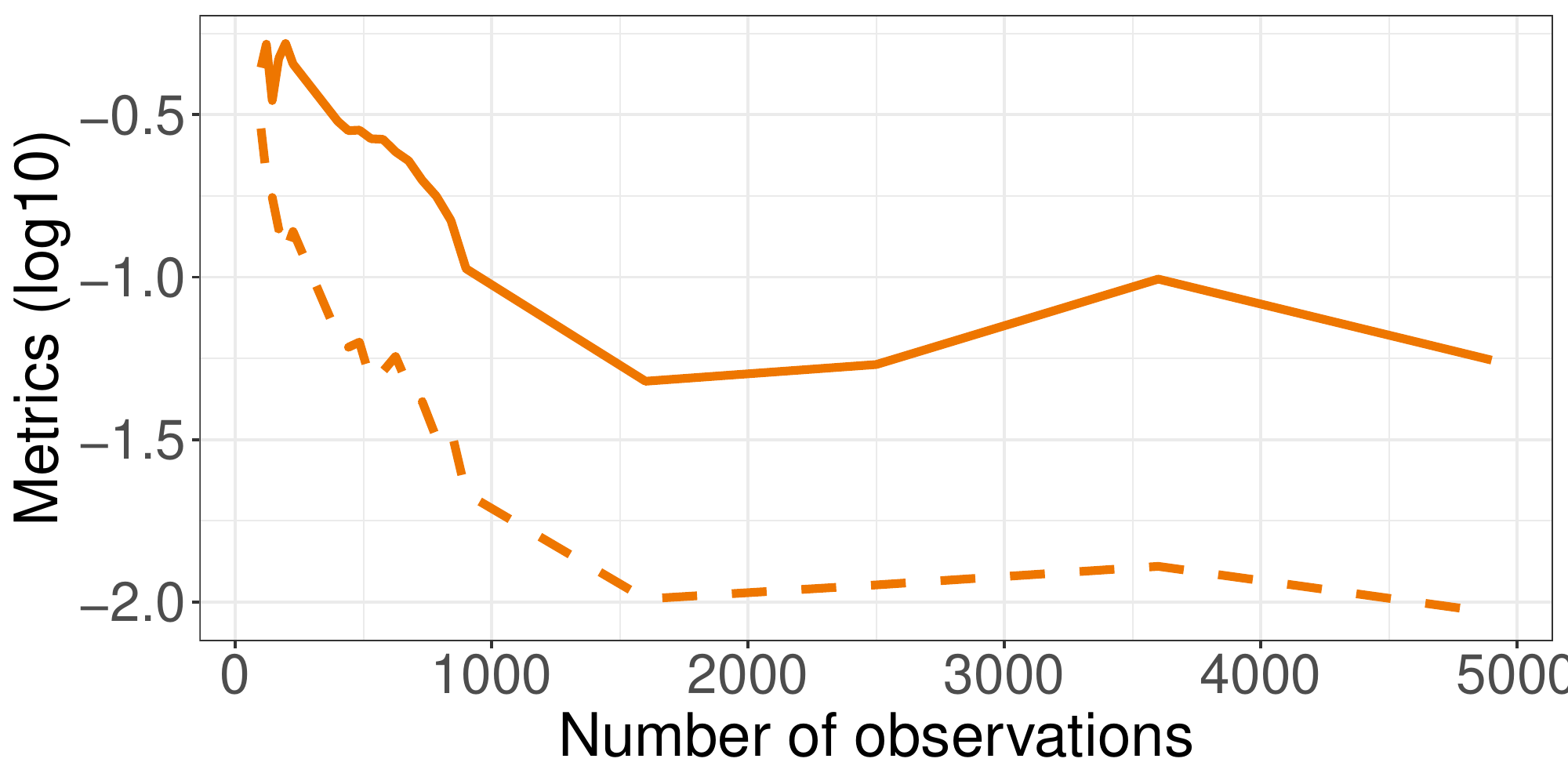}
\caption{Statistical performance (Normalized Sup Norm) of the method using $\uplambda_\text{opt}$ (dashed) and $\uplambda_{\text{EBIC}}$ (solid) for the estimation of $f_1$ (right) and $f_2$ (left) obtained from 10 replications.}
\label{fig:metrics_lambda_f1}
\end{center}
\end{figure}

The corresponding performance is shown \textcolor{black}{on the right part of} Figure \ref{fig:metrics_lambda_f1} for $N=201$. This figure displays the average of the most stringent metric (Normalized Sup Norm) obtained from 10 different samplings of $x_1, \ldots, x_n$ for each $n$. We observed from this figure that the Normalized Sup Norm reaches $10^{-1.75}$ (resp. $10^{-1.5}$) for $\uplambda_\text{opt}$ (resp. $\uplambda_{\text{EBIC}}$) which represents a normalized maximum absolute error of $2\%$ (resp. $3\%$). Once again, these results show that the choice of $\uplambda$ does not alter the performance of our approach.

\subsection{Extension to the two-dimensional case.}\label{sec:two-dim-case}

In this section, we will extend the previous method 
for estimating a two-dimensional function $f$ from  the observations $(Y_i)_{1\leq i \leq n}$ defined in Model \eqref{eq:model} when $d=2$.

Here $\S$ is defined as the cartesian product of two compact sets $\S_1$ and $\S_2$ of $\R$. More precisely, we will consider $(x_{11}, \ldots, x_{1n_1})$ belonging to $\S_1$ and similarly $(x_{21}, \ldots, x_{2n_2})$ belonging to $\S_2$ the $n_1$ and $n_2$ observation values for the first and second variables of $f$ at which $f$ is evaluated, respectively. 
Thus, the set of observations belonging to $\S$ will be defined as: 
\begin{align*}
&\bigl((x_{11},x_{21}), (x_{11},x_{22}), \ldots, (x_{11},x_{2n_2}), (x_{12},x_{21}), (x_{12},x_{22}), \dots, (x_{12},x_{2n_2}), \ldots, (x_{1n_1},x_{2n_2})\bigr) \\ &= \bigl((x_{1k},x_{2\ell})\bigr)_{1 \leq k\leq n_1, 1 \leq \ell \leq n_2}.
\end{align*}

For estimating $f$, we shall approximate it by a linear combination of multidimensional B-splines defined in \cite[p. 162-163]{hastie01statisticallearning} as the tensor product of the B-splines of order $M$ introduced in Section \ref{sec:one_dim_case}. More precisely, $f(x)=f(x_1,x_2)$
will be approximated by:
\begin{equation}
\sum_{i=1}^{Q_1}\sum_{j=1}^{Q_2}\gamma_{ij} \, B_{1,i,M}(x_1)B_{2,j,M}(x_2),
\label{eq:multidim_bsplines}
\end{equation}
where $B_{1,i,M}$ and $B_{2,j,M}$ are the B-spline basis of order $M$ defined in \eqref{eq:Bspline_recurrence} for the first and second dimension, respectively. In \eqref{eq:multidim_bsplines}, $Q_1 = q + K_1 +1 $, $Q_2 = q + K_2 + 1 $ with $K_1$ and $K_2$ the number of knots defined in the B-spline basis of the first and second variables, respectively and $M =q+1$.

\subsubsection{Creation of a candidate set of knots.}\label{sec:knots_2d}

The idea is to consider the two dimensions independently and thus, by fixing one dimension at a time, the problem can be rewritten
as an estimation problem in the one-dimensional framework.

First, we shall consider the knot selection of the B-spline basis of the first dimension by fixing the second dimension to a certain value of $x_{2}$ belonging to $\{x_{21}, \ldots, x_{2n_2}\}$. Thus, we can apply the polynomial trend filtering method described in Section \ref{sec:one_dim_case} and get the grid of penalization parameters $\left(\uplambda_{(1,i),k}\right)_{1 \leq k \leq s_i}$, with $i$ belonging to $\{1, \ldots, n_2\}$ and $s_i$ corresponds to the number of penalization parameters.
The index $1$ in $(1,i)$ denotes the first dimension and $i$ indexes the $i$th value of $\{x_{21}, \ldots, x_{2n_2}\}$. For each value $\uplambda_{(1,i),k}$  with $k$ belonging to $\{1, \ldots, s_i\}$, we can get the corresponding selected knots by following the procedure described in Section \ref{sec:choice_lambda}: after calculating $\a(\uplambda_{(1,i),k})$ as in \eqref{eq:Alambda}, we can determine the set of knots $\widehat{\t}_{1, \uplambda_{(1,i),k}}$ as in \eqref{eq:knot_sequence}. In order to take into account all the information obtained for each value of $x_2$, we gather the selected knots into a single vector depending on the value of the penalization parameters. 
Nevertheless, because not all the vectors $\left(\uplambda_{(1,i),k}\right)_{1 \leq k \leq s_i}$ have exactly the same values $\uplambda_{(1,i),k}$ and the same number of penalization parameters $s_i$ when $i$ varies, we shall define the set of \textit{equivalent regularization parameters} $\widetilde{\Lambda}_{1}$ and the minimal number of penalization parameters $s_{min_1}$: 
\begin{equation}\label{eq:equivalent_set_smin}
\widetilde{\Lambda}_{1} = \left\{\widetilde{\uplambda}_{1,1}, \ldots, \widetilde{\uplambda}_{1, s_{min_1}}\right\} \quad \text{and} \quad s_{min_1} = \min_{1\leq i \leq n_2} s_i \; ,
\end{equation}
where  
\begin{equation}\label{eq:equivalent_lambdas}
\widetilde{\uplambda}_{1,k} = \left(\uplambda_{(1,i),k}\right)_{1\leq i\leq n_2}, \quad 1\leq k \leq s_{min_1}.
\end{equation}

In \eqref{eq:equivalent_lambdas}, $\widetilde{\uplambda}_{1,k}$ can be seen as the vector of parameters which penalize \eqref{eq:genlasso} at an equivalent strength for each fixed value of $x_2$. We can therefore get the vector of selected knots $\widehat{\t}_{1,\widetilde{\uplambda}_{1,k}}$ for the first dimension by grouping together and ordering all the corresponding selected knots of $\widetilde{\uplambda}_{1,k}$. 

We proceed the same way to get the set of equivalent parameters $\widetilde{\Lambda}_{2}$ for the second dimension by fixing this time the value of $x_1$ and with $s_{min_2}$ defined similarly as in \eqref{eq:equivalent_set_smin} for $i$ belonging to $\{1, \ldots, n_1\}$.
Analogously as in  \eqref{eq:equivalent_set_smin} and \eqref{eq:equivalent_lambdas}, we have:
\begin{equation*}\label{eq:equivalent_lambdas_t_dim2}
\widetilde{\Lambda}_{2} = \left\{\widetilde{\uplambda}_{2,1}, \ldots, \widetilde{\uplambda}_{2, s_{min_2}}\right\}  \quad \text{and} \quad \widetilde{\uplambda}_{2,\ell} = \left(\uplambda_{(2,i),\ell}\right)_{1\leq i\leq n_1}, \quad 1\leq \ell \leq s_{min_2}.
\end{equation*}
Moreover, as well as for the first dimension, the vector of selected knots for the second dimension for each $\widetilde{\uplambda}_{2,\ell}$ is defined as  $\widehat{\t}_{2,\widetilde{\uplambda}_{2, \ell}}$, $\ell$ belonging to $\{1, \ldots, s_{min_2}\}$ .

In the following, let us consider two generic penalization parameters $\widetilde{\uplambda}_{1}$ belonging to $\widetilde{\Lambda}_{1}$ and  $\widetilde{\uplambda}_{2}$ belonging to $\widetilde{\Lambda}_{2}$. 
Thus, we can define the candidate sets of knots  for both dimensions $\widehat{\t}_{1,\widetilde{\uplambda}_1}$ and $\widehat{\t}_{2,\widetilde{\uplambda}_2}$. 
We must now determine which combination of penalization parameters $\widetilde{\uplambda}_{1}$ and $\widetilde{\uplambda}_{2}$ and hence, which combination of selected knots for the first and second dimension, allows us to get an optimal estimator of $f$.

\subsubsection{Choice of the penalization parameters of the regularized method.}

In order to choose the penalization parameters leading to the best selection of knots, we consider the following EBIC criterion which can be seen as the adaptation to the two-dimensional case of the one defined in \eqref{eq:EBIC}:
\begin{equation}
\text{EBIC}\left(\widetilde{\uplambda}_1, \widetilde{\uplambda}_2\right)= \text{SS}\left(\widetilde{\uplambda}_1, \widetilde{\uplambda}_2\right) + \widetilde{Q}_1\widetilde{Q}_2\log{n} + 2\log{(q+n_1+1)(q+n_2+1) \choose \widetilde{Q}_1\widetilde{Q}_2}.
\label{eq:EBIC_2D}
\end{equation}
where $\widetilde{Q}_1 = q + K_{\widetilde{\uplambda}_1} + 1$ and $\widetilde{Q}_2 = q + K_{\widetilde{\uplambda}_2} + 1$, $K_{\widetilde{\uplambda}_1}$ and $K_{\widetilde{\uplambda}_2}$ being the number of selected knots with the parameters $\widetilde{\uplambda}_1$ and $\widetilde{\uplambda}_2$ for the first and second dimension, respectively.

In \eqref{eq:EBIC_2D}, $\text{SS}\left(\widetilde{\uplambda}_1, \widetilde{\uplambda}_2\right)$ is defined as: 
\begin{equation*}
\text{SS}\left(\widetilde{\uplambda}_1, \widetilde{\uplambda}_2\right)= \Big\|\Y - \widehat{\Y}\left(\widetilde{\uplambda}_1, \widetilde{\uplambda}_2\right)\Big\|_2^2 \, ,
\label{eq:SS_lambda_2d}
\end{equation*}
where 
\begin{equation}
\widehat{\Y}\left(\widetilde{\uplambda}_1, \widetilde{\uplambda}_2\right) = \B\left(\widetilde{\uplambda}_1, \widetilde{\uplambda}_2\right) \widehat{\gammaM},
\label{eq:Bspline_regression_2d}
\end{equation}
and $\B\left(\widetilde{\uplambda}_1, \widetilde{\uplambda}_2\right)$ is defined as:
\begin{equation}\label{eq:Blambda_2D}
\B\left(\widetilde{\uplambda}_1, \widetilde{\uplambda}_2\right) = \B\left(\widetilde{\uplambda}_1\right) \otimes \B\left(\widetilde{\uplambda}_2\right),
\end{equation}
$E \otimes F$ denoting the Kronecker product of the matrices $E$ and $F$. In \eqref{eq:Blambda_2D}, $\B\left(\widetilde{\uplambda}_1\right)$ is a $n_1\times \widetilde{Q}_1$ matrix having as $i$th column $\left(B_{1,i,M}(x_{1k})\right)_{1\leq k \leq n_1}$, $i$ belonging to $\{1, \ldots, \widetilde{Q}_1\}$ and $\B\left(\widetilde{\uplambda}_2\right)$ is a $n_2\times \widetilde{Q}_2$ matrix having as $j$th column $\left(B_{2,j,M}(x_{2\ell})\right)_{1\leq \ell \leq n_2}$, $j$ belonging to $\{1, \ldots, \widetilde{Q}_2\}$.

In \eqref{eq:Bspline_regression_2d}, $\widehat{\gammaM} = (\widehat{\gamma}_{ij})_{1\leq i \leq \widetilde{Q}_1, 1\leq j \leq \widetilde{Q}_2}$ is obtained using the following least-square criterion:
\begin{equation}
\widehat{\gammaM} = \argmin_{\gammaM \in \R^{\widetilde{Q}_1\widetilde{Q}_2}}{\Big\|\Y - \B\left(\widetilde{\uplambda}_1, \widetilde{\uplambda}_2\right)\gammaM\Big\|_2^2} \,.
\label{eq:least_square_criterion_2D} 
\end{equation}

As in Equation \eqref{eq:lambda_EBIC}, we shall define $\widetilde{\uplambda}_{1, \text{EBIC}}$ and $\widetilde{\uplambda}_{2, \text{EBIC}}$ the penalization parameters which verify:
\begin{equation*}
\left(\widetilde{\uplambda}_{1, \text{EBIC}},\widetilde{\uplambda}_{2, \text{EBIC}}\right)  = \argmin_{\widetilde{\uplambda}_1 \in \widetilde{\Lambda}_{1}, \,\widetilde{\uplambda}_2 \in \widetilde{\Lambda}_{2}}\left\{\text{EBIC}\left(\widetilde{\uplambda}_1, \widetilde{\uplambda}_2\right)\right\}.
\label{eq:lambda_EBIC_2D}
\end{equation*}

Hence, the final estimator of $f$ is defined as: 
\begin{equation*}
\widehat{f}(x_1, x_2) = \widehat{f}_{\widetilde{\uplambda}_{1, \text{EBIC}},\widetilde{\uplambda}_{2, \text{EBIC}}}(x_1, x_2),
\label{eq:estimated_model_ebic_2D}
\end{equation*}
with $\widehat{f}_{\widetilde{\uplambda}_{1}, \widetilde{\uplambda}_{2}}$ defined as: 
\begin{equation}
\widehat{f}_{\widetilde{\uplambda}_{1}, \widetilde{\uplambda}_{2}}(x)= \widehat{f}_{\widetilde{\uplambda}_{1}, \widetilde{\uplambda}_{2}}(x_1, x_2)= \sum_{i=1}^{\widetilde{Q}_1}\sum_{j=1}^{\widetilde{Q}_2}\widehat{\gamma}_{ij} \, B_{1,i,M}(x_1)B_{2,j,M}(x_2).
\label{eq:estimated_model_2d}
\end{equation}
In \eqref{eq:estimated_model_2d},  $\widehat{\gammaM} = (\widehat{\gamma}_{ij})_{1\leq i \leq \widetilde{Q}_1, 1\leq j \leq \widetilde{Q}_2}$ is obtained as in \eqref{eq:least_square_criterion_2D}.

\subsubsection{Illustration of our method on a simple case.}
\begin{figure}[ht]
\begin{center}
\includegraphics[width=7cm, trim=0 1.2cm 0 0.5cm, clip]{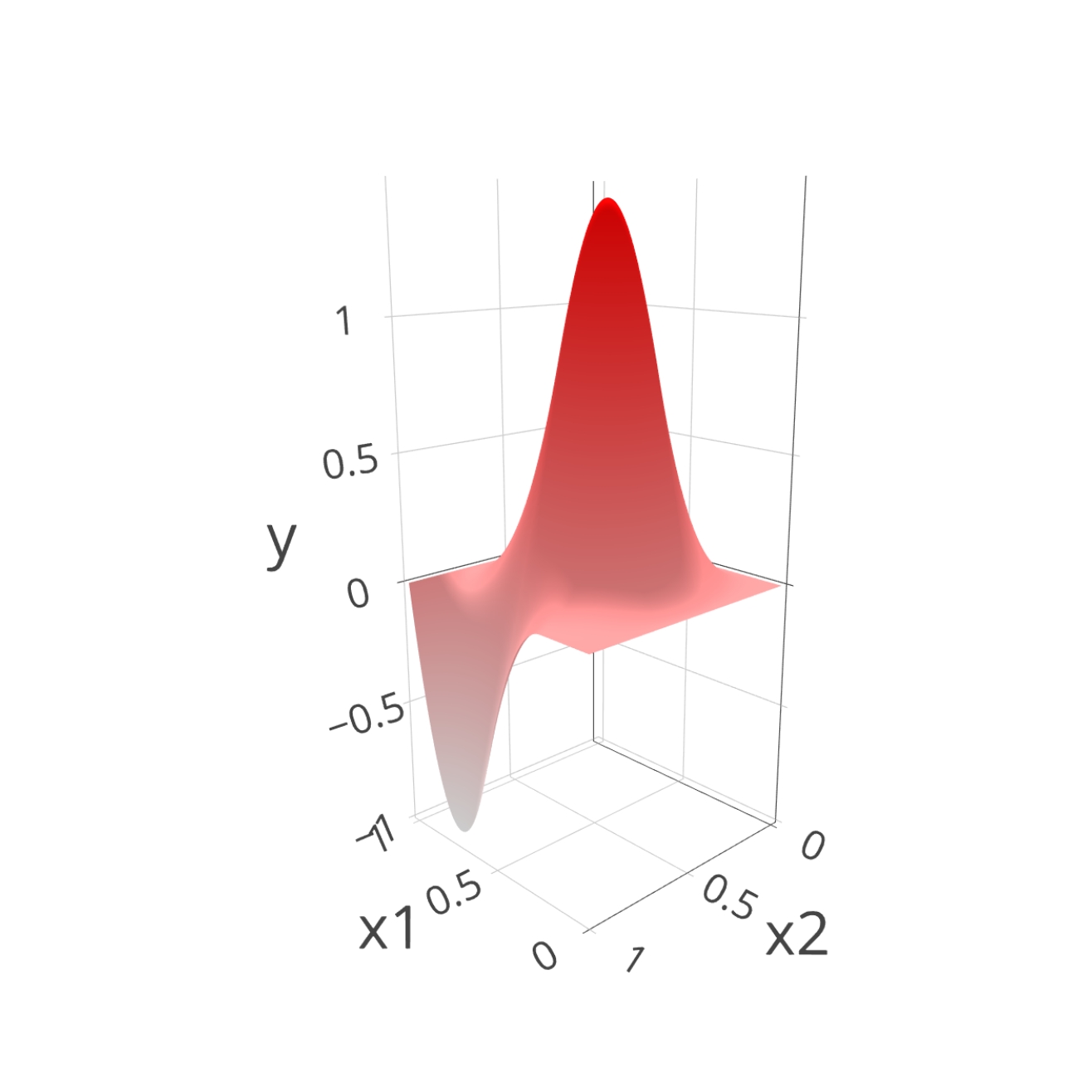}
\includegraphics[width = 7cm, trim=0 1.2cm 0 0.5cm, clip]{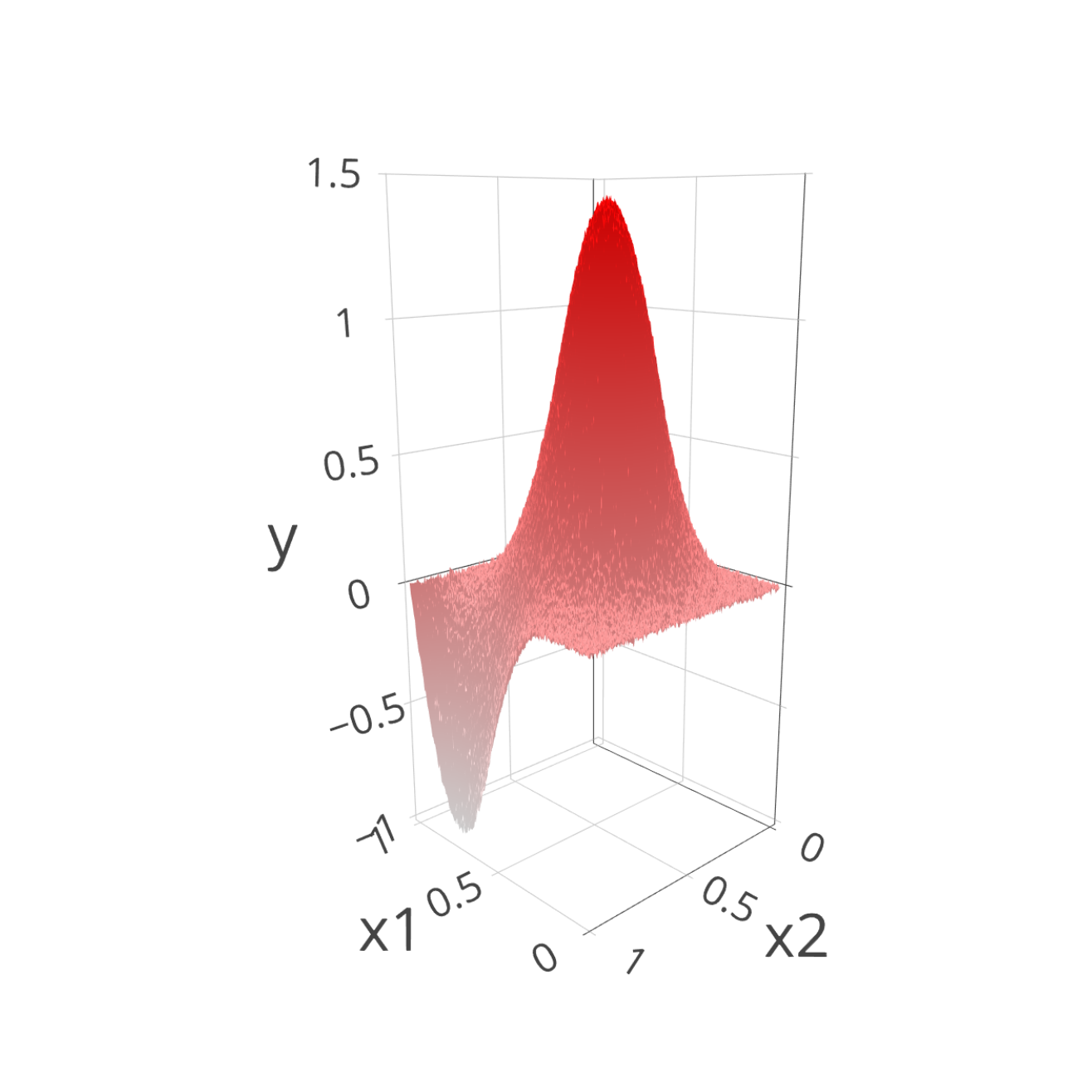}
\caption{Function $f_2$ to estimate (left) and a noisy set of observations $Y_1,\dots,Y_{40401}$ with $\sigma = 0.01$ (right).}
\label{fig:functions_test_2D}
\end{center}
\end{figure}

To illustrate the extension of our method to the two-dimensional case, we propose estimating a function $f= f_2$ which is  a linear combination of tensor product of quadratic B-splines ($M=3$) with $\t_1 = (0.24, 0.545)$  and $\t_2 = (0.395, 0.645)$:
\begin{equation}\label{eq:f2}
f_2(x_1, x_2) = 2.3B_{1,3,3}(x_1)B_{2,3,3}(x_2) - 1.5B_{1,4,3}(x_1)B_{2,5,3}(x_2), \quad (x_1, x_2) \in [0,1]^2, 
\end{equation}
where $B_{i,j,M}$ is defined in \eqref{eq:multidim_bsplines} with $\t_1$ and $\t_2$ the knots involved in the definition of $B_{1,j,M}$ and $B_{2,j,M}$, respectively.
We shall apply our method to a noisy set of observations $\Y = (Y_1, \ldots, Y_n)$ where the $Y_i$ are defined in \eqref{eq:model} and the $\varepsilon_i$ are i.i.d Gaussian centered random variables with $\sigma = 0.01$. The set of knots $\t_1$ and $\t_2$ are a part of the observation set $\{x_{11}, \ldots, x_{1n_1}\}$ and $\{x_{21}, \ldots, x_{2n_2}\}$, respectively.  The corresponding $(Y_i)_{1\leq i\leq n}$ (resp. $(f_2(x_{1,k},x_{2,\ell}))_{1\leq k\leq n_1,1\leq \ell\leq n_2}$) are displayed in the right (resp. left) part of Figure \ref{fig:functions_test_2D} for $n=n_1n_2= 201^2 = 40401$.

In order to assess the performance of our knot selection procedure, we shall use the Hausdorff distance defined in \eqref{eq:hausdorff_dist} for each dimension independently. The results for the first and second part of the Hausdorff distance and the number of selected knots for both dimensions are displayed in the boxplots of Figure \ref{fig:boxplot_ebic_f2} of the Appendix for $\widetilde{\uplambda}_1 = \widetilde{\uplambda}_{1,\text{EBIC}}$ and $\widetilde{\uplambda}_2 = \widetilde{\uplambda}_{2,\text{EBIC}}$ and from 10 different samplings of $x_{11}, \ldots, x_{1n_1}$ and  $x_{21}, \ldots, x_{2n_2}$. New observation points are then randomly added to the current observation sets in order to have an increasing number of observations. We can see from this figure that from $n = 1600$ and so from $40$ observation points by dimension, the second part of the Hausdorff distance is close to 0 which means that the estimated knots are near from the real ones. 
The numbers of selected knots required to get these results are between 5 and 10. 
Similarly as for the one-dimensional case, we compare these results with those obtained for $\widetilde{\uplambda}_1 = \widetilde{\uplambda}_{1,\text{opt}}$ and $\widetilde{\uplambda}_2 = \widetilde{\uplambda}_{2,\text{opt}}$, two optimal parameters defined as:
\begin{equation*}
\left(\widetilde{\uplambda}_{1, \text{opt}},\widetilde{\uplambda}_{2, \text{opt}}\right)  = \argmin_{\widetilde{\uplambda}_{1} \in \widetilde{\Lambda}_{1}, \,\widetilde{\uplambda}_{2} \in \widetilde{\Lambda}_{2}}\left\{\text{Normalized sup norm}\left(\widetilde{\uplambda}_{1}, \widetilde{\uplambda}_{2}\right)\right\}
\end{equation*}
with Normalized sup norm being defined in \eqref{eq:norm_sup_norm} depending here on the values of $\widetilde{\uplambda}_1$ and $\widetilde{\uplambda}_2$, with $\widehat{f}_\uplambda$ becoming $\widehat{f}_{\widetilde{\uplambda}_1, \widetilde{\uplambda}_2}$ defined in \eqref{eq:estimated_model_2d} and  $x_k$ belonging to the set:
\begin{align}\label{eq:x1_xN}
& \{x_1, \ldots, x_N\} = \nonumber\\ & \Bigl\{(x_{11},x_{21}), (x_{11},x_{22}), \ldots, (x_{11},x_{2N_2}), (x_{12},x_{21}), (x_{12},x_{22}), \dots, (x_{12},x_{2N_2}), \ldots, (x_{1N_1},x_{2N_2)}\Bigr\}.
\end{align}
%
$N$ is the cardinality of $\{x_1, \ldots, x_N\}$ and is such that $N = N_1N_2$. We can see from Figure \ref{fig:boxplot_lambda_opt_f2} of the Appendix, where the results are displayed, that they are comparable to those found for $\widetilde{\uplambda}_{1,\text{EBIC}}$ and $\widetilde{\uplambda}_{2,\text{EBIC}}$.
This means that our choice of the penalization parameters does not alter the performance of our approach.
The corresponding performance is shown \textcolor{black}{on the right part} of Figure \ref{fig:metrics_lambda_f1} for $N=40401$ and from 10 different samplings of $x_{11}, \ldots, x_{1n_1}$ and  $x_{21}, \ldots, x_{2n_2}$. 
The most stringent metric (Normalized Sup Norm) reaches $10^{-2}$ (resp. $10^{-1.4}$) for $\uplambda_{1,\text{opt}}$ and  $\uplambda_{2,\text{opt}}$ (resp. $\uplambda_{1, \text{EBIC}}$ and  $\uplambda_{2, \text{EBIC}}$) which represents a normalized maximum absolute error of $1\%$ (resp. $4\%$). Once again, these results show that the choice of  $\widetilde{\uplambda}_1$ and $\widetilde{\uplambda}_2$ does not alter the performance of our approach. We can see for both penalization parameters, the optimal and the ones from the EBIC criterion, that the performance reaches a plateau from $n=1600$ for $K_{\widetilde{\uplambda}_{1,\text{opt}}} =  K_{\widetilde{\uplambda}_{1,\text{EBIC}}} = 6$ and $K_{\widetilde{\uplambda}_{2,\text{opt}}} = K_{\widetilde{\uplambda}_{2,\text{EBIC}}} = 9$, which is on a par with what has been found for the number of selected knots for the one-dimensional case in Figures \ref{fig:boxplot_lambda_ebic_f1} and \ref{fig:boxplot_lambda_opt_f1} for $n\geq40$.


%
%


\section{Numerical experiments}\label{sec:numexp}

In this section, we will study the behavior of our method using the EBIC criterion called GLOBER for Generalized LassO for knot selection in multivariate B-splinE Regression and implemented in the \texttt{glober} R package when the variance of the noise $\sigma^2$ increases
  and when the observation set changes. 

To assess the efficiency of our method, we will compare it to state-of-the-art approaches: Gaussian Processes (GP) described in \cite{rasmussen2006gp} and implemented in the Python package \texttt{scikit-learn}, Multivariate Adaptive Spline Regression (MARS) introduced in \cite{friedman1991mars} and implemented in the R package \texttt{earth} and Deep Neural Networks (DNN) implemented in the R package \texttt{keras}.

For the GP, we chose the squared exponential covariance function as defined in \cite{savino2022active}. For the MARS approach, we used the default settings proposed in the \texttt{earth} package. It has to be noticed that for the two-dimensional case the interaction terms are included in the model in order not to penalize it. 
The architecture of the DNN was chosen arbitrarily since our goal is not to optimize it in this paper. More precisely, we used a 2-hidden-layered structure composed of 10 neurons per layer. The activation function of the hidden layers was the RELU function since it is one of the most used functions. In order to train this DNN, we used the stochastic gradient descent method Adam as the optimizer and the Mean Squared Error (MSE) as the loss function. According to the analysis of loss function curves during a pre-processing step, we trained our DNN over 300 epochs for functions of $d=1$ and 50 epochs for functions of $d=2$ to avoid overfitting.


\subsection{Influence of $\sigma$ on the statistical performance of the method}

We first investigate the influence of the level of noise on the performance of GLOBER.
To do so, we applied our method to observations corrupted with two different levels of noise and we computed the average Normalized Sup Norm in \eqref{eq:norm_sup_norm}, for 10 different samplings of the observations. In both cases ($d=1$ or 2), the set of knots used to define the underlying function to estimate belongs to the observation set.
\subsubsection{One-dimensional case ($d=1$)} 
\begin{figure}[ht]
\begin{center}
\includegraphics[width=6cm]{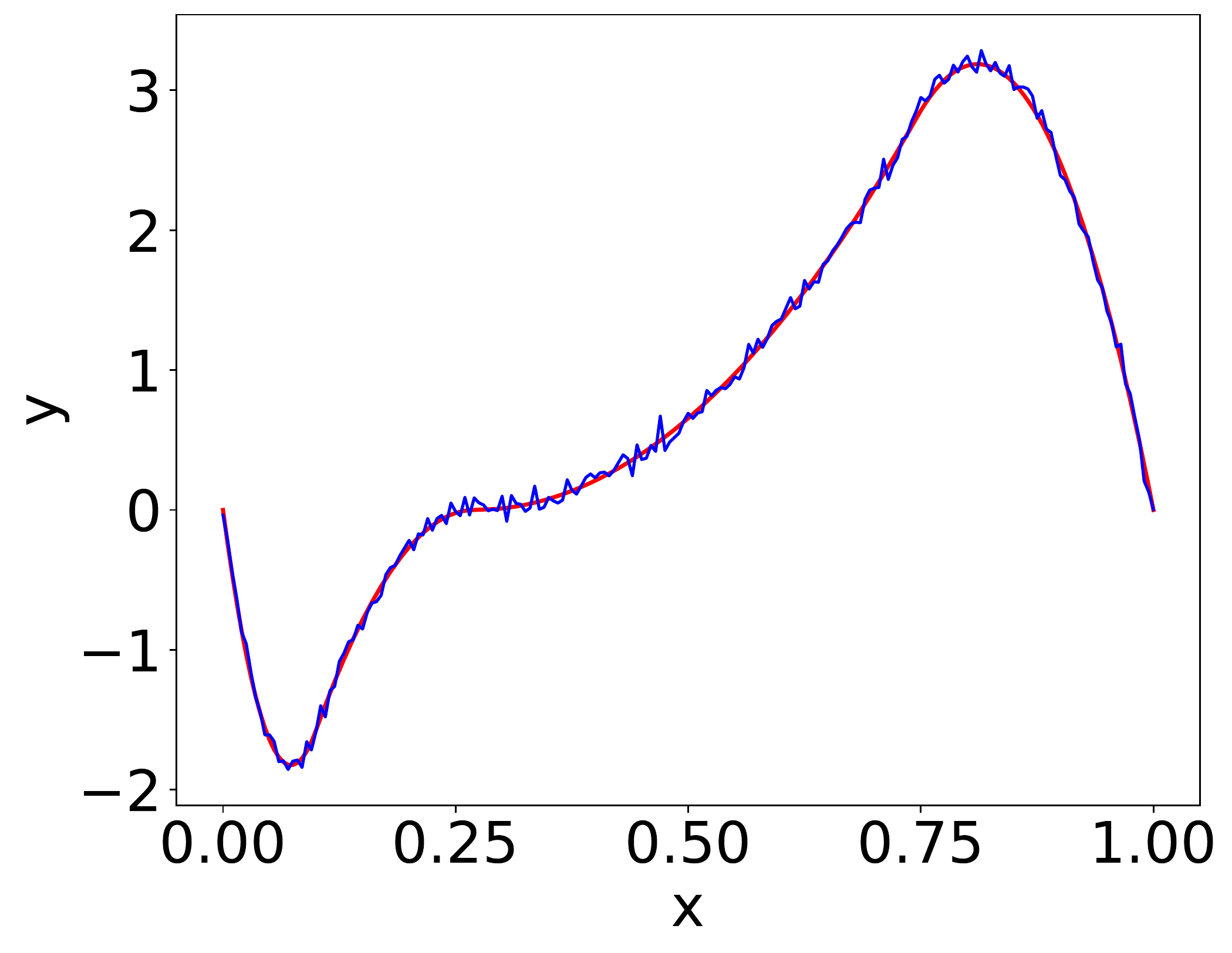}
\includegraphics[width=6cm]{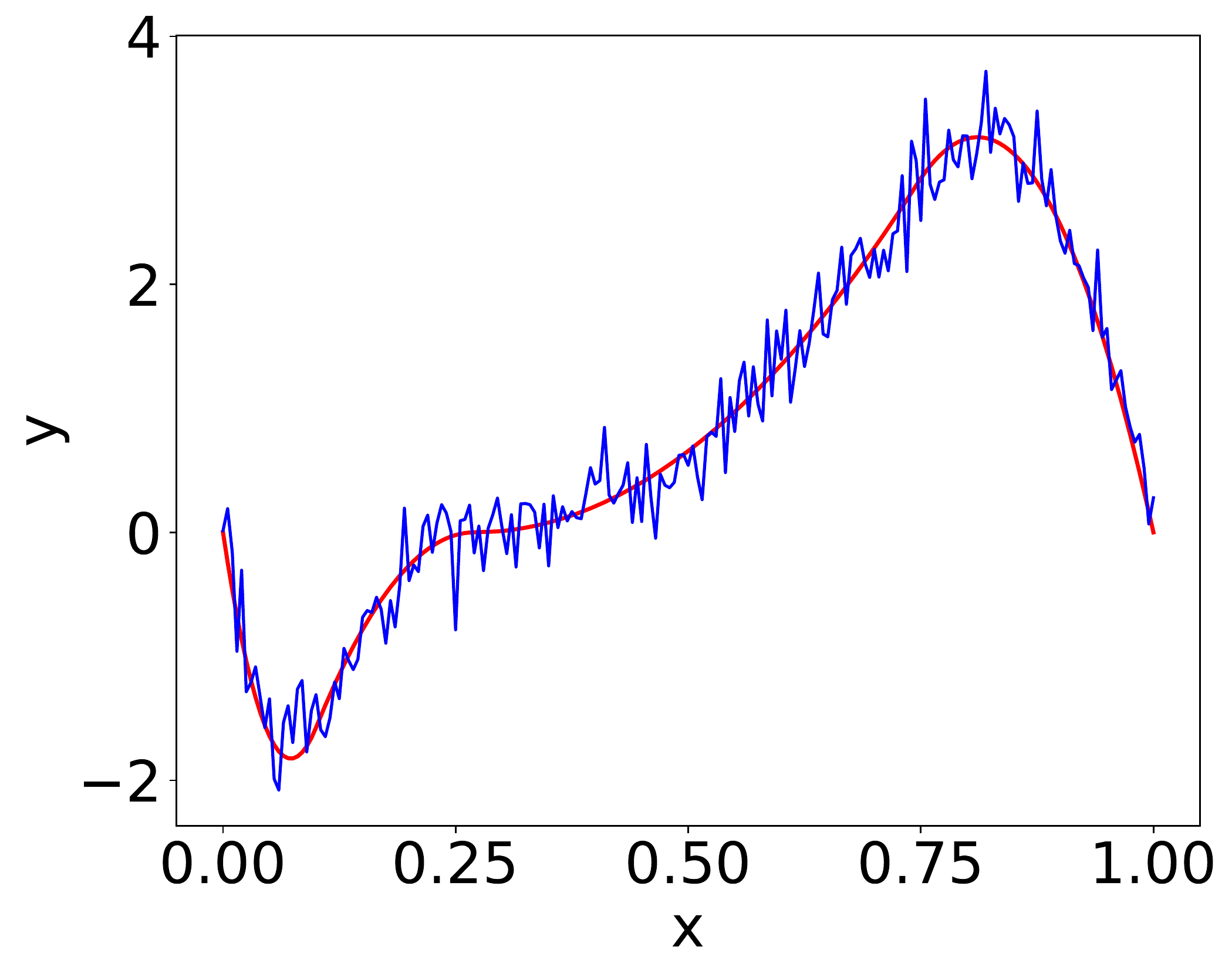}
\caption{Function $f_1$ to estimate with a noisy set of observations $Y_1,\dots,Y_{201}$ of $\sigma = 0.05$ (left) and $\sigma = 0.25$ (right).}\label{fig:illustration_sigma_1D}
\end{center}
\end{figure}

We first study the estimation of the function $f_1$ defined in (\ref{eq:f1}) from a noisy set of observations. The corresponding $(Y_i)_{1\leq i\leq n}$ for $\sigma = 0.05$ (resp. $\sigma = 0.25$) and $n=201$ are displayed in the left (resp. right) part of Figure \ref{fig:illustration_sigma_1D}.

\begin{figure}[ht]
\begin{center}

\includegraphics[width = 16cm]{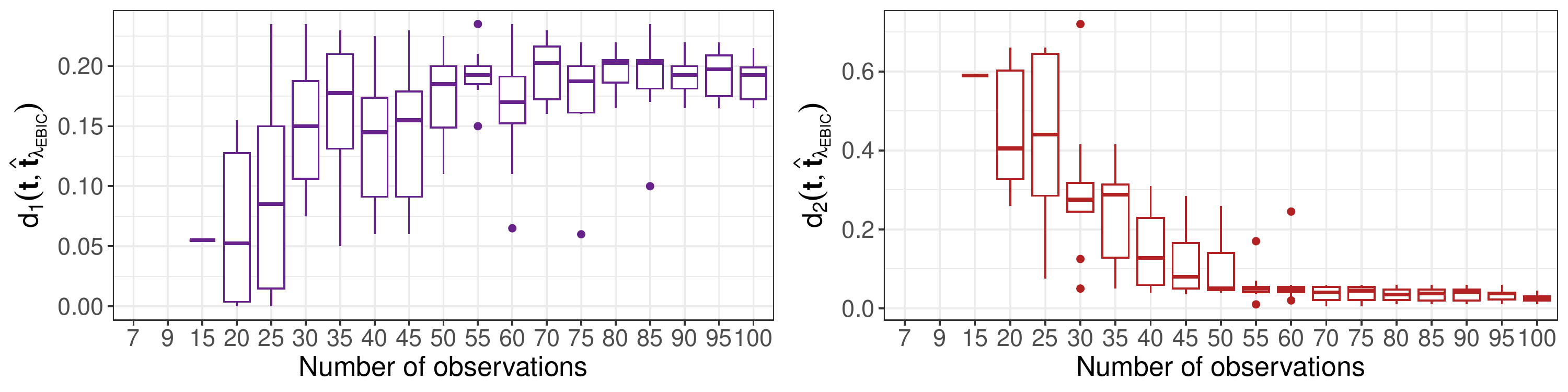}
\includegraphics[width = 8cm]{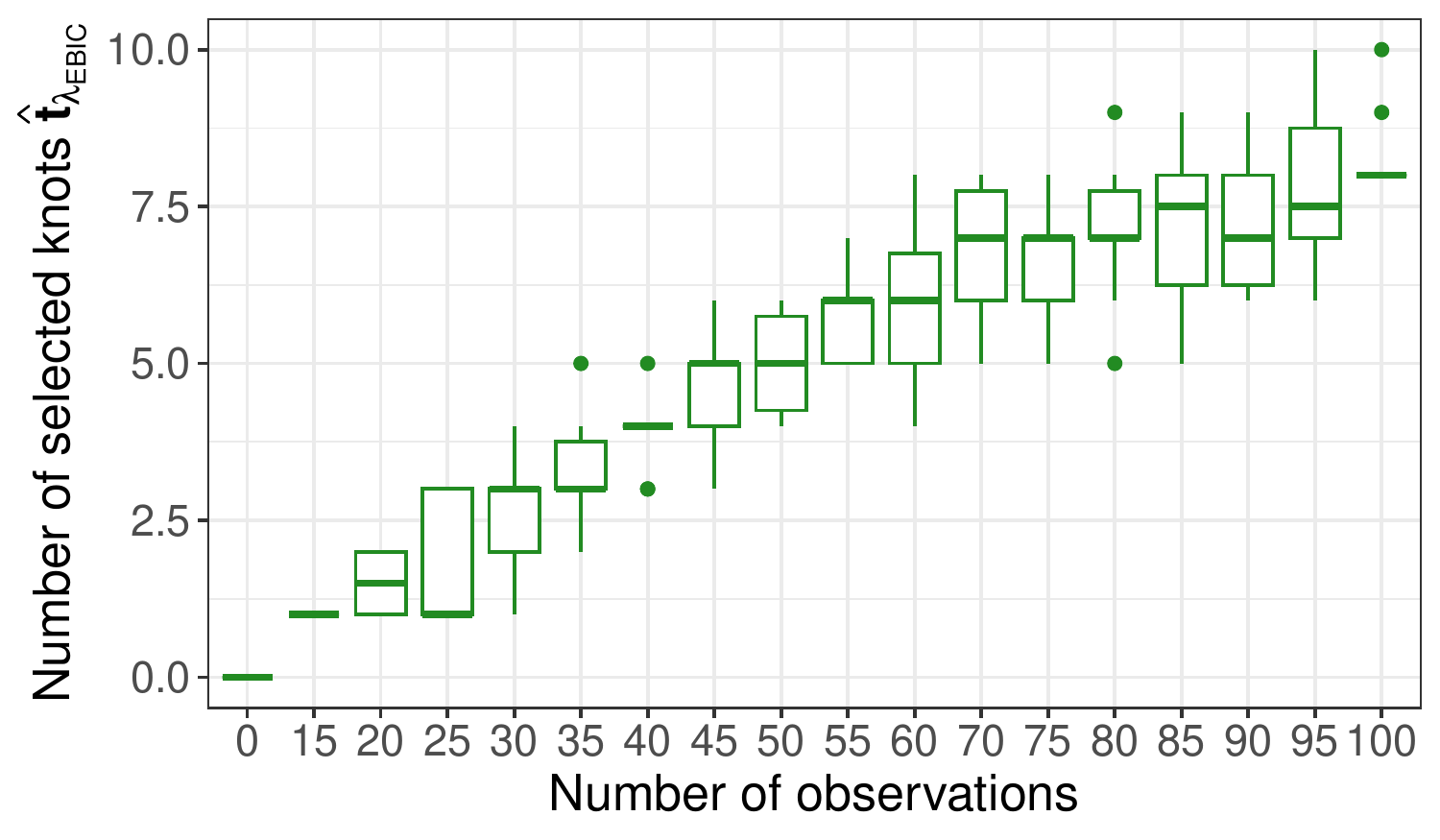}
\caption{Similar to Figure \ref{fig:boxplot_lambda_ebic_f1} with $\uplambda = \uplambda_{\text{EBIC}}$ for estimating $f_1$ when $\sigma=0.25$.}
\label{fig:boxplot_f1_sigma_25}
\end{center}
\end{figure}

The two parts of the Hausdorff distance between the real knots and the estimated ones as well as the number of selected knots obtained for the noisiest observation set ($\sigma = 0.25$) are displayed in Figure \ref{fig:boxplot_f1_sigma_25}. We can see from these results that even the highest value of $\sigma$ ($\sigma= 0.25$)  does not alter the second part of the Hausdorff distance $d_2$ and that the number of selected knots remains the same as the one previously found for $\sigma=0.1$. 

\begin{figure}[ht]
\begin{center}

\includegraphics[width=0.48\textwidth,height=4cm]{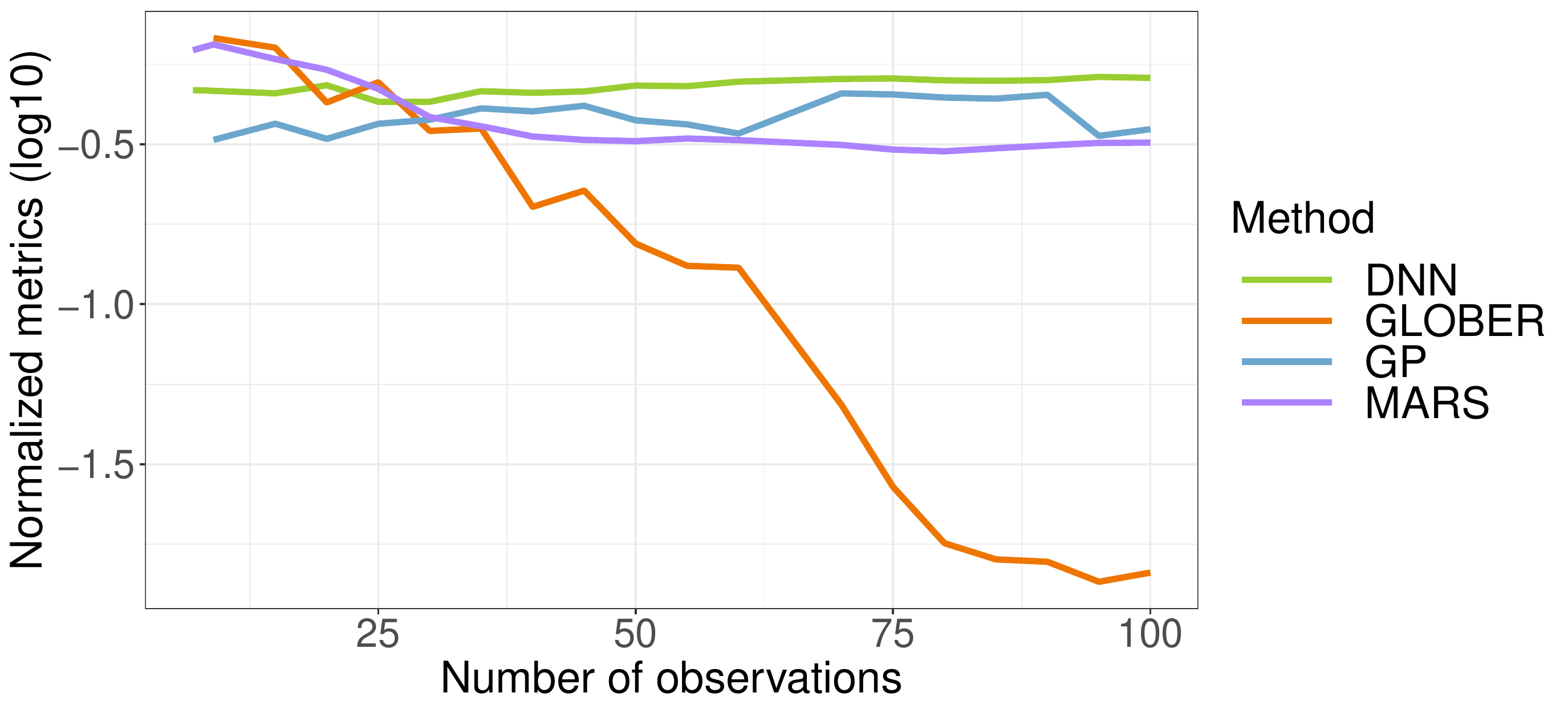}
\includegraphics[width=0.48\textwidth,height=4cm]{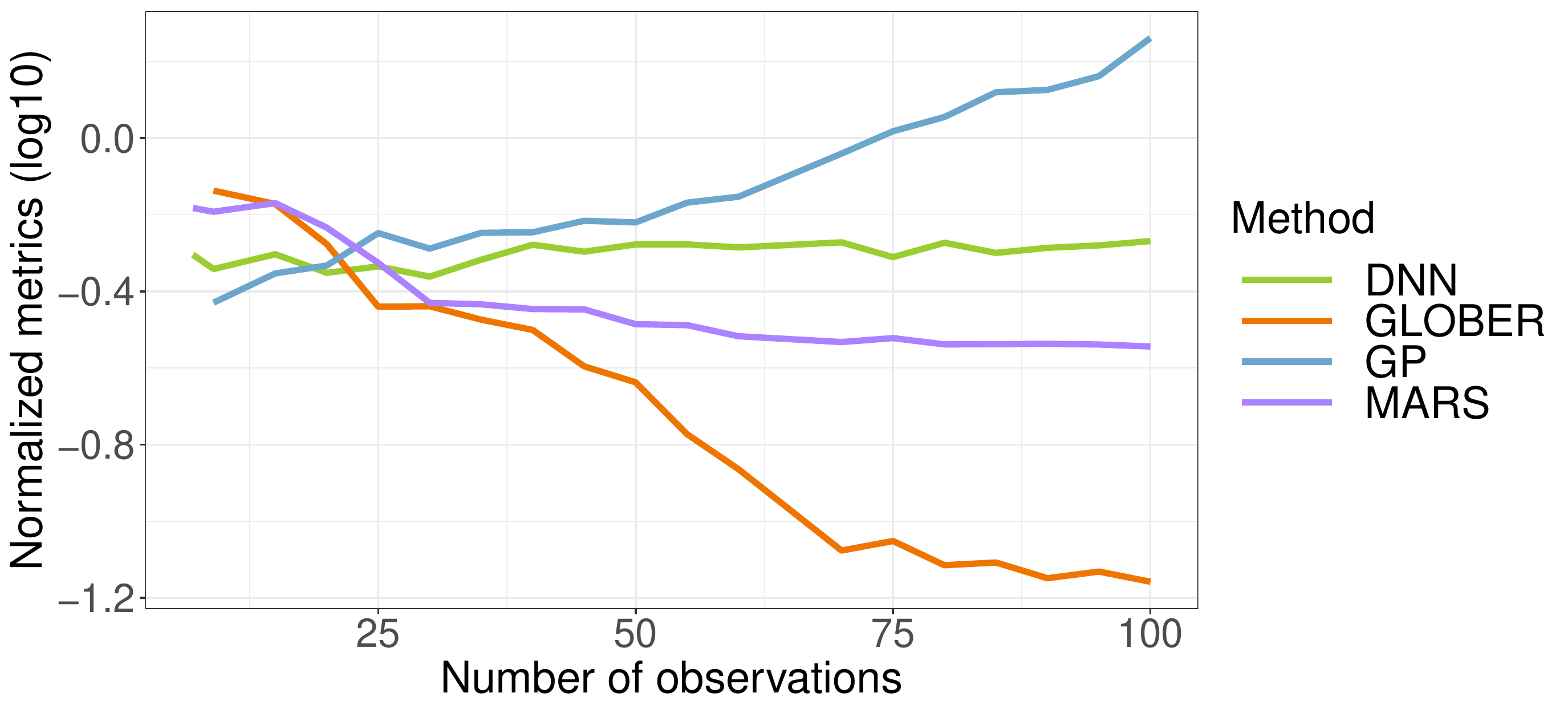}
\caption{Statistical performance (Normalized Sup Norm) of GLOBER for estimating $f_1$ when $\sigma = 0.05$ (left) and $\sigma = 0.25$ (right) and of the state-of-the-art methods obtained from 10 replications.}\label{fig:sigma_val_1D}

\end{center}
\end{figure}

The corresponding results for the statistical performance defined in (\ref{eq:norm_sup_norm}) for $n$ varying from 7 to 100 are displayed in Figure \ref{fig:sigma_val_1D} for $N=201$.
We can see from this figure that the level of noise deteriorates the performance of every method. However, our approach still has high levels of precision since the Normalized Sup Norm varies from $10^{-1.75}$ to $10^{-1.25}$ for $n=100$ which allows it to outperform the other methods.

\subsubsection{Two-dimensional case ($d=2$)}
\begin{figure}[ht]
\begin{center}
\includegraphics[width=7cm, trim=0 1.2cm 0 0.5cm, clip]{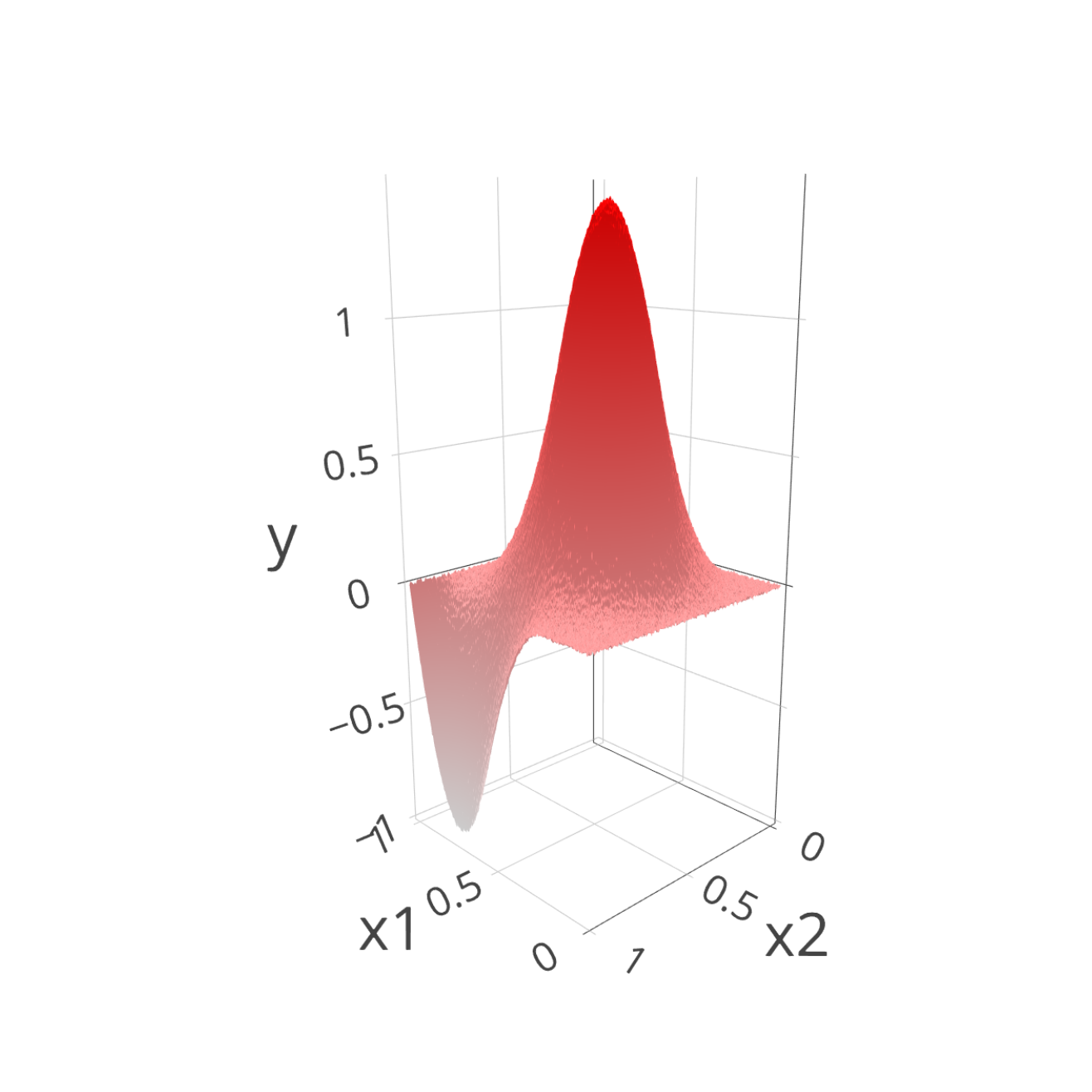}
\includegraphics[width=7cm,  trim=0 1.2cm 0 0.5cm, clip]{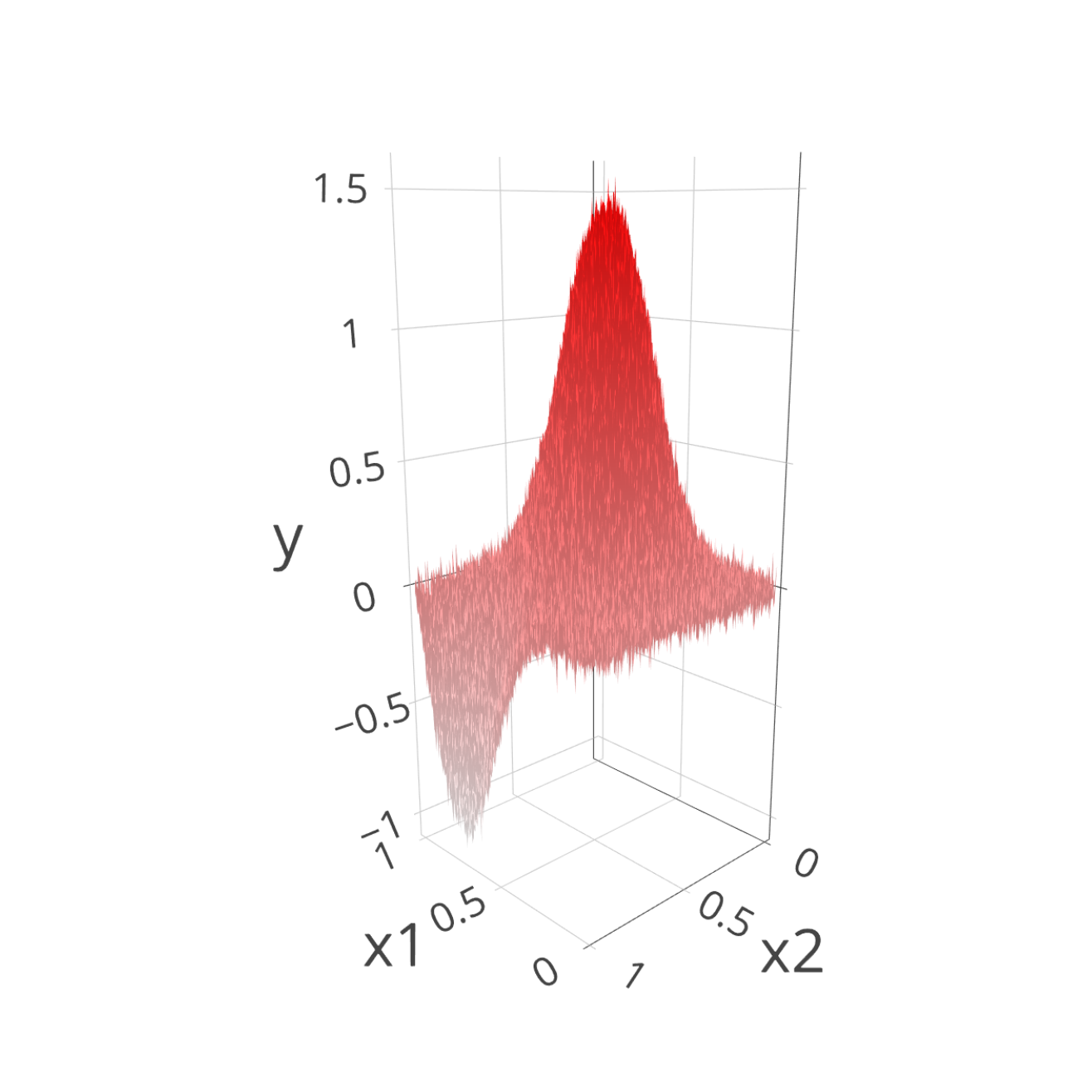}
\caption{Function $f_2$ to estimate with a noisy set of observations $Y_1,\dots,Y_{40401}$ with $\sigma = 0.005$ (left) and $\sigma = 0.05$ (right).}\label{fig:illustration_sigma_2D}

\end{center}
\end{figure}

In this part, we focus on the estimation of the function $f_2$ for $d=2$ from a noisy set of observations $(Y_i)_{1\leq i\leq n}$ obtained with $\sigma = 0.005$ (resp. $\sigma = 0.05$) and $n=40401$. The corresponding $Y_i$s are displayed in the left (resp. right) part of Figure \ref{fig:illustration_sigma_2D}. 

The two parts of the Hausdorff distance between the estimated and the real knots as well as the number of selected knots obtained for the noisiest observation set ($\sigma=0.05$) are shown in Figure \ref{fig:boxplot_f2_sigma_05} of the Appendix. As for the one-dimensional case, we can see that the results are similar to those previously obtained with $\sigma=0.01$ in Figure \ref{fig:boxplot_lambda_opt_f2} of the Appendix: the number of selected knots is the same and the second part of the Hausdorff distance tends to 0.

Figure \ref{fig:sigma_val_2D} of the Appendix displays the average of the statistical performance obtained from 10 random samplings of the set of observations for $N=40401$ where the statistical measure is defined in (\ref{eq:norm_sup_norm}). We can see that even though an alteration of the performance is visible, our approach still outperforms the other methods since the normalized metric keeps decreasing, contrary to the DNNs and the MARS approaches which seem to reach a plateau and the Gaussian Processes which led to very poor accuracy on noisy observations. 
Once again, our method remains robust with highly noisy observation sets both in the one-dimensional and in the two-dimensional case.

\subsection{Influence of the sampling of the observation set}
We now assess the influence of the sampling of the observation set on the performance of our approach. To do so, we apply our method on randomly chosen observations and we calculate the average Normalized Sup Norm defined in \eqref{eq:norm_sup_norm} on 10 different samplings of the observations. In such situations, the knots used to define the function to estimate are not necessarily
included in the set of observations and in this case, cannot thus be chosen as knots of the B-spline basis. Then, we compare it to the case where the set of observations necessarily contains the set of knots used to define the underlying function $f$ to estimate.

\subsubsection{One-dimensional case ($d=1$)}
\begin{figure}[ht]
\begin{center}

\includegraphics[width=14cm]{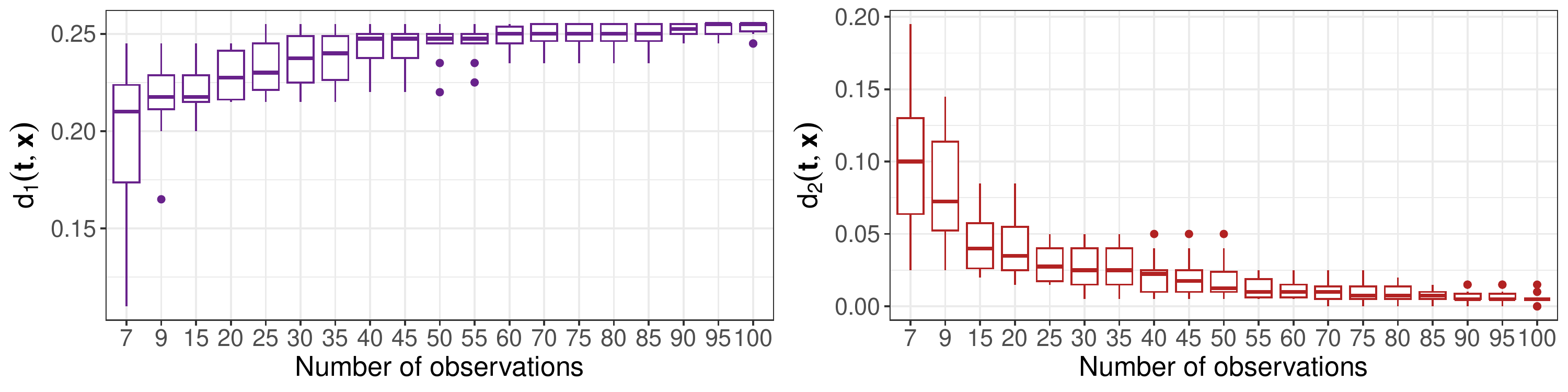}
\includegraphics[width=14cm]{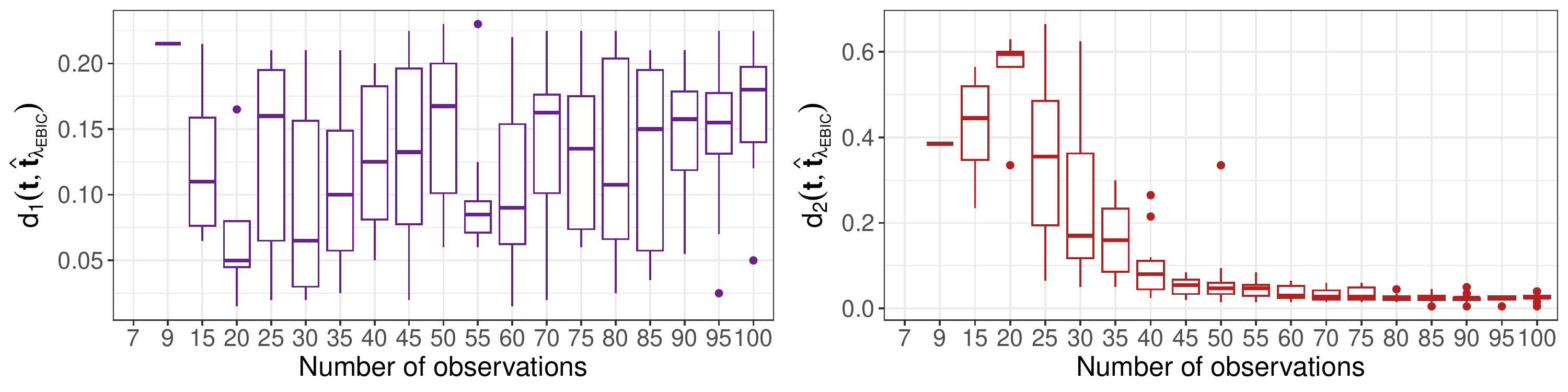}
\caption{Boxplots of the first part of the Hausdorff distance as a function of $n$: $d_1(\t,\x)$ (top left) and $d_1(\t,\widehat{\t}_{\uplambda_\text{EBIC}})$ (bottom left) and for the second part of the Hausdorff distance as a function of $n$: $d_2(\t,\x)$ (top right) and $d_2(\t,\widehat{\t}_{\uplambda_\text{EBIC}})$ (bottom right) for estimating $f_1$ when the observation set is randomly chosen and $\sigma = 0.05$.}\label{fig:hausdorff_random_sampling_f1}
\end{center}
\end{figure}

We first assess the estimation of the function $f_1$ defined in \eqref{eq:f1} from a noisy set of observations obtained with $\sigma = 0.05$. Figure \ref{fig:hausdorff_random_sampling_f1} shows the Hausdorff distance between the set of knots $\t$ of the function $f_1$ and the observation set $\x$ and the Hausdorff distance between the set of knots $\t$ and its estimation with our method in the case where the knots are not necessarily included in the observation set. We can see from this figure that for each element of the observation set, there exists at least one knot at a distance smaller than 0.25. Moreover, for each knot there exists at least one point of the observation set at a distance very close to 0 for large enough $n$. In addition, we can see from the plot on the bottom right that for large enough $n$, there exists for each knot an estimated one which is very close.

In the case where $\t$ belongs to $\x$, the results are displayed in Figure \ref{fig:hausdorff_inObs_f1} of the Appendix. The boxplot on the top left shows the same results for the distance $d_1(\t, \x)$ as for the random sampling and the boxplot on the top right confirms that $\t$ belongs to $\x$ since $d_2(\t, \x)=0$ at every value of $n$. Furthermore, we can see similar results on the bottom left and bottom right boxplots since distances $d_1(\t, \widehat{\t}_{\uplambda_\text{EBIC}})$ and $d_2(\t, \widehat{\t}_{\uplambda_\text{EBIC}})$ have exactly the same behavior as for the random sampling case.

\begin{figure}[h!]
\begin{center}
\includegraphics[width=7cm]{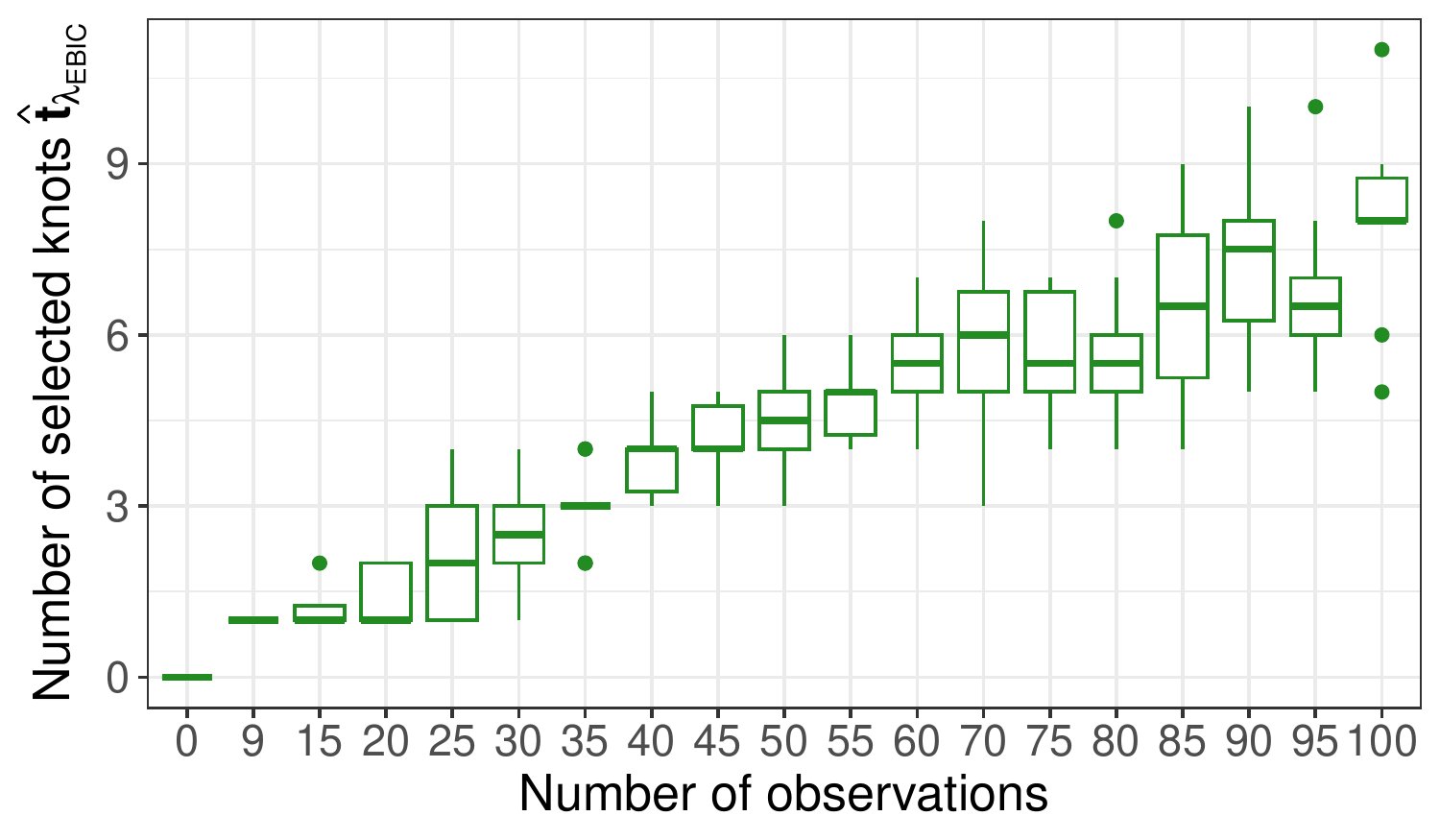}
\includegraphics[width=7cm]{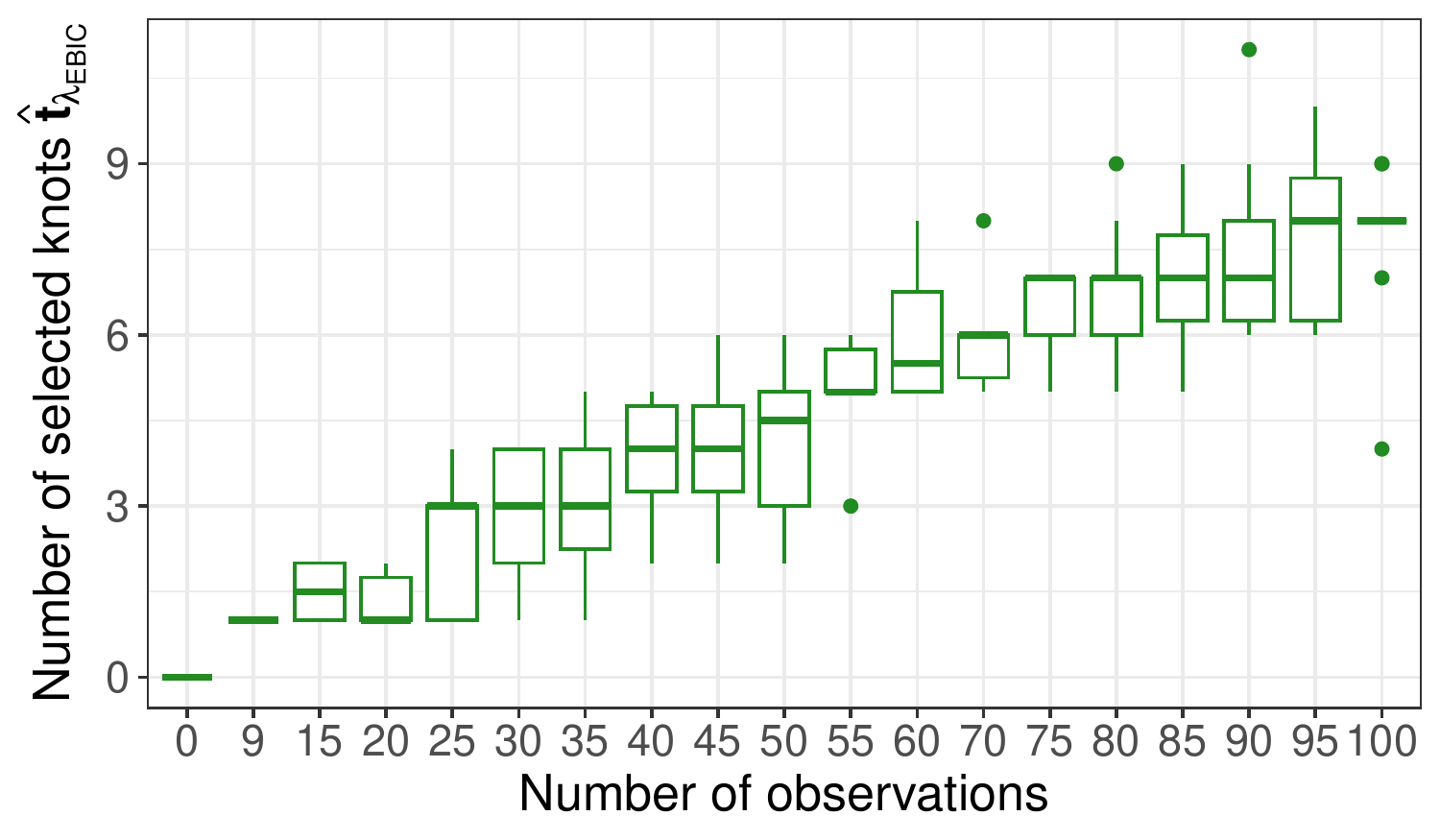}
\caption{Number of estimated knots as a function of $n$ for the estimation of $f_1$ with GLOBER from a random sampling of observations (left) and when $\t$ belongs to $\x$ (right) with $\sigma = 0.05$.}\label{fig:number_selected_knots_f1}

\end{center}
\end{figure}

Finally, the number of selected knots displayed in Figure \ref{fig:number_selected_knots_f1} shows comparable results between the random sampling of the observations and when $\t$ belongs to $\x$.  Therefore, the random sampling of the observations does not seem to affect the knot selection of our method.

\begin{figure}[h!]
\begin{center}
\includegraphics[width=0.48\textwidth,height=4cm]{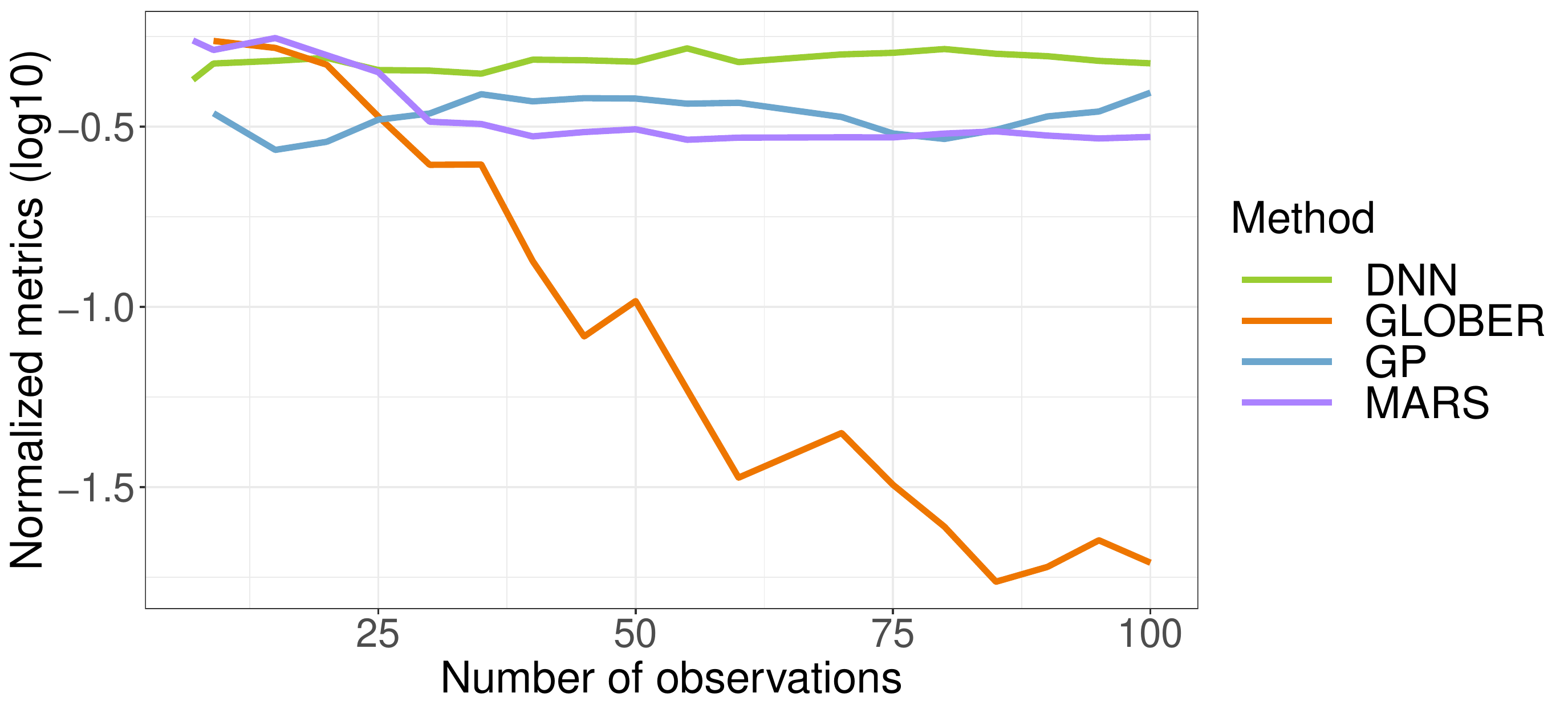}
\includegraphics[width=0.48\textwidth,height=4cm]{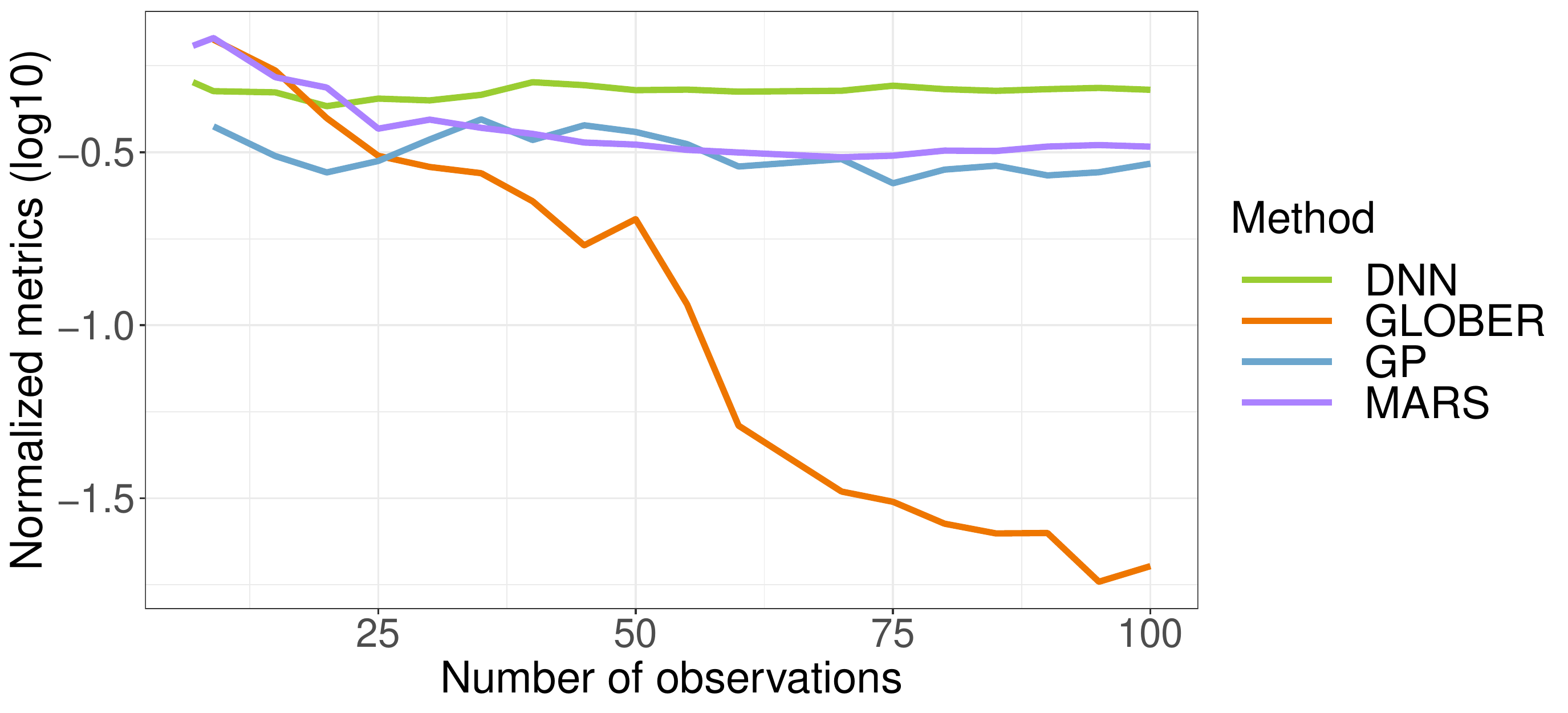}
\caption{Statistical performance (Normalized Sup Norm) of GLOBER for estimating $f_1$ from a random sampling of the observation set (left) and with $\t$ belonging to the observation set (right) with $\sigma = 0.05$. Comparison to the performance of the state-of-the-art methods obtained from 10 replications.}\label{fig:sampling_1d}
\end{center}
\end{figure}

The results of the statistical performance of our method for the estimation of the function $f_1$ are displayed in Figure \ref{fig:sampling_1d} for $N=201$.
We can clearly see that the random sampling of the observation set does not deteriorate the performance of our method in comparison to the case where the observation set contains all the knots: the value of the Normalized Sup Norm reaches in both cases $10^{-1.75}$ for $n=100$ and our method still outperforms the other ones.

\subsubsection{Two-dimensional case ($d=2$)}
Similarly to the previous part, we study the estimation of the function $f_2$ from a noisy set of observations obtained with $\sigma = 0.01$. Figure \ref{fig:hausdorff_sampling_2D} of the Appendix shows the Hausdorff distance between the set of knots  $\t_1$ (resp. $\t_2$) of the function $f_2$ and the observation set $\x_1 = \{x_{11}, \ldots, x_{1n_1}\}$ (resp. $\x_2 = \{x_{21}, \ldots, x_{2n_2}\}$) for the first (resp. second) dimension. It also displays the Hausdorff distance between the set of knots $\t_1$ (resp. $\t_2$) and its estimation with our method 
$\widehat{\t}_{1,\widetilde{\uplambda}_1}$ (resp. $\widehat{\t}_{2,\widetilde{\uplambda}_2}$), with  $\widetilde{\uplambda}_1 = \widetilde{\uplambda}_{1,\text{EBIC}}$ (resp. $\widetilde{\uplambda}_2 = \widetilde{\uplambda}_{2,\text{EBIC}}$) for the first (resp. second) dimension in the case where the knots do not necessarily belong to the observation set. We can see from this figure that for each element of the observation set, there exists at least one knot at a distance smaller than 0.45 (resp. 0.38) for the points of the first dimension (resp. second dimension). Furthermore, for each knot of each dimension there exists at least one point of the observation set of the corresponding dimension at a distance very close to 0 when $n$ is large enough (at least 20 points per dimension).
For the case where $\t_1$ and $\t_2$ belong to $\x_1$ and $\x_2$, respectively, we can see from Figure \ref{fig:hausdorff_inObs_f2} of the Appendix that the distances $d_1(\t_1, \x_1)$ and $d_1(\t_2, \x_2)$ have the same behavior as for the random sampling. Moreover, the boxplots on the top right shows that $d_2(\t_1, \x_1)=0$ and $d_2(\t_2, \x_2)=0$ at every $n$ value which confirms that the sets of knots belong to the observation set. Similar results can be observed for the evolution of the distances 
$d_1(\t_1, \widehat{\t}_{1,\widetilde{\uplambda}_1})$ (resp. $d_1(\t_2, \widehat{\t}_{2,\widetilde{\uplambda}_2}))$ and $d_2(\t_1, \widehat{\t}_{1,\widetilde{\uplambda}_1})$ (resp. $d_2(\t_2, \widehat{\t}_{2,\widetilde{\uplambda}_2}))$ for the random sampling and when the knots belong to the observation set. Indeed, in both cases we can see from the plot on the bottom right of Figures \ref{fig:hausdorff_sampling_2D} and \ref{fig:hausdorff_inObs_f2} of the Appendix that for large enough $n$, there exists for each
knot of each dimension an estimated one which is very close. Moreover, the number of selected knots displayed in Figure \ref{fig:number_selected_knots_f2} of the Appendix shows comparable results between the random sampling of the observations and when $\t$ belongs to $\x$.  Therefore, the random sampling of the observations does not seem to affect the knot selection of our method for $d=2$. \\

As it is the case for $d=1$, we can see from Figure \ref{fig:sampling_2d} of the Appendix where the statistical performance of our method for estimating $f_2$ for $N=40401$ are displayed that the performance
  of our approach is not altered by the sampling of the observation set. More precisely,  the value of the Normalized Sup Norm reaches in both cases $10^{-1.5}$ for $n=1600$ and our method still outperforms the other approaches which seem to reach a constant value.

\subsection{Numerical performance}
\begin{figure}[h!]
\begin{center}
\includegraphics[width=6.5cm]{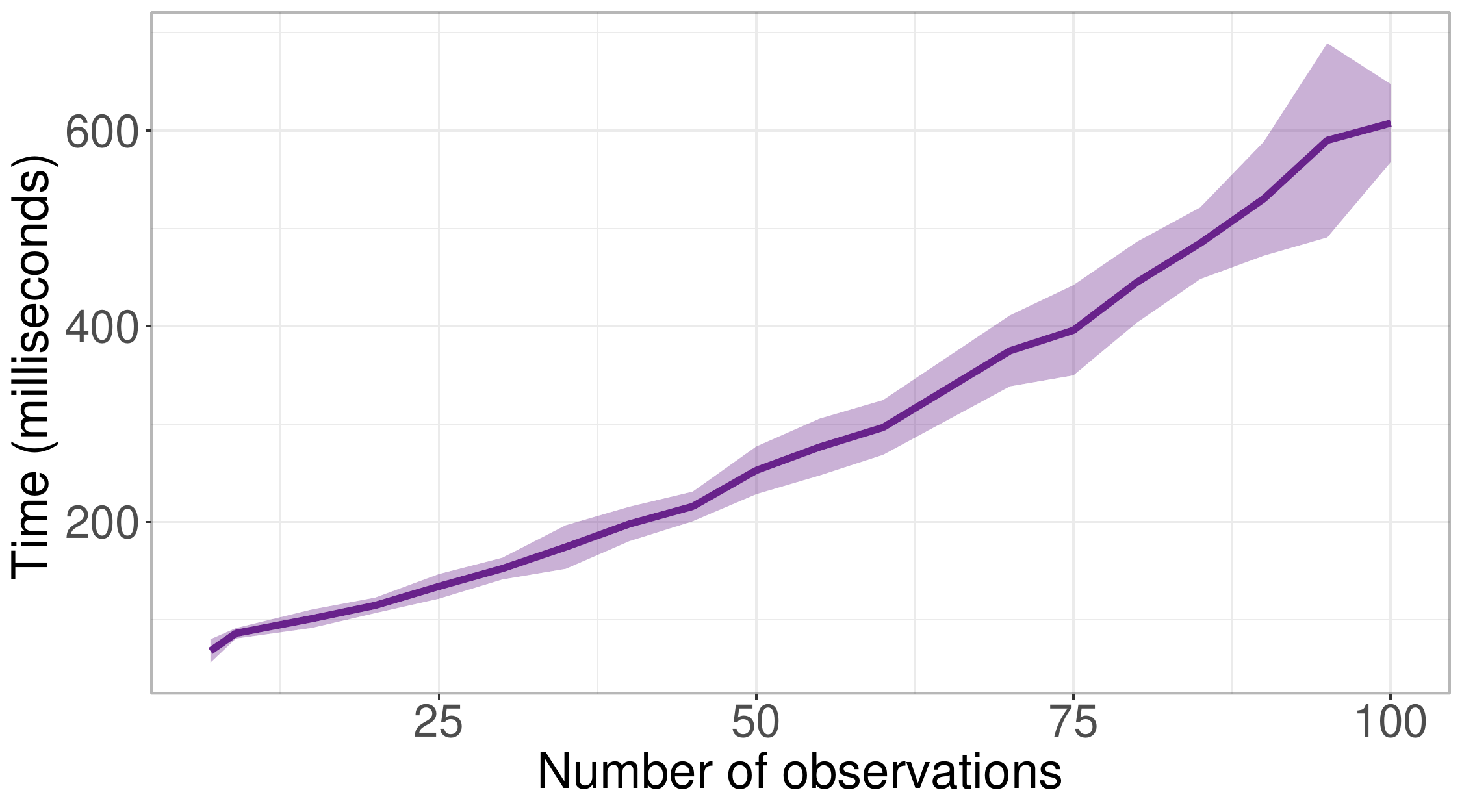}
\hspace{3em}
\includegraphics[width=6.5cm]{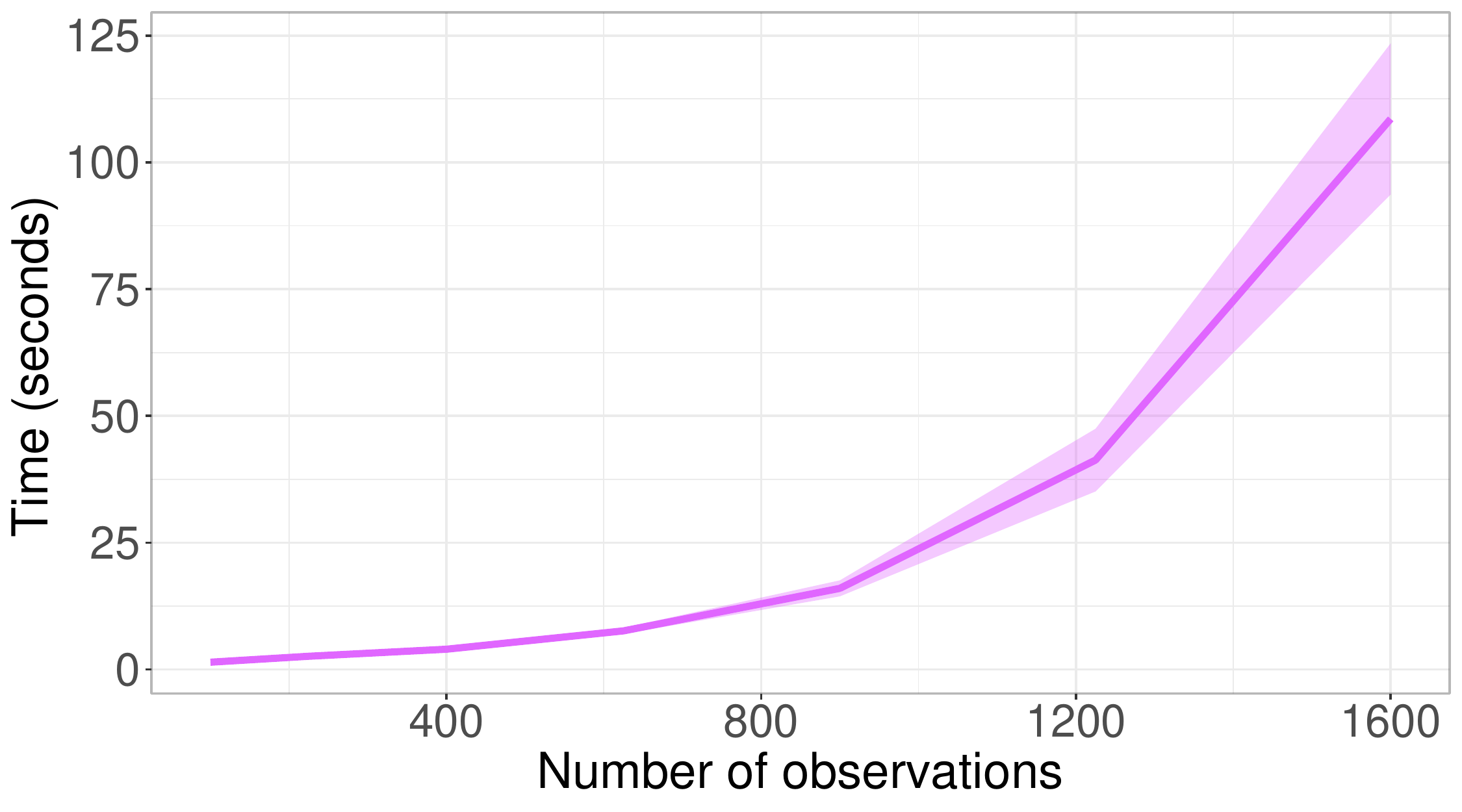}
\caption{Average and standard deviation of the execution time measured in milliseconds for $d=1$ (left) and in seconds for $d=2$ (right) on 30 independent executions.}\label{fig:time_execution}
\end{center}
\end{figure}

The goal of this section is to investigate the computational times of our approach GLOBER implemented in the \texttt{glober} R package available on the CRAN as a function of the number of observation points.
The timings were obtained on a workstation with 31.2GB of RAM and Intel Core i7 (3.8GHz) CPU. The average computational times and their standard deviation obtained from 30 independent executions are displayed in Figure \ref{fig:time_execution}. We can see from this figure that it only takes 600ms to estimate the underlying function with our approach in the one-dimensional. In the two-dimensional case, the computational time is larger but even in this case the computational burden of our approach is low since it only takes 110 seconds for processing 1600 observations.

\section{Application to geochemical systems}\label{sec:real}

In this section, we apply our method on simple chemical problems with one or two input concentrations which correspond to the estimation of a function $f$ for $d=1$ or $d=2$, respectively. We consider hereafter the estimation of the amount of a "Salt" mineral as a function of the concentrations of its constituents Sp$_{a}^{+}$ and Sp$_{b}^{-}$ as in \cite{savino2022active}. For this example, the thermodynamic constants of the halite salt (NaCl) were considered because there are only two constitutive
elements and they do not depend on the pH of the solution.
Following the law of mass action, the dissolution reaction of this mineral writes:
\[\mathrm{
   Salt \rightleftharpoons Sp_{a}^{+} + Sp_{b}^{-}.
}\]
At equilibrium, the activity of these elements a$_{Sp_{a}^{+}}$ and a$_{Sp_{b}^{-}}$ obey the solubility product
\[\mathrm{
   K_{Salt} = a_{Sp_{a}^{+}} a_{Sp_{b}^{-}} = 10^{1.570}.
}\]
In the following, the increasing number of observations is obtained by randomly adding new points from a given grid of $N$ points to the pre-existing observation sets. Since the functions to estimate come from chemical processes and are known to be smooth, we propose taking $q\geq 2$. Nevertheless, we observed that using $q=3$ does not yield better results compared to $q=2$. This is the reason why we will present in the sequel results obtained for $q=2$. 

\subsection{One-dimensional application ($d=1$)}

\begin{figure}[h!]
\begin{center}
\includegraphics[width=6cm]{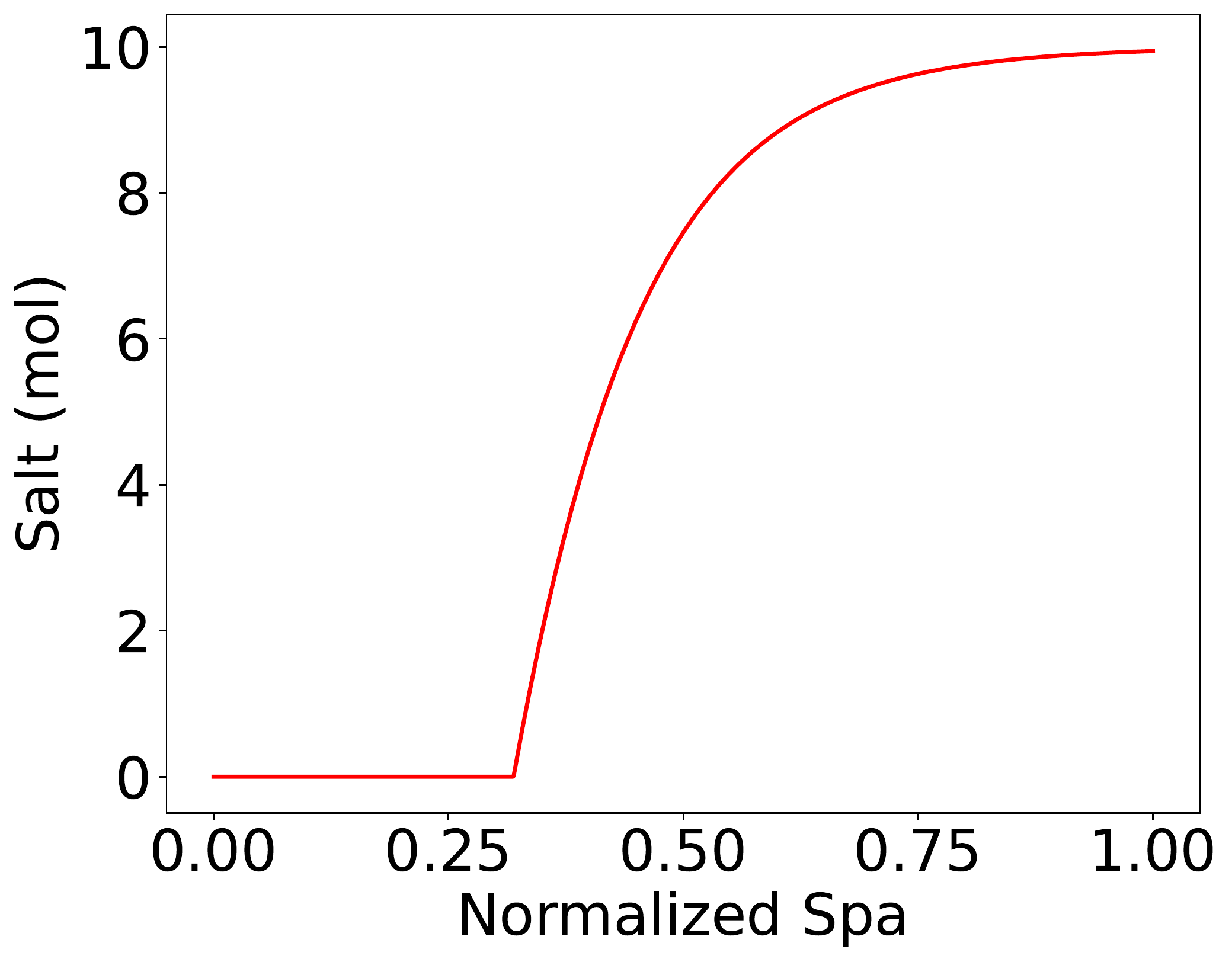}
\includegraphics[width=9cm,height=4.6cm]{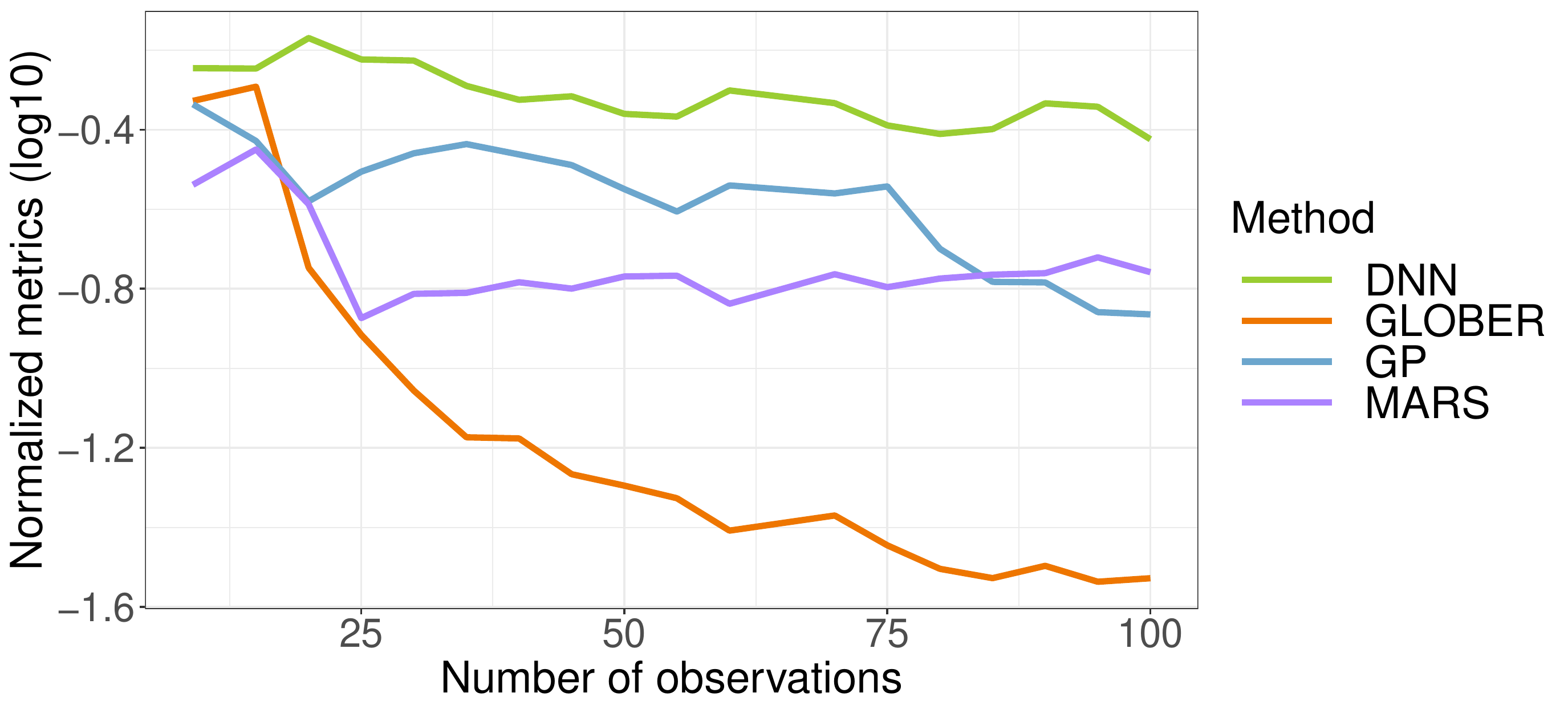}
\caption{Left: Function $f_3$ to estimate when $d=1$ with $Y_1,\dots,Y_{1140}$. Right: Statistical performance (Normalized Sup Norm) of GLOBER and of the state-of-the-art methods for estimating $f_3$. The average values are obtained from 10 replications.}\label{fig:halite1d}
\end{center}
\end{figure}

\begin{figure}[h!]
\begin{center}
\includegraphics[width=5.5cm]{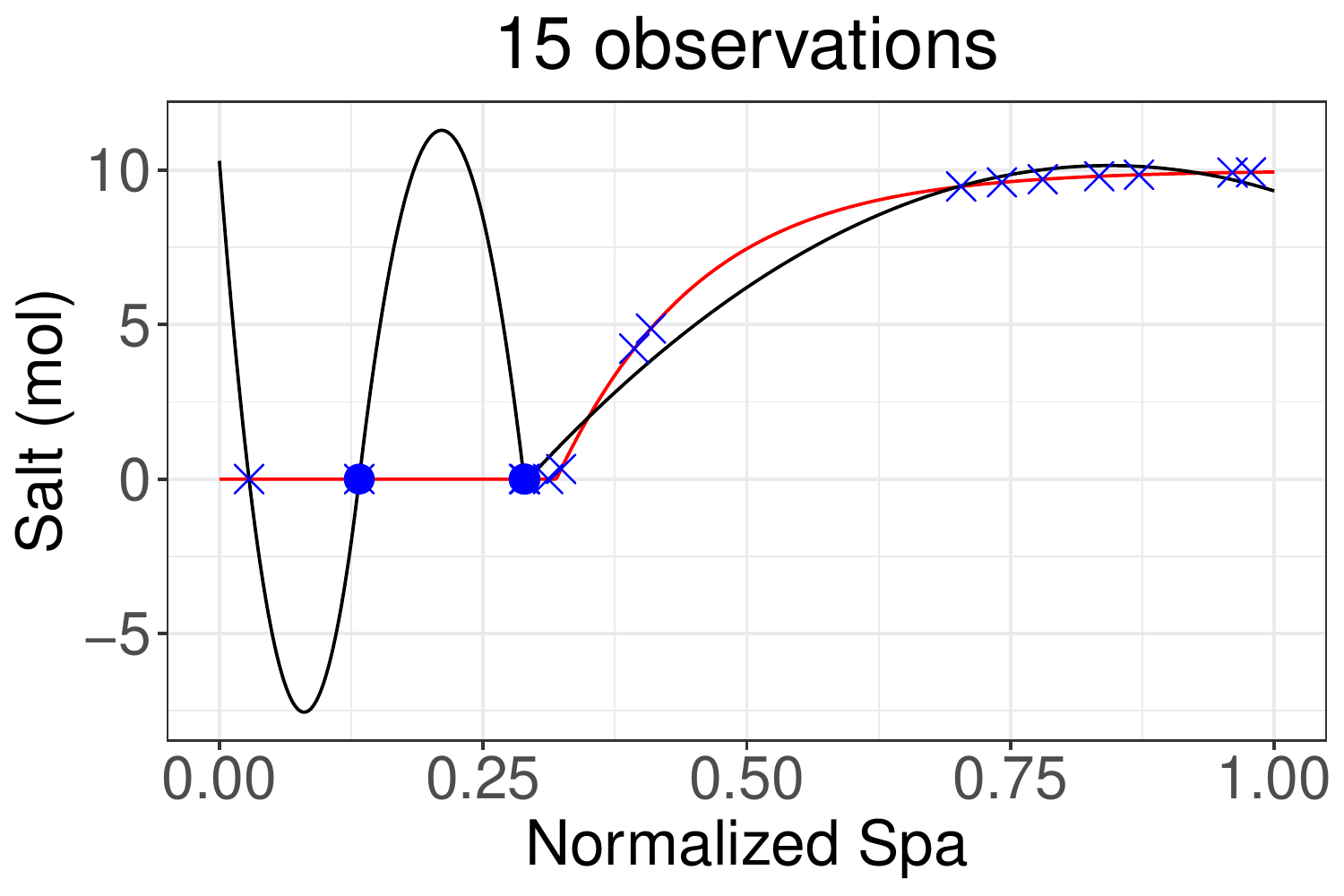}
\includegraphics[width=5.5cm]{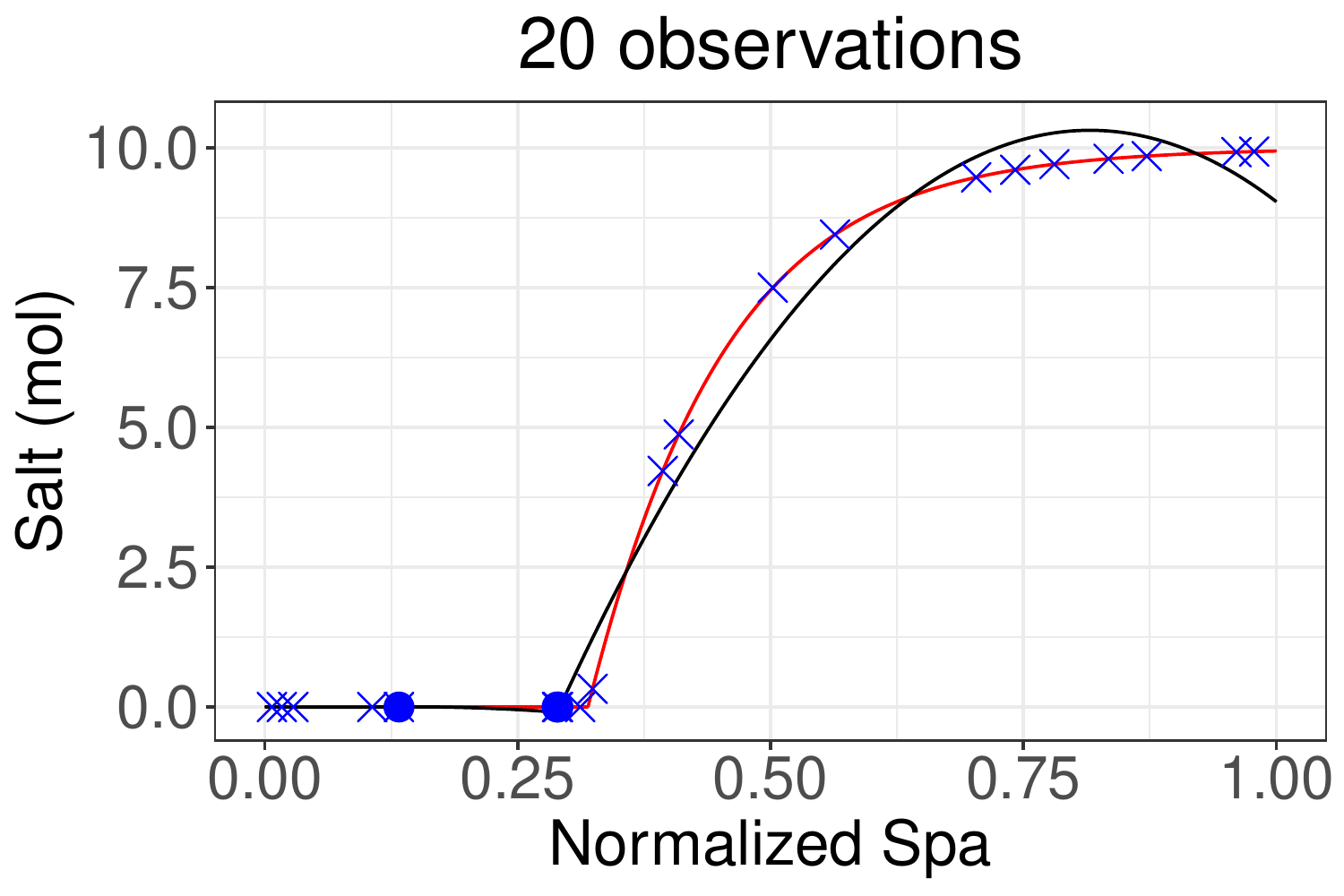}
\includegraphics[width=5.5cm]{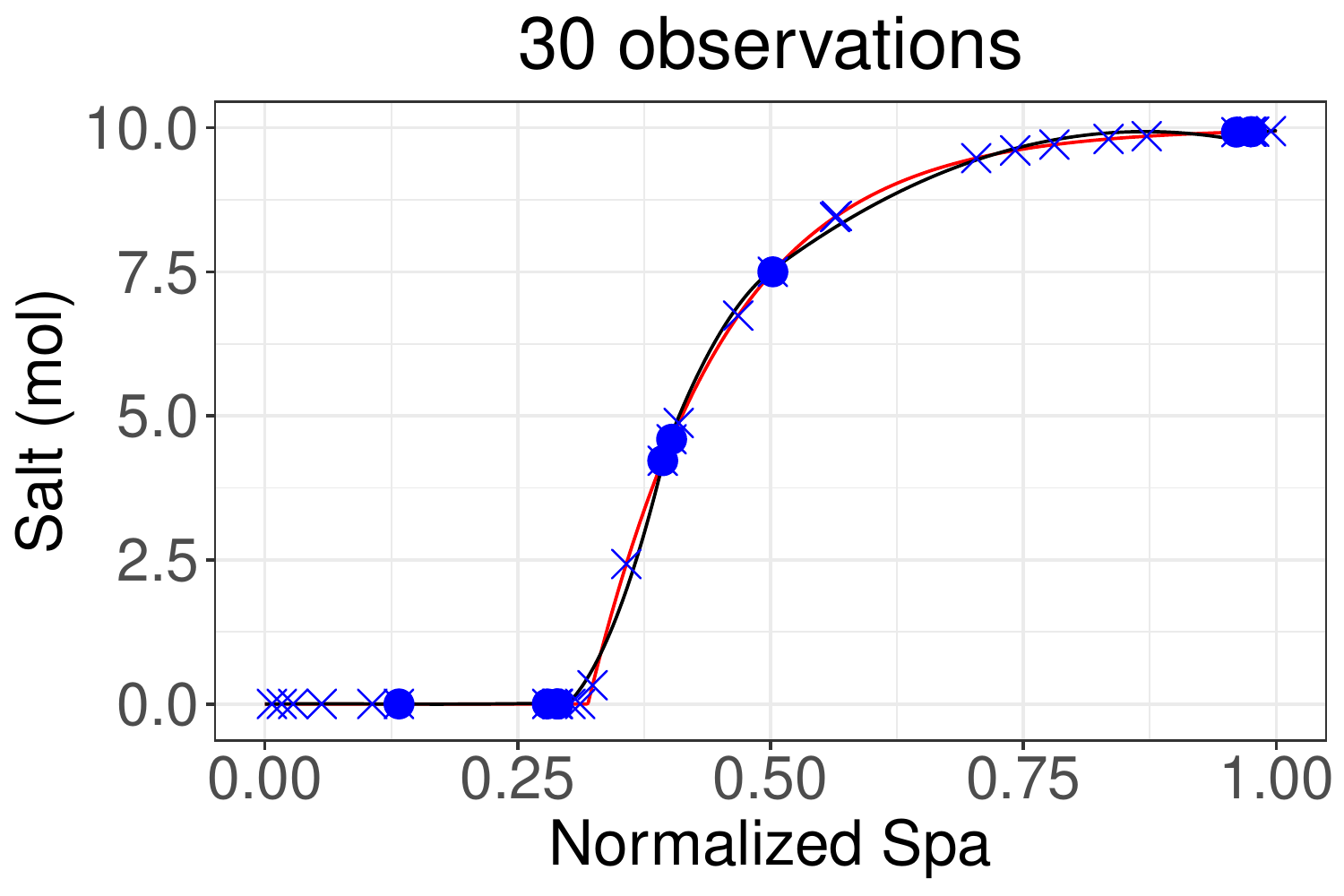}
\includegraphics[width=5.5cm]{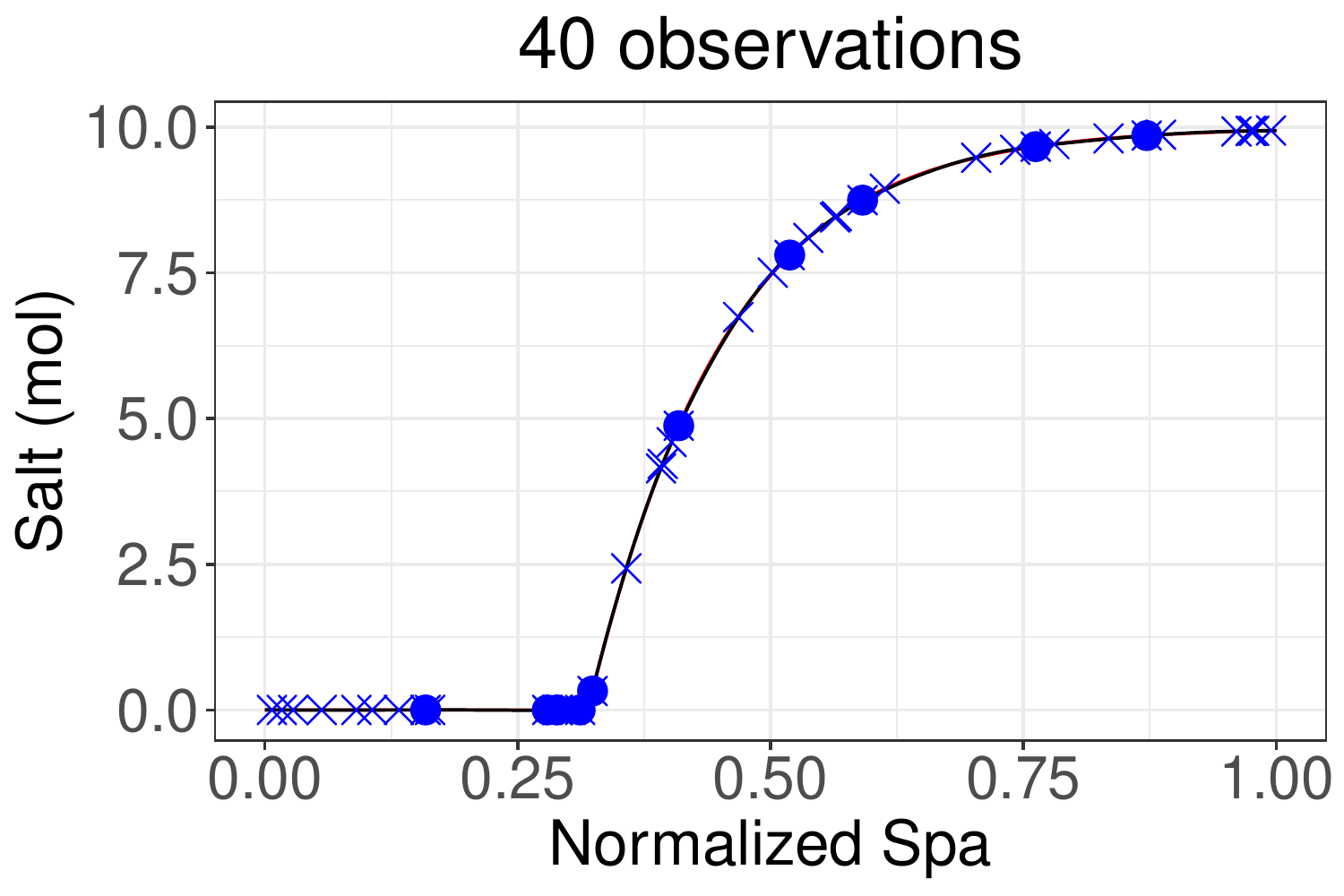}
\caption{Illustration of the estimation of the amount of Salt depending on the normalized concentration of Sp$_{a}^{+}$ for 15 (top left) 20 (top right) 30 (bottom left) and 40 observations (bottom right). The red curve describes the true underlying function $f_3$ to estimate, the black curve corresponds to the estimation with GLOBER, the blue crosses are the observation points and the blue bullets are the observation points chosen as estimated knots.}\label{fig:illustration_halite1D}
\end{center}
\end{figure}

The amount of Salt is first calculated with PHREEQC as in \cite{Parkhurst2013} as a function of the normalized concentration of Sp$_{a}^{+}$ so that it belongs to $[0,1]$. The corresponding function $f_3$ to estimate is defined as:
\begin{equation*}
\text{Salt} = f_3(\text{Normalized Sp}_{a}^{+})
\end{equation*} 
and is displayed on the left part of Figure \ref{fig:halite1d}. Our method is used to estimate this function by using a set of a varying number of observation points ($7 \leq n \leq 100$) and is then compared to the state-of-the-art methods described in Section \ref{sec:numexp}. 
Figure  \ref{fig:illustration_halite1D} displays the illustration of the estimation of $f_3$ for a varying number of $n$. We can see by adding new points to the observation set (blue crosses) that GLOBER has better chance to choose knots (blue bullets) among observation points which depict more precisely changes in the underlying curve. However, it seems that 40 observations are enough to have a good approximation of our function $f_3$ since the estimation (black curve) fits quite perfectly the function to estimate (red curve).

The corresponding statistical performance is displayed on the right side of Figure \ref{fig:halite1d} for $N =1140$.
We can see from this figure that our method outperforms the other approaches and allows us to get a very good accuracy of the amount of Salt since the average Normalized Sup Norm reaches $10^{-1.5}$ for only $n=100$. 

\subsection{Two-dimensional application ($d=2$)}\label{sec:dolomite2d}

Our method is then applied to a more complex chemical problem which derives from the calcite dissolution
and precipitation study described in \cite{kolditz12}. The corresponding thermodynamic data for aqueous species and minerals are available in the Phreeqc.dat distributed with PHREEQC. The compositional system actually solved consists of 14 species in solution, 2 mineral components, 8 geochemical reactions and 2 mineral dissolution-precipitation reactions. However, in this paper, we will only consider the dolomite precipitation: 
\[\mathrm{Dolomite \rightleftharpoons Ca^{2+} + Mg^{2+} + 2 CO_{3}^{2-}, \quad logK_{10}=-17.09.}\]
The amount of dolomite is computed with PHREEQC as a function of the total elemental concentrations  (C, Ca, Cl, Mg), the pH (as $\mathrm{-log(H^+}$)) and the amount of calcite.
In this article, we will only consider the dolomite precipitation for C=5$\times 10^{-4}$ mol/kgw, Cl=$2\times 10^{-3}$ mol/kgw, pH=10, calcite=0 mol in order to reduce the problem to a two-dimensional case. Thus, the function $f_4$ to be estimated is defined as:
 \begin{equation*}
\text{Dolomite} = f_4(\text{Normalized Ca}, \text{Normalized Mg}).
\end{equation*} 

\begin{figure}[h!]
\begin{center}

\includegraphics[width=5.5cm, trim= 0.5cm 1cm 0 3cm, clip]{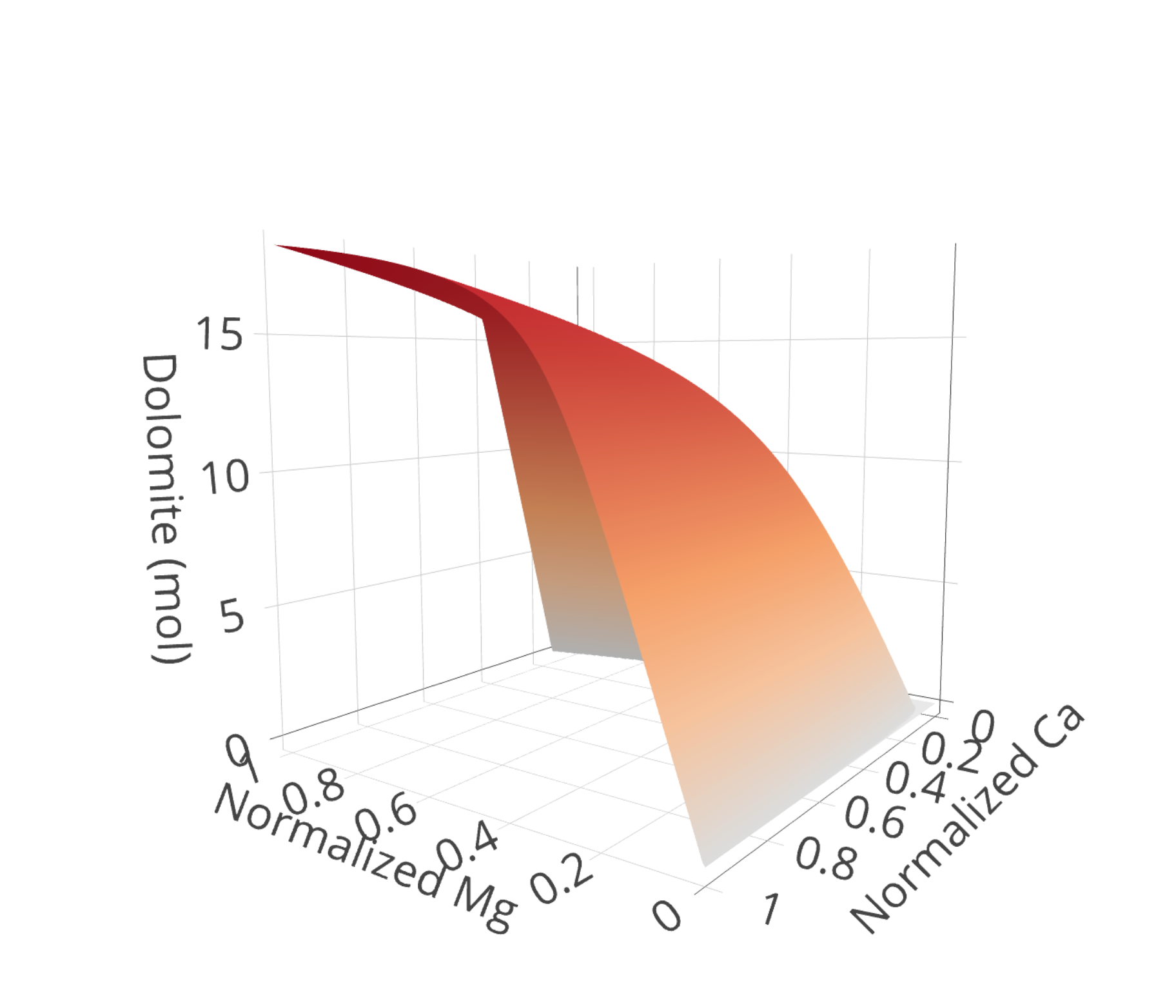}
\includegraphics[width=7.6cm]{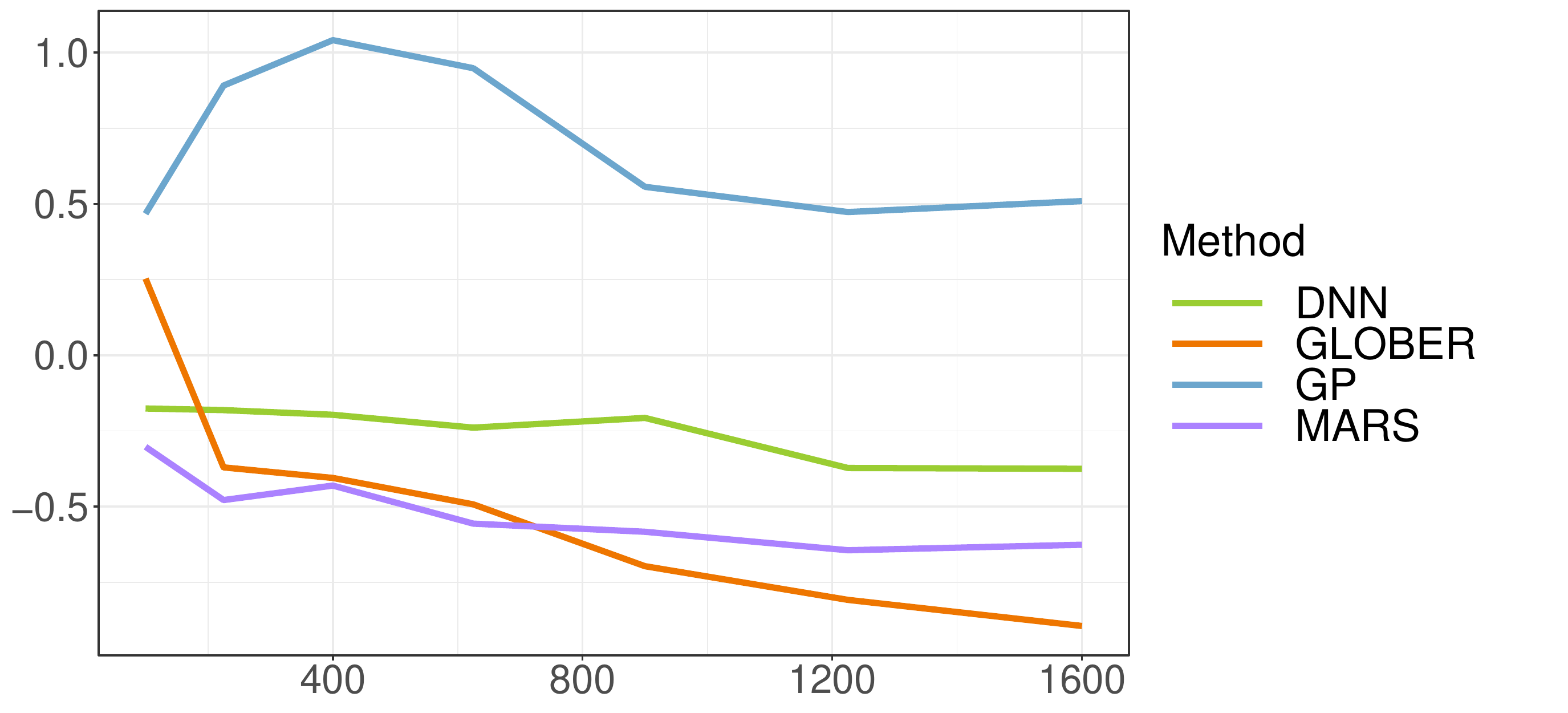}
\caption{Left: Function $f_4$ to estimate when $d=2$ with $Y_1,\dots,Y_{40000}$. Right: Statistical performance (Normalized Sup Norm) of GLOBER and of the state-of-the-art methods for estimating $f_4$. The average values are obtained from 10 replications.}\label{fig:dolomite2D}

\end{center}
\end{figure}

We seek to estimate $f_4$ by applying our method to an increasing number of observations ($100 \leq n \leq 1600$) which corresponds to an increasing number of points per dimension ($10 \leq n_1, n_2 \leq 40$ with $n=n_1n_2$). The resulting illustration is displayed in Figure \ref{fig:illustration_dolomite2D} of the Appendix and shows an
improvement of the fitting of GLOBER (green curve) to the underlying curve (red curve) by adding new points to the observation set (orange bullets) in order to get a perfect overlapping for $n \geq 900$. In the right part of Figure \ref{fig:dolomite2D}, the statistical performance of GLOBER compared to the state-of-the-art approaches is displayed for $N=40000$. Similarly to what has been shown for the precipitation of Salt, our method gives satisfactory results as the Normalized Sup Norm reaches $10^{-1}$. Moreover, our method still outperforms the other ones for which the statistical metric seems to rapidly reach a constant value.

\section{Extension to higher dimensional and more general observation settings}\label{sec:extension}

Here, let us consider the case where $x_i \in \R^d$ with $d \geq 2$ and $f$ is still a function to estimate from \eqref{eq:model}. In this section, the grid of $N$ points and the observation sets are no longer constrained to result from a cartesian product of $d$ compact sets, such as in Section \ref{sec:two-dim-case}. 

\subsection{Adaptation of the knot selection method by using clustering}

In order to avoid finding the knots associated to a given dimension for each fixed value of the others, we propose in this section to cluster
  the $n$ observation points by using a $k$-means approach. More precisely, for each dimension $j$, the $n$ observation points are gathered into clusters based
  on their multidimensional characteristics excluding dimension $j$. Then, the one-dimensional approach described in Section \ref{sec:one_dim_case}
  is applied to the $Y$'s
  associated to the $x$'s belonging to this cluster to find the knots in dimension $j$.
The number $k$ of clusters is chosen as the integer part of $n$ divided by 25 to guarantee a large enough amount of points per cluster.
Thus, since sets of knots are found for each $\widetilde{\uplambda}_j$ of $\widetilde{\Lambda_j}$ we gather the knots $\widehat{\t}_{\widetilde{\uplambda}_j}$
  into clusters. The final knot sets are built by keeping the median value of each cluster of knots.
The number of knot clusters is chosen as the value where there is a change in the slope
of the within-sum of squares. Finally, the EBIC criterion defined in \eqref{eq:EBIC_2D} is used to find the final combination of knot sets.

\subsection{Case where $d=2$}
\begin{figure}
\includegraphics[width=7.6cm]{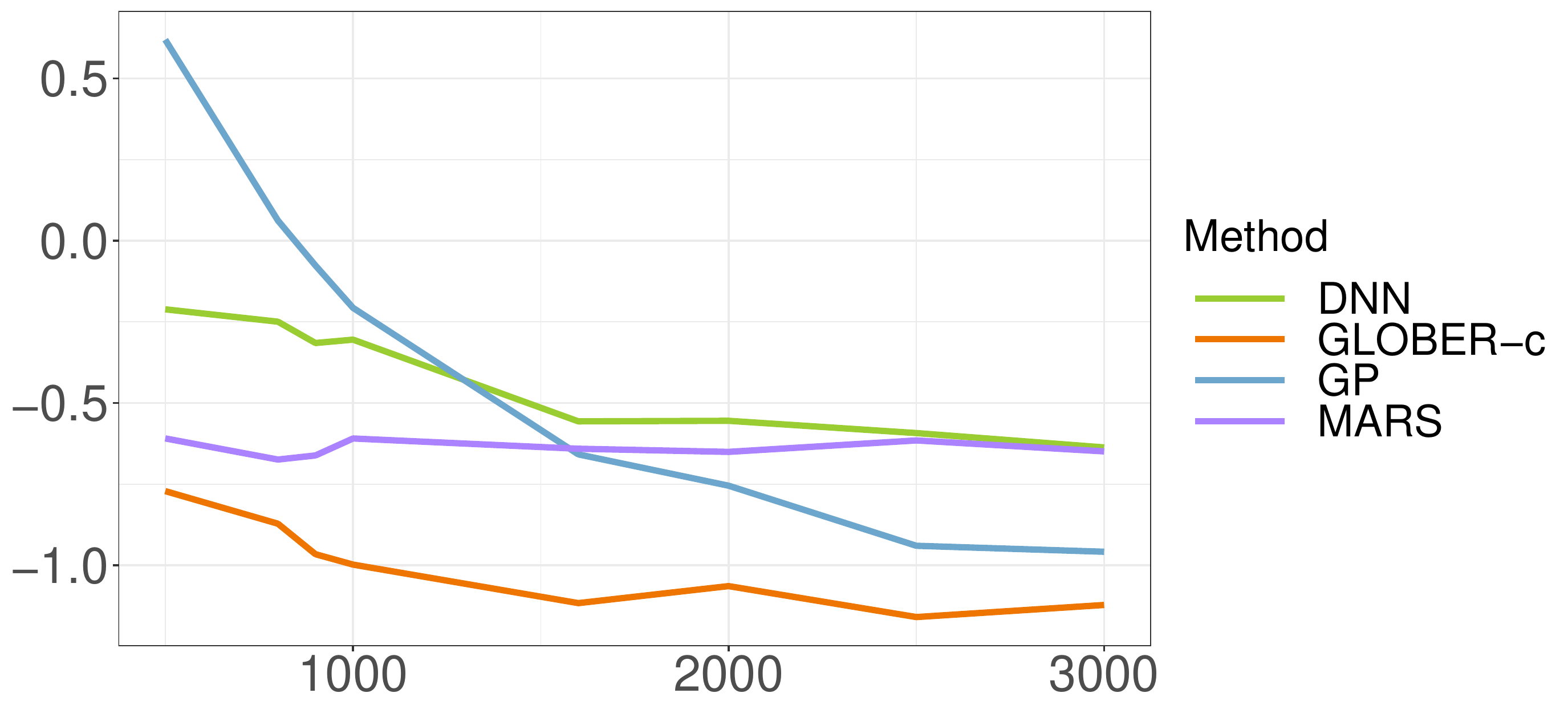}
\includegraphics[width=7.6cm]{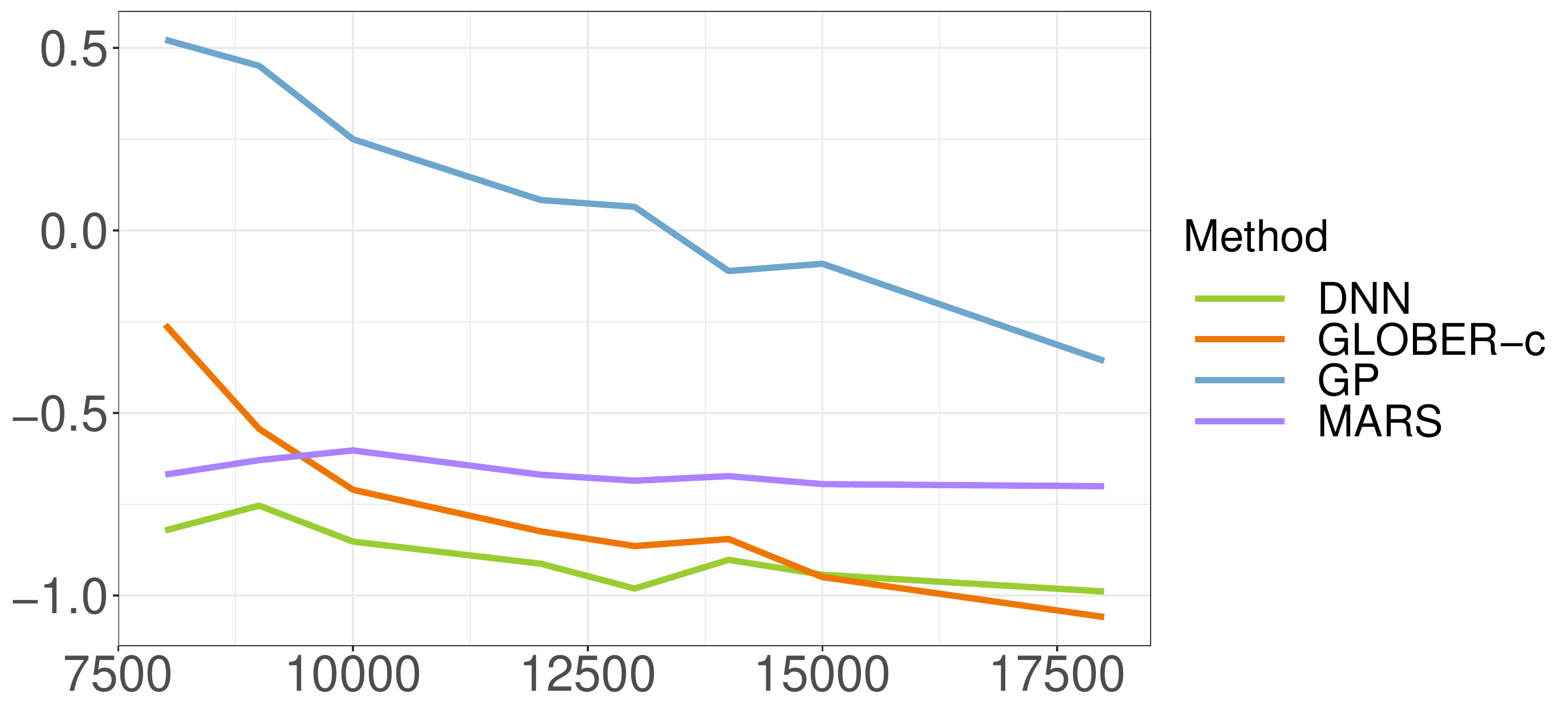}
\caption{Statistical performance (Normalized Sup Norm) of GLOBER-c and of the state-of-the-art methods for estimating $f_4$ (left) and $f_5$ (right). The average values are obtained from 10 replications.}
\label{fig:dolomite2d_extension}
\end{figure}

This method was first validated on the two-dimensional framework application by comparing the results obtained for the clusterized version called GLOBER-c on the Dolomite precipitation case described in Section \ref{sec:dolomite2d}. The obtained results are displayed in the left part of
  Figure \ref{fig:dolomite2d_extension}.
  We can see from this figure that compared to the results obtained in Section \ref{sec:dolomite2d} with the non-clusterized version, the maximal absolute error is smaller than $10^{-1}$ for $n=1600$ and that our method still outperforms the three others. 

\subsection{Case where $d=3$}

In this section, we apply our method to an extension of the geochemical system described in Section \ref{sec:dolomite2d}. Our goal is to estimate $f_5$ which describes the relationship between the precipitation of Dolomite and three chemical elements:
 \begin{equation*}
\text{Dolomite} = f_5(\text{Normalized C}, \text{Normalized Ca}, \text{Normalized Mg}).
\end{equation*} 
The obtained results are displayed in the right part of Figure \ref{fig:dolomite2d_extension}.
We can see from this figure that our method is the only one that displays a maximal absolute error below $10^{-1}$. Moreover, it outperforms the three others when
the number of observations is larger than 15000 observations. 

\newpage

\vspace{1mm}

\section{Appendix}\label{sec:appendix}

\begin{figure}[ht]
\begin{center}
\includegraphics[width = 16cm]{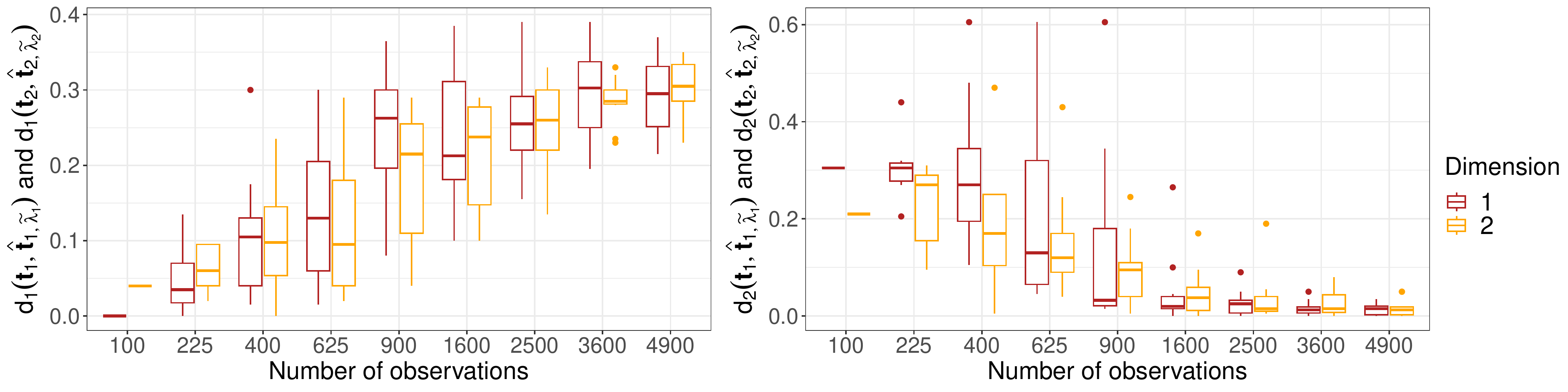}
\includegraphics[width = 8cm]{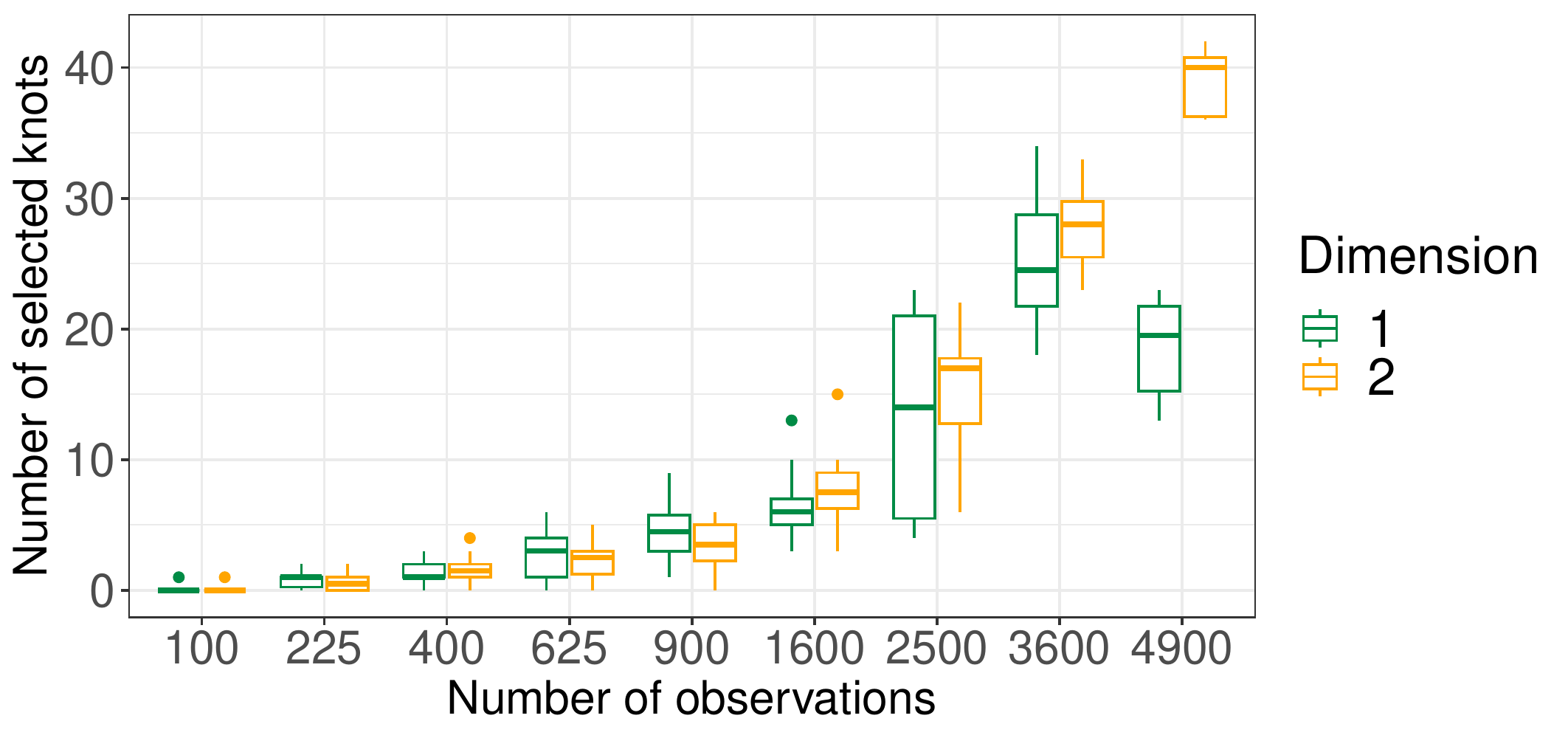}
\caption{Top left: Boxplots for the first part of the Hausdorff distance ($d_1$)  as a function of $n=n_1n_2$ and top right: boxplots for the second part of the Hausdorff distance ($d_2$)  as a function of $n=n_1n_2$ between the two sets of knots $\t_1$ and $\widehat{\t}_{1,\widetilde{\uplambda}_1}$ and $\t_2$ and $\widehat{\t}_{2,\widetilde{\uplambda}_2}$ for the first and second dimension, respectively. Bottom: number of estimated knots  as a function of $n=n_1n_2$ by choosing $\widetilde{\uplambda}_1 = \widetilde{\uplambda}_{1,\text{EBIC}}$ and $\widetilde{\uplambda}_2 = \widetilde{\uplambda}_{2,\text{EBIC}}$ for the estimation of $f_2$.}
\label{fig:boxplot_ebic_f2}
\end{center}
\end{figure}

\begin{figure}[ht]
\begin{center}

\includegraphics[width = 16cm]{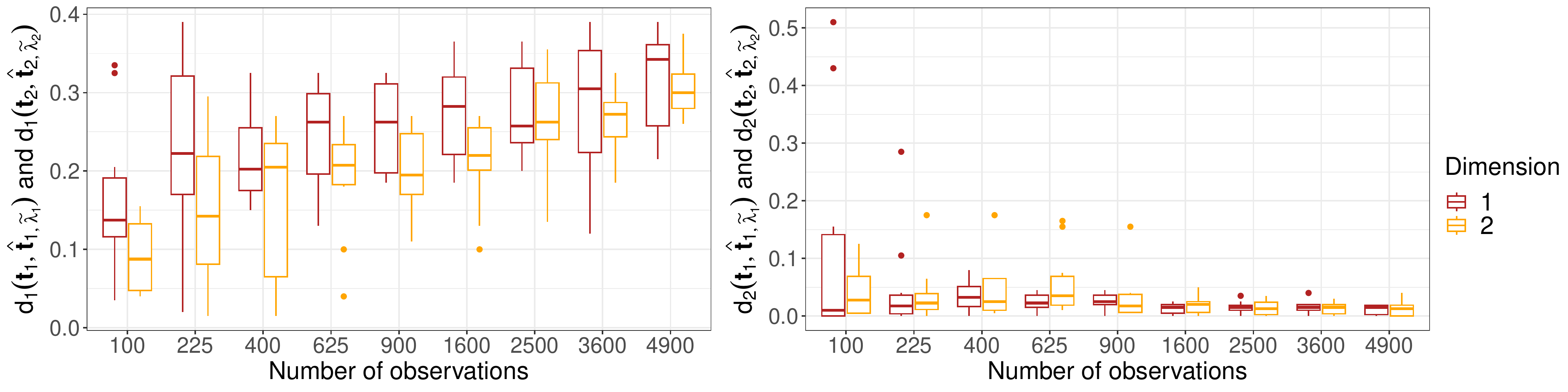}
\includegraphics[width = 8cm]{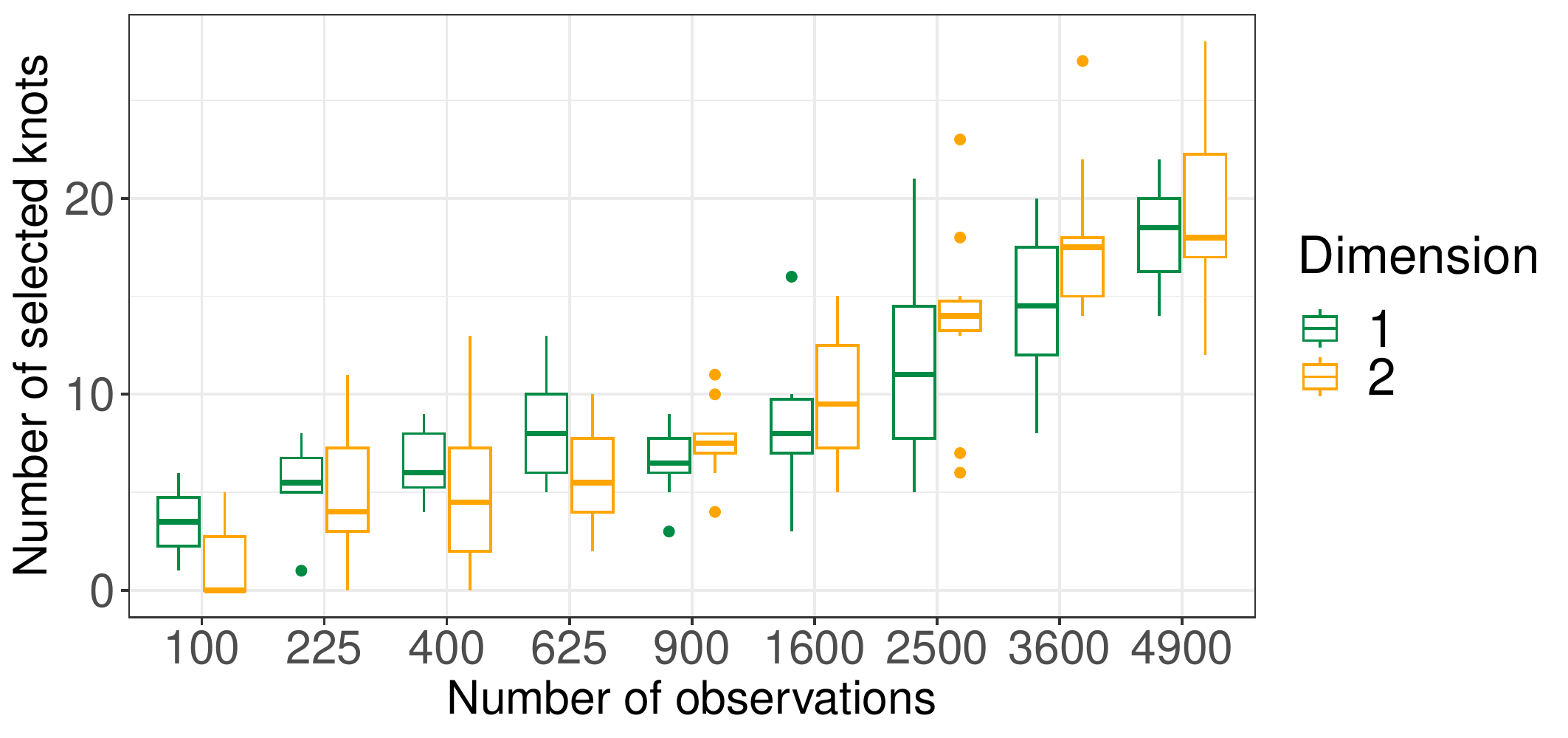}
\caption{Similar to Figure \ref{fig:boxplot_ebic_f2} by choosing $\widetilde{\uplambda}_1 = \widetilde{\uplambda}_{1,\text{opt}}$ and $\widetilde{\uplambda}_2 = \widetilde{\uplambda}_{2,\text{opt}}$ for the estimation of $f_2$ with $\sigma=0.01$. }
\label{fig:boxplot_lambda_opt_f2}

\end{center}
\end{figure}

\begin{figure}[ht]
\begin{center}
\includegraphics[width = 16cm]{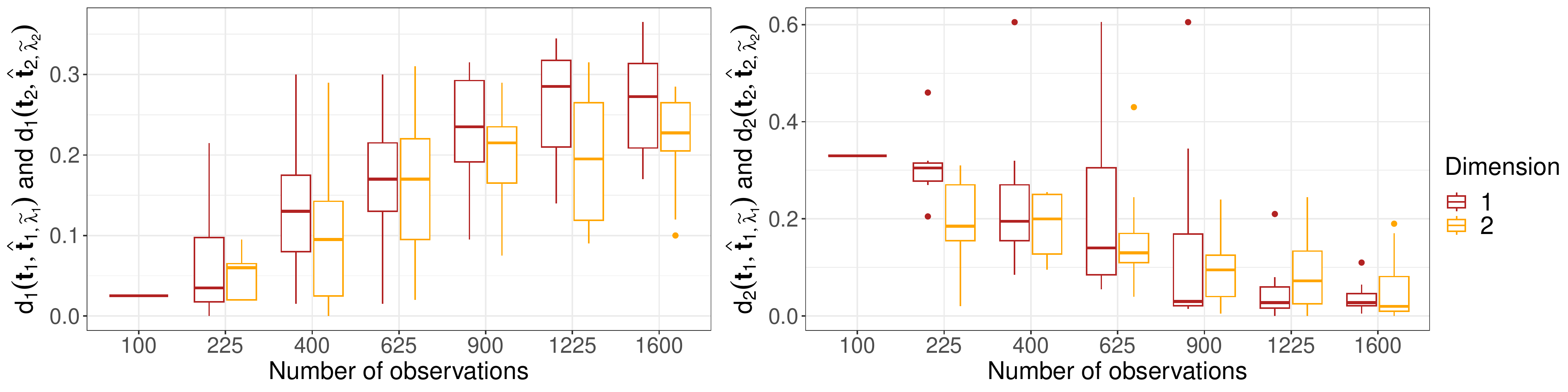}
\includegraphics[width = 8cm]{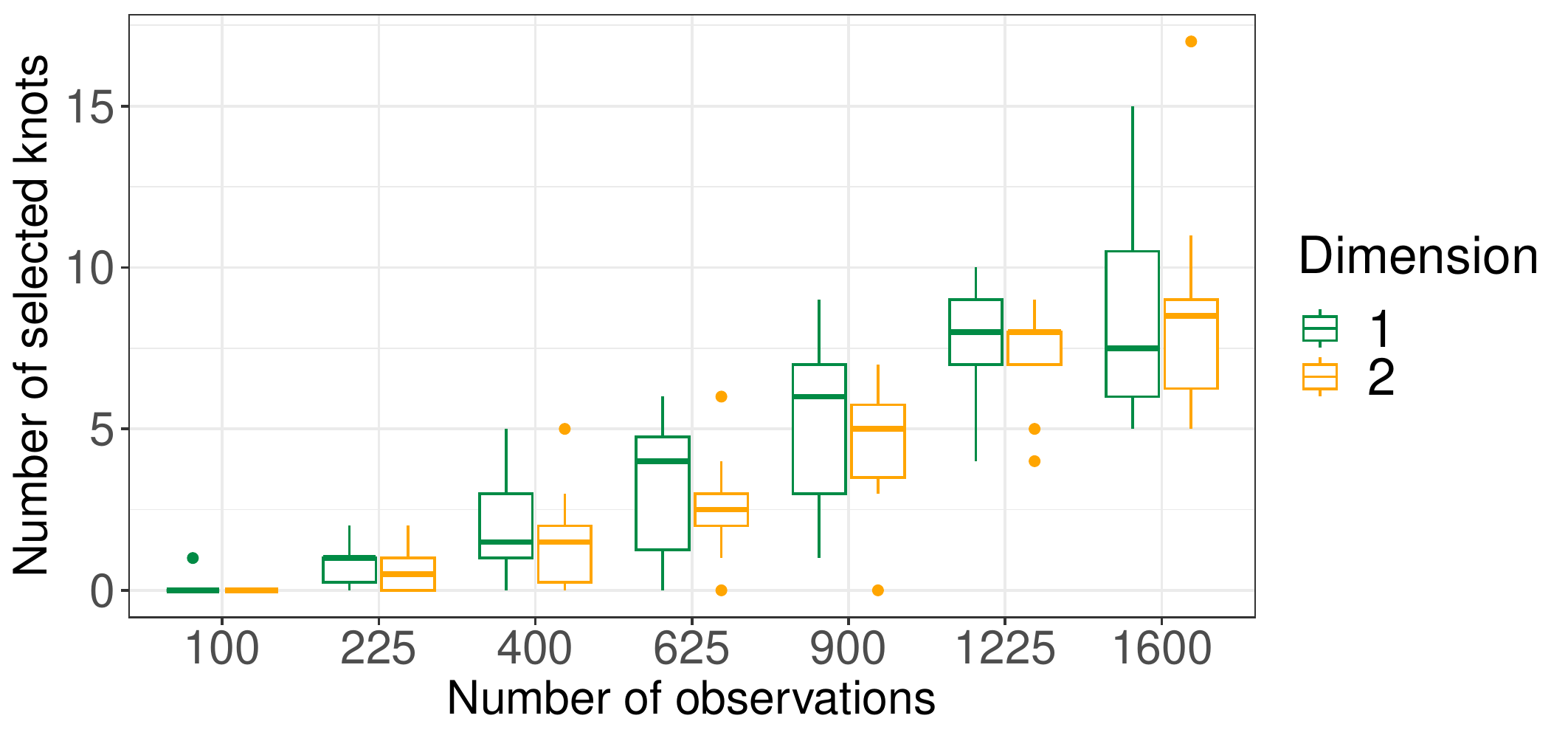}
\caption{Similar to Figure \ref{fig:boxplot_ebic_f2} by choosing $\widetilde{\uplambda}_1 = \widetilde{\uplambda}_{1,\text{EBIC}}$ and $\widetilde{\uplambda}_2 = \widetilde{\uplambda}_{2,\text{EBIC}}$ for the estimation of $f_2$ with $\sigma=0.05$.}\label{fig:boxplot_f2_sigma_05}
\end{center}

\end{figure}

\begin{figure}[ht]
\begin{center}

\includegraphics[width=0.49\textwidth,height=4cm]{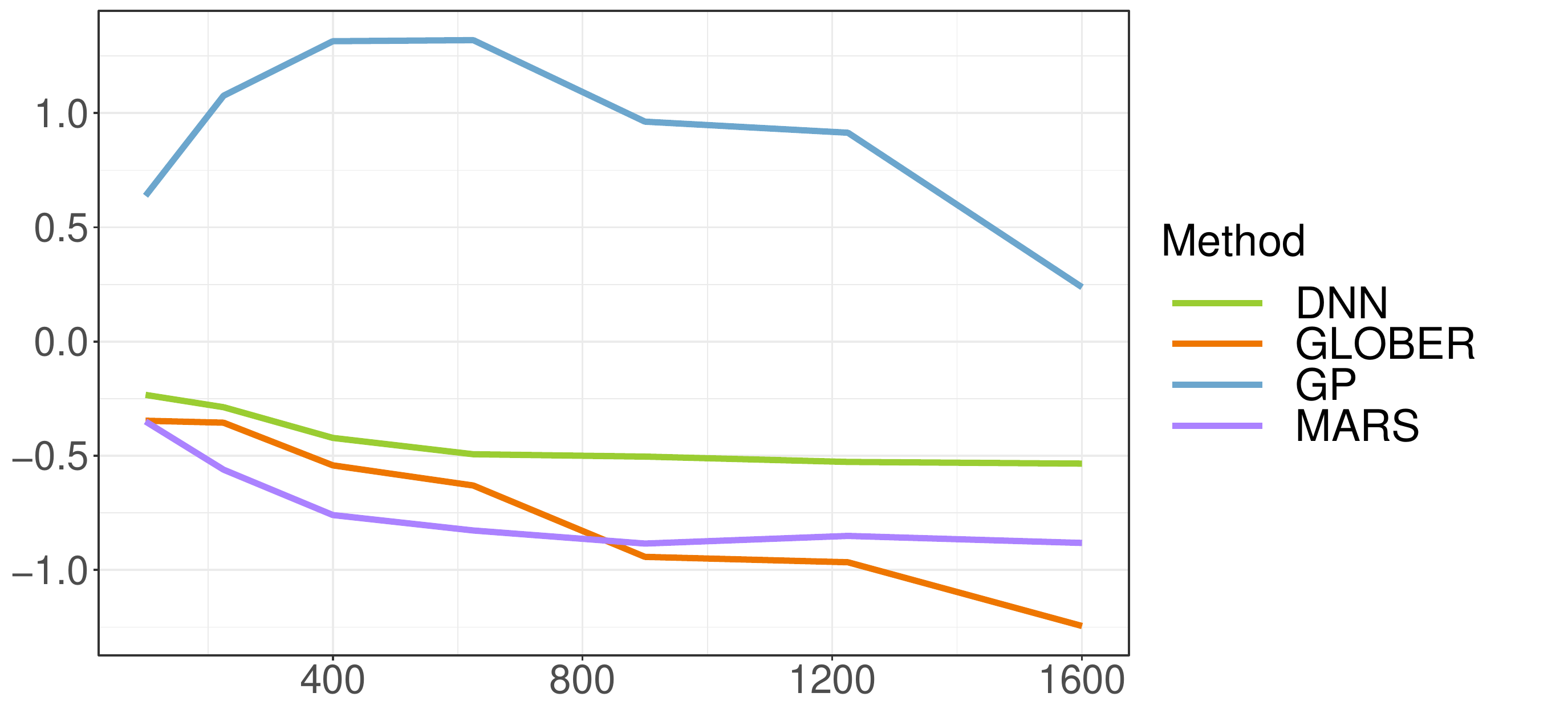}
\includegraphics[width=0.49\textwidth,height=4cm]{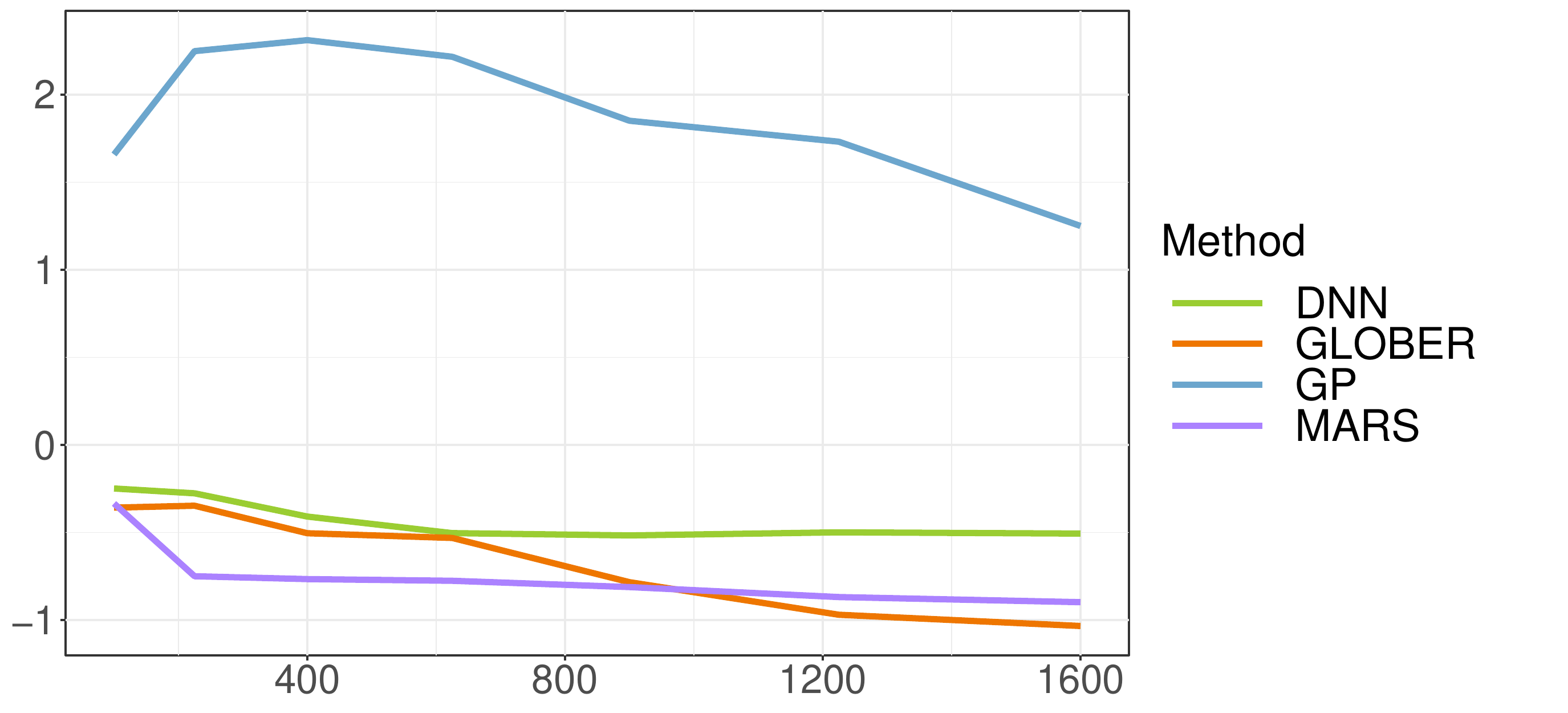}
\caption{Statistical performance (Normalized Sup Norm) of our method using the EBIC criterion for $\sigma = 0.005$ (left) and $\sigma = 0.05$ (right) and of the state-of-the-art methods obtained from 10 replications.}\label{fig:sigma_val_2D}

\end{center}
\end{figure}

\begin{figure}[h!]
\begin{center}
\includegraphics[width=14cm]{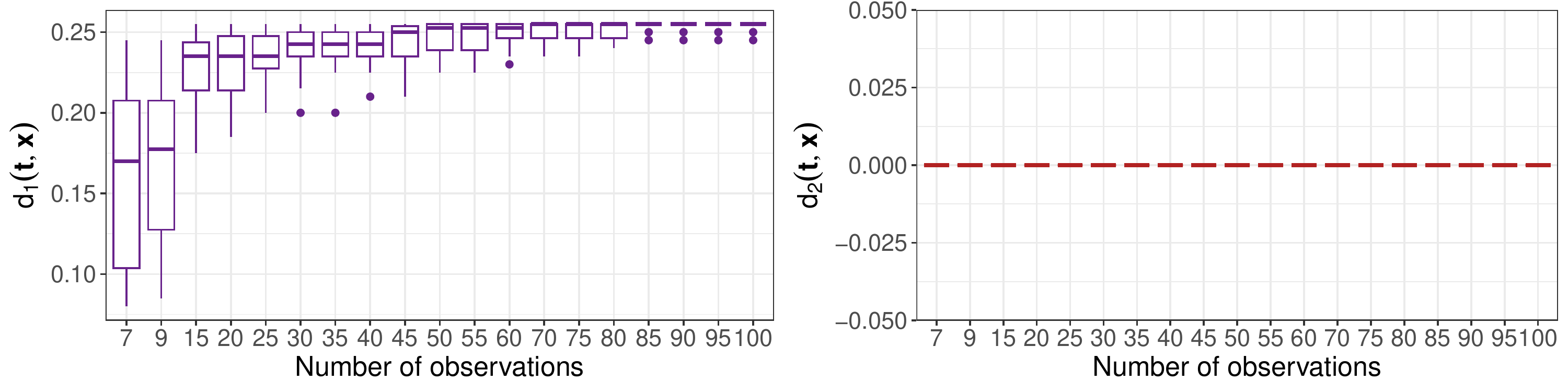}
\includegraphics[width=14cm]{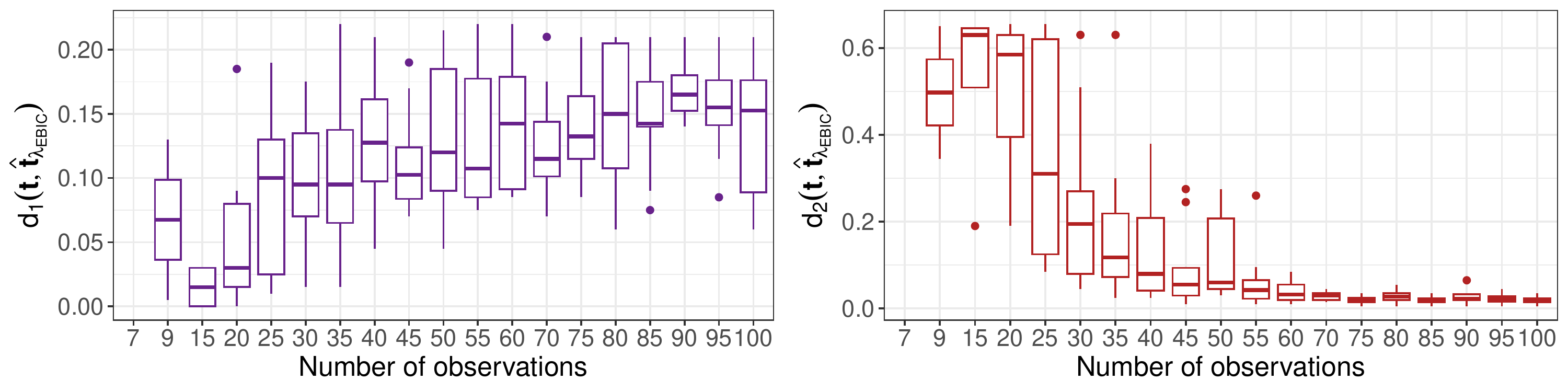}
\caption{Similar to Figure \ref{fig:hausdorff_random_sampling_f1} for the estimation of $f_1$ with $\t$ belonging to $\x$ and $\sigma = 0.05$.}\label{fig:hausdorff_inObs_f1}
\end{center}
\end{figure}

\begin{figure}[ht]
\begin{center}
\includegraphics[width = 14cm]{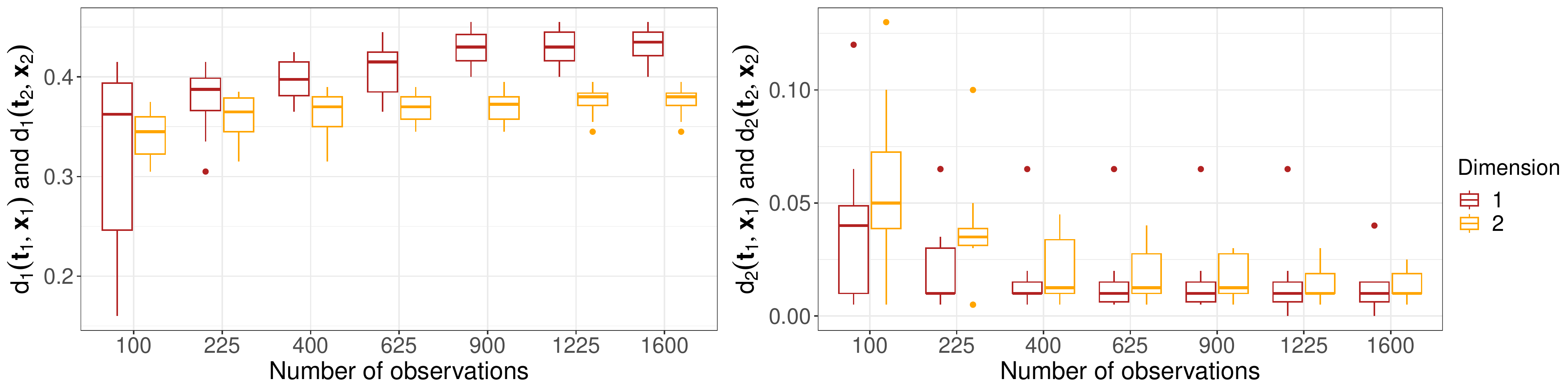}
\includegraphics[width = 14cm]{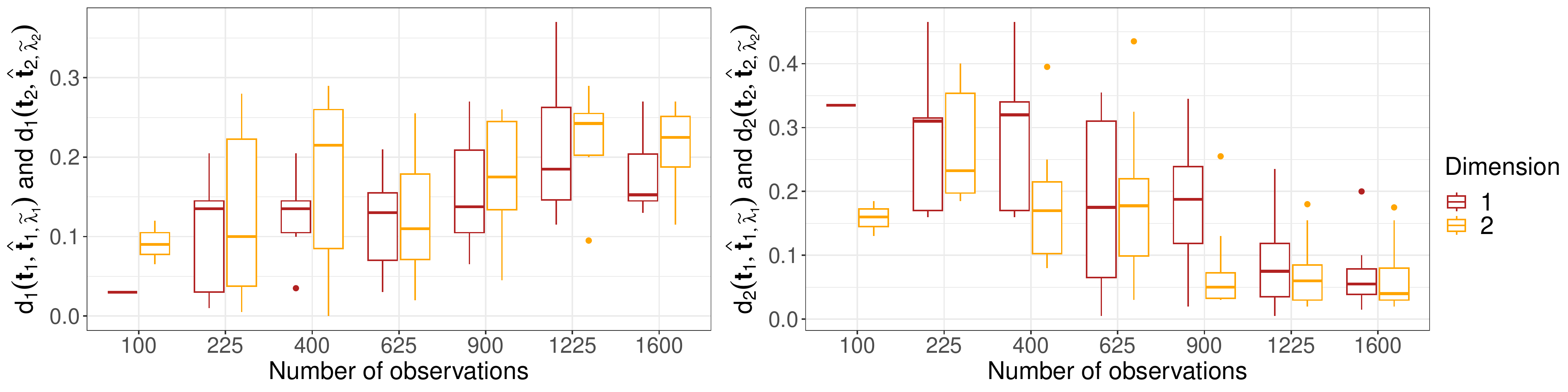}
\caption{Boxplots of the first part of the Hausdorff distance as a function of $n$: $d_1(\t_1, \x_1)$ (resp. $d_1(\t_2, \x_2)$) (top left) and $d_1(\t_1, \widehat{\t}_{1,\widetilde{\uplambda}_1})$ (resp. $d_1(\t_2, \widehat{\t}_{2,\widetilde{\uplambda}_2})$) (bottom left) and for the second part of the Hausdorff distance as a function of $n$: $d_2(\t_1, \x_1)$ (resp. $d_2(\t_2, \x_2)$) (top right) and  $d_2(\t_1, \widehat{\t}_{1,\widetilde{\uplambda}_1})$ (resp. $d_2(\t_2, \widehat{\t}_{2,\widetilde{\uplambda}_2})$) (bottom right) for the first (resp. second) dimension, for the estimation of $f_2$ by choosing $\widetilde{\uplambda}_1 = \widetilde{\uplambda}_{1,\text{EBIC}}$ and $\widetilde{\uplambda}_2 = \widetilde{\uplambda}_{2,\text{EBIC}}$ with a random sampling of the observation set with $\sigma = 0.01$. }
\label{fig:hausdorff_sampling_2D}
\end{center}
\end{figure}

\begin{figure}[ht]
\begin{center}

\includegraphics[width = 14cm]{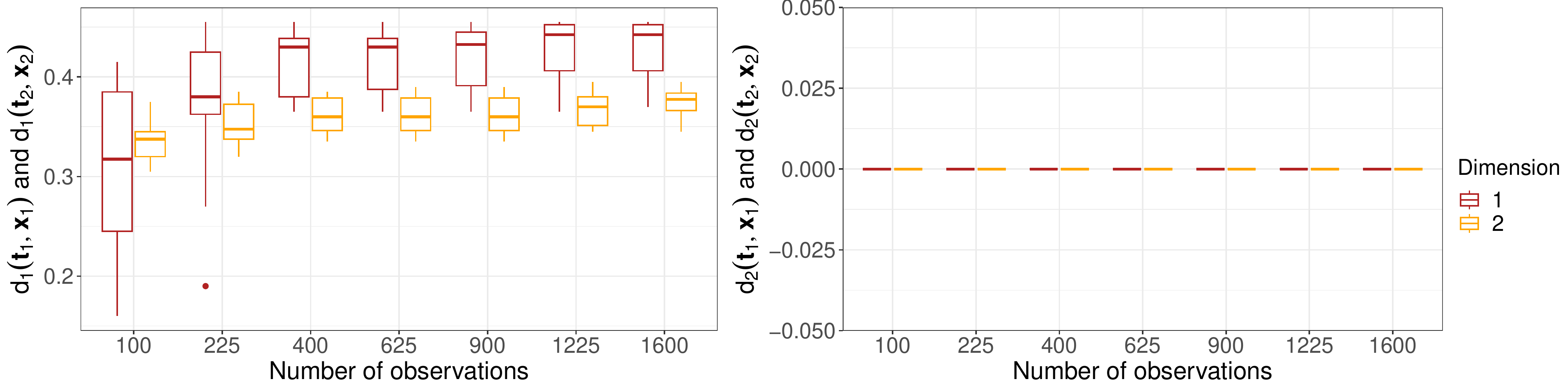}
\includegraphics[width = 14cm]{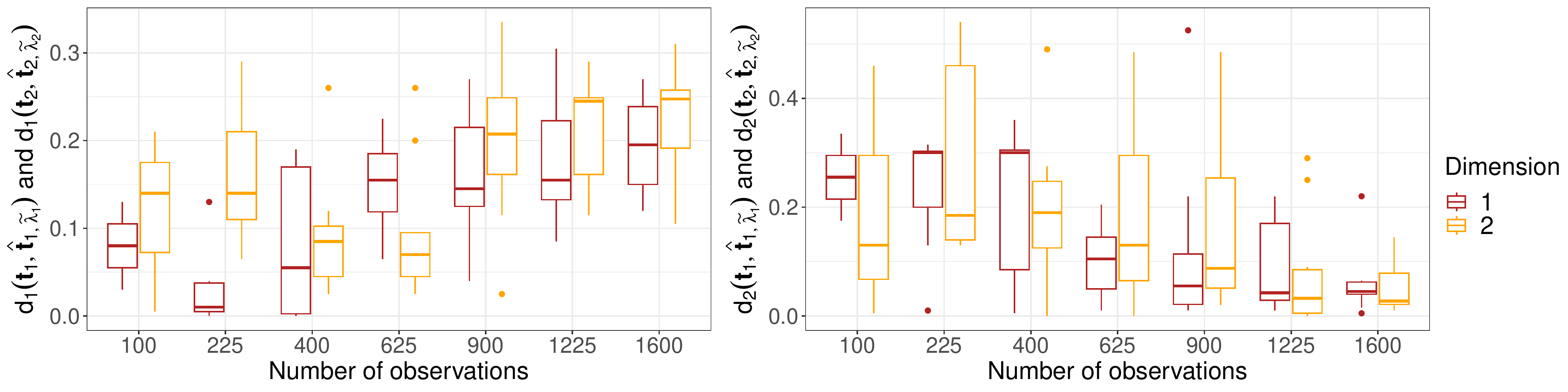}
\caption{Similar to Figure \ref{fig:hausdorff_sampling_2D} for the estimation of $f_2$ with $\t_1$ belonging to $\x_1$ (resp. $\t_2$ belonging to $\x_2$) and $\sigma = 0.01$. }

\label{fig:hausdorff_inObs_f2}
\end{center}
\end{figure}

\begin{figure}[ht]
\begin{center}
\includegraphics[width = 7cm]{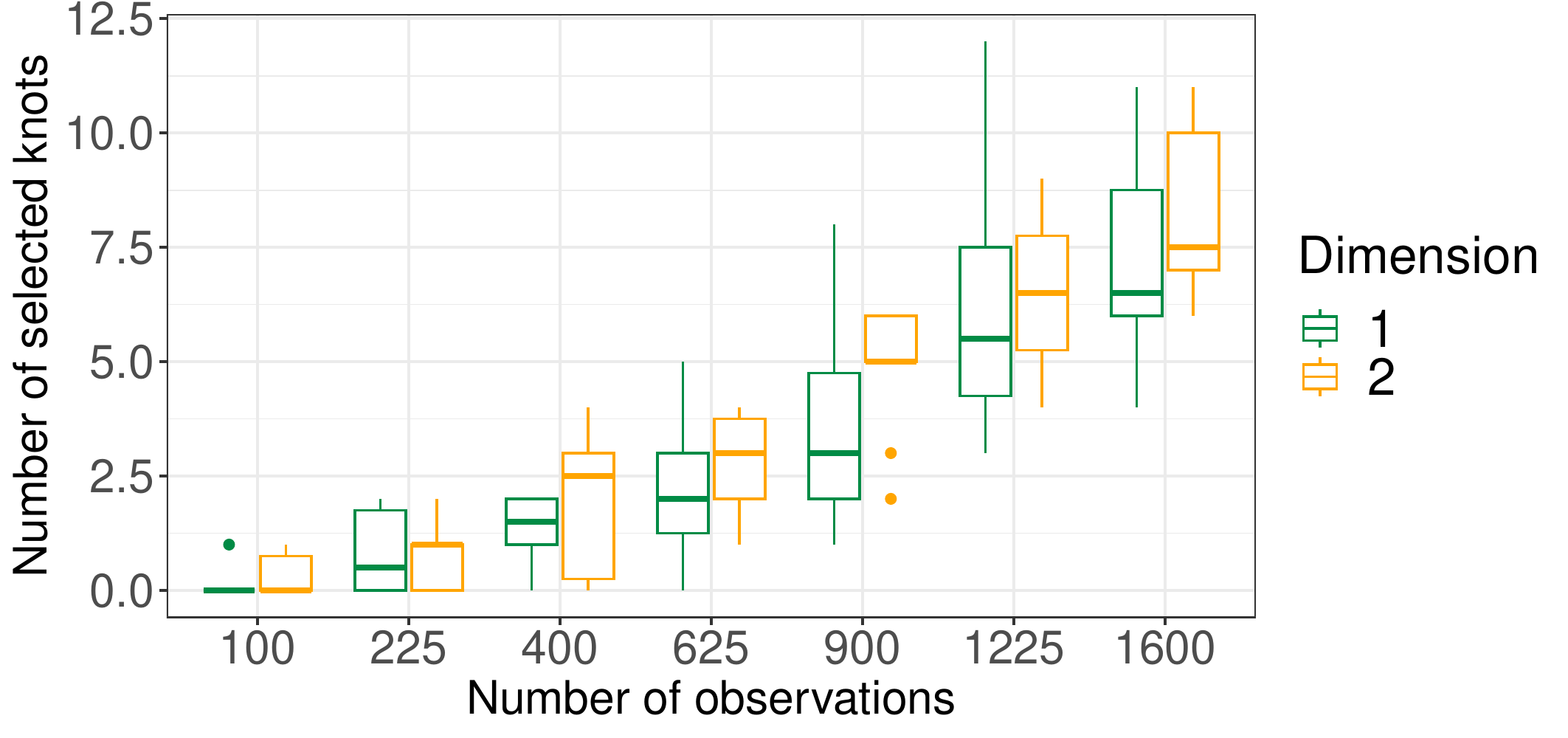}
\includegraphics[width = 7cm]{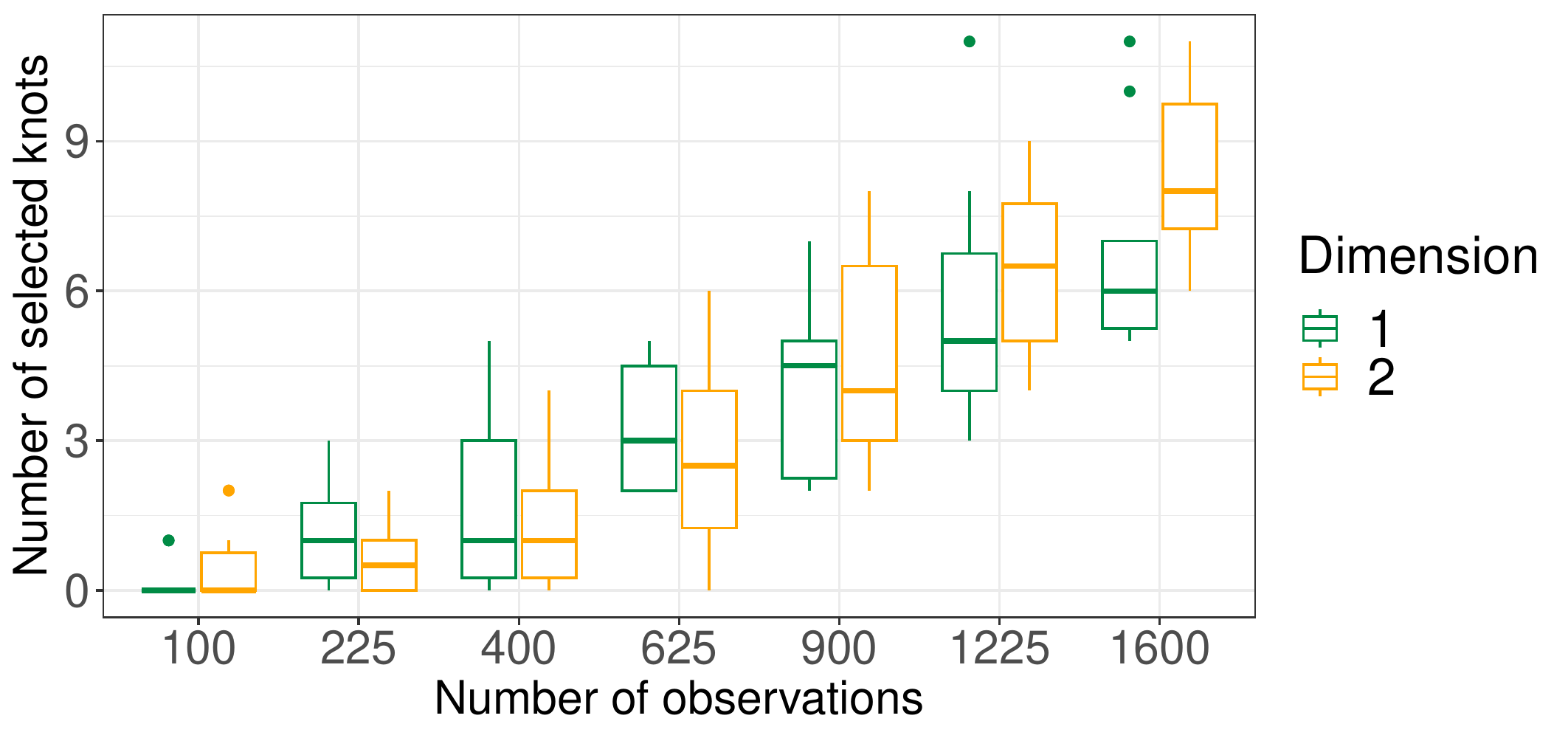}
\caption{Left: number of estimated knots as a function of $n$ for the estimation of $f_2$ with GLOBER from a random sampling of observations (left) and when $\t_1$ and $\t_2$ belong to $\x_1$ and $\x_2$, respectively (right) with $\sigma = 0.01$. }
\label{fig:number_selected_knots_f2}
\end{center}
\end{figure}

\begin{figure}[h!]
\begin{center}
\includegraphics[width=0.48\textwidth,height=4cm]{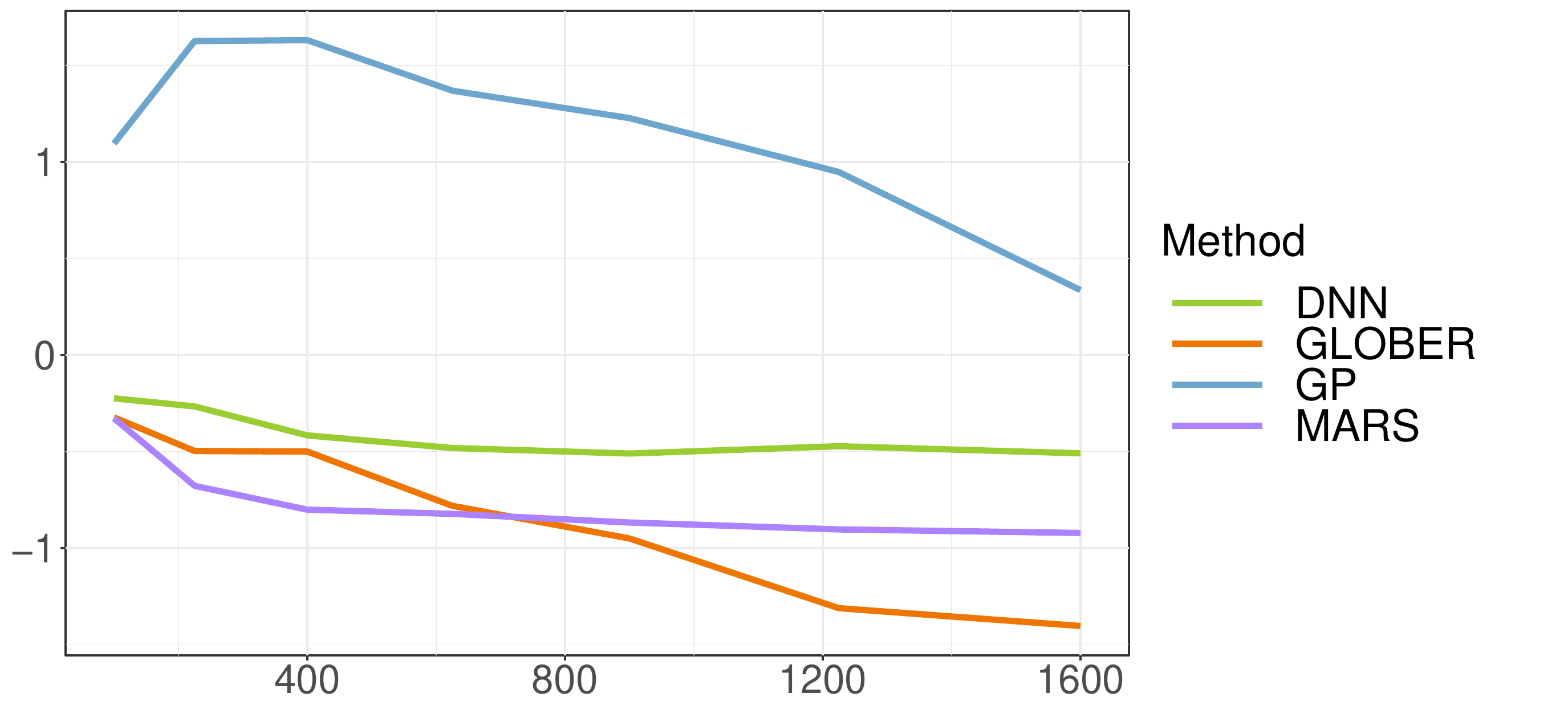}
\includegraphics[width=0.48\textwidth,height=4cm]{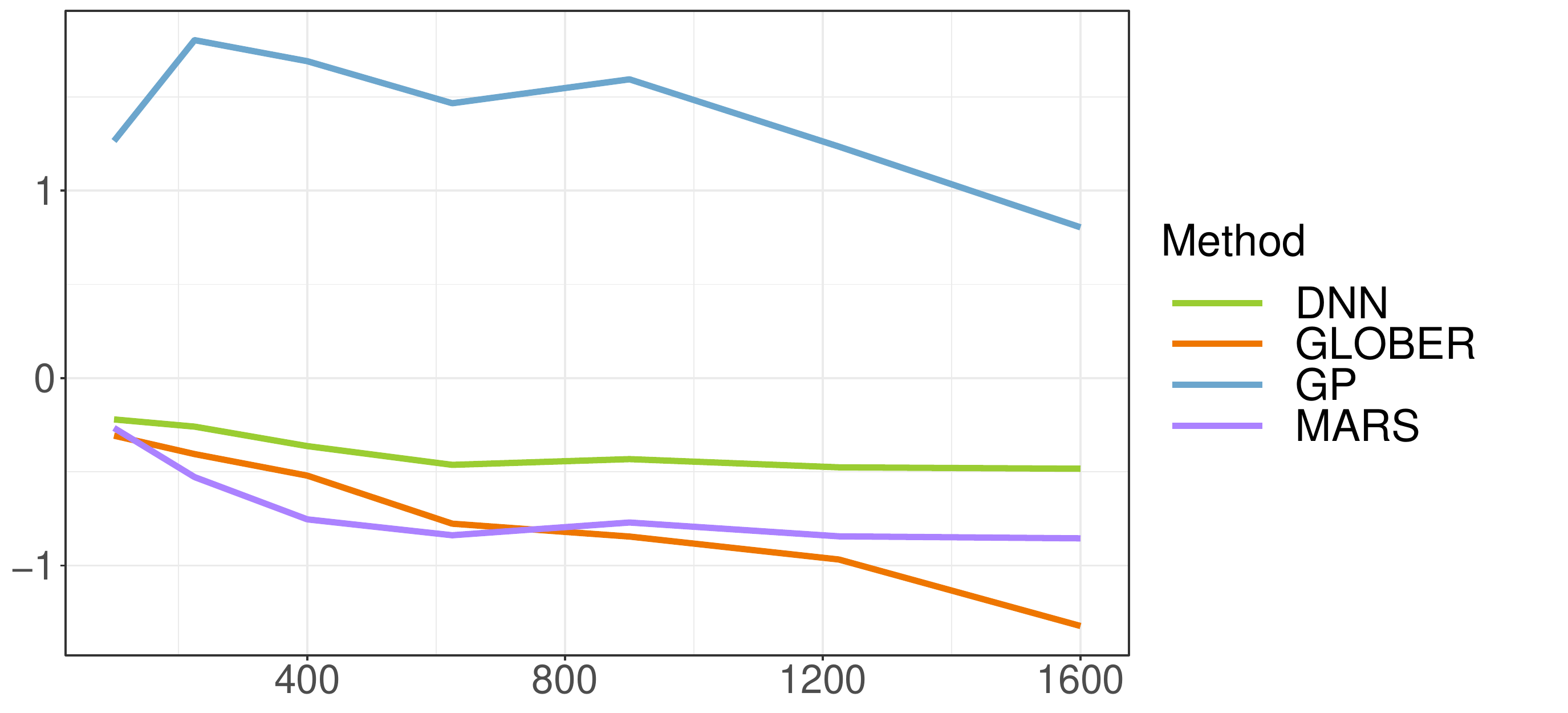}
\caption{Statistical performance (Normalized Sup Norm) of GLOBER from a random sampling of the noisy observation set (left) and with $\t_1$ and $\t_2$ belonging to the observation set (right) with  $\sigma = 0.01$. Comparison to the performance of state-of-the-art methods obtained from 10 replications. }\label{fig:sampling_2d}
\end{center}
\end{figure}

\begin{figure}[h!]
\begin{center}

\includegraphics[width=5.5cm, trim= 0.5cm 1cm 0 2cm, clip]{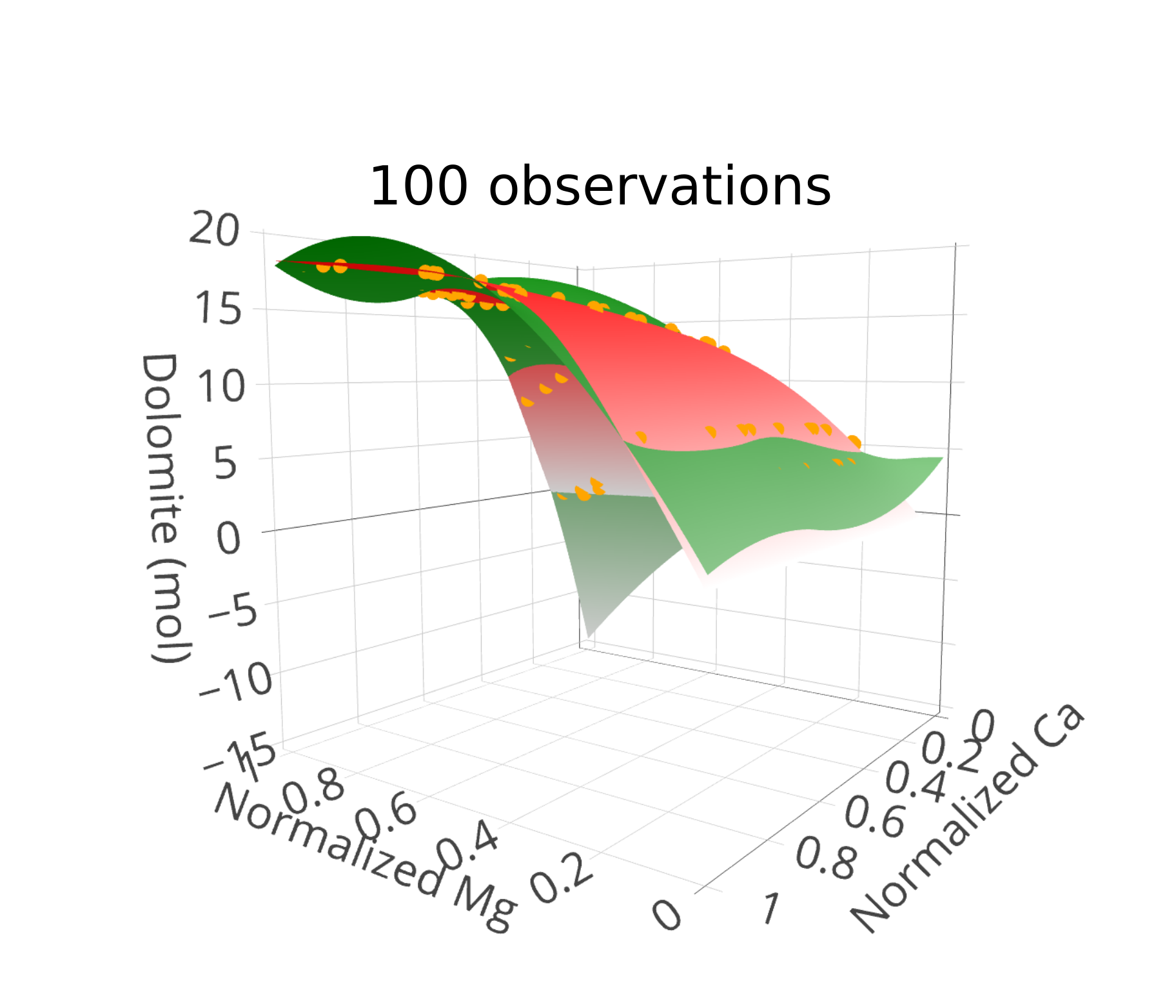}
\includegraphics[width=5.5cm, trim= 0.5cm 1cm 0 2cm, clip]{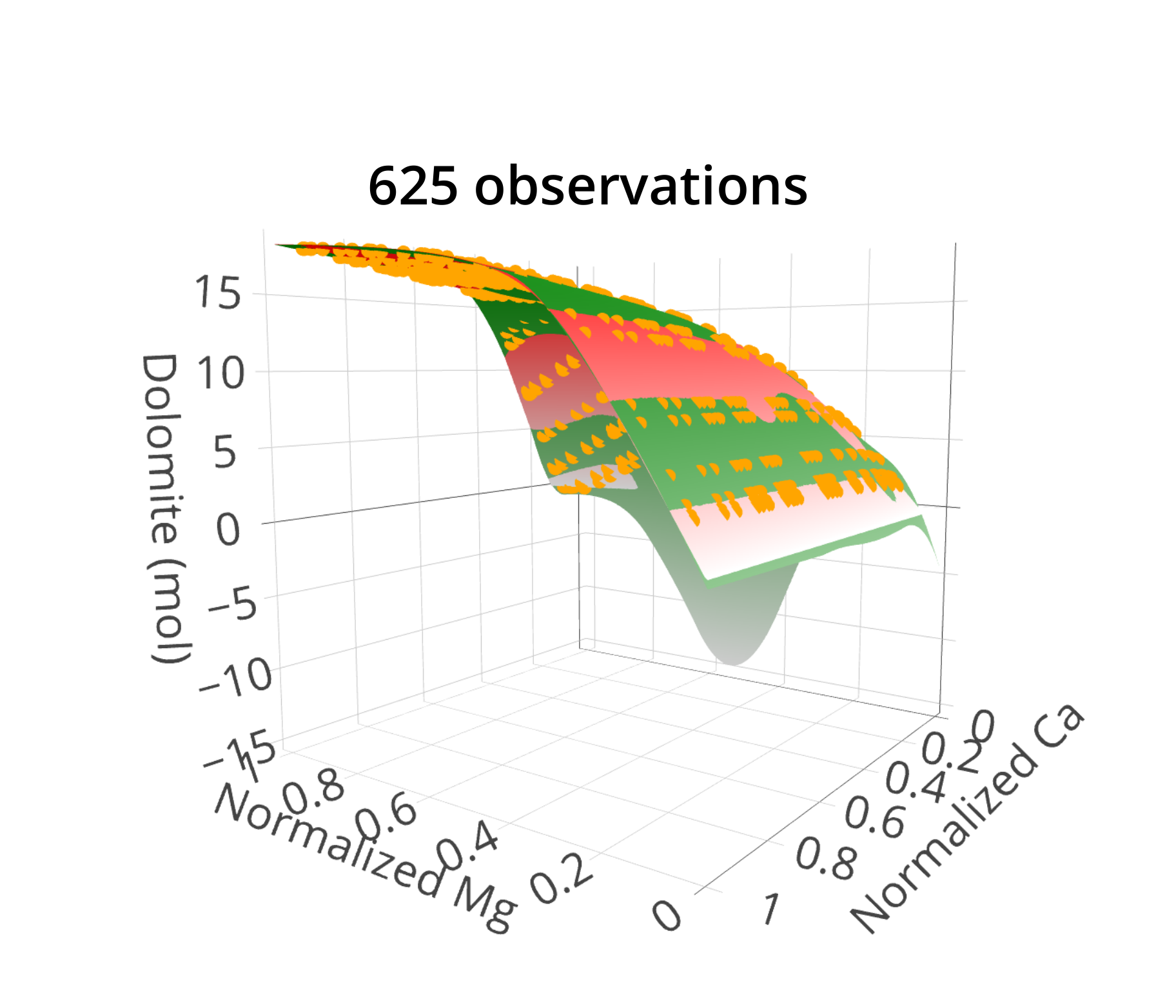}
\includegraphics[width=5.5cm, trim= 0.5cm 1cm 0 1.5cm, clip]{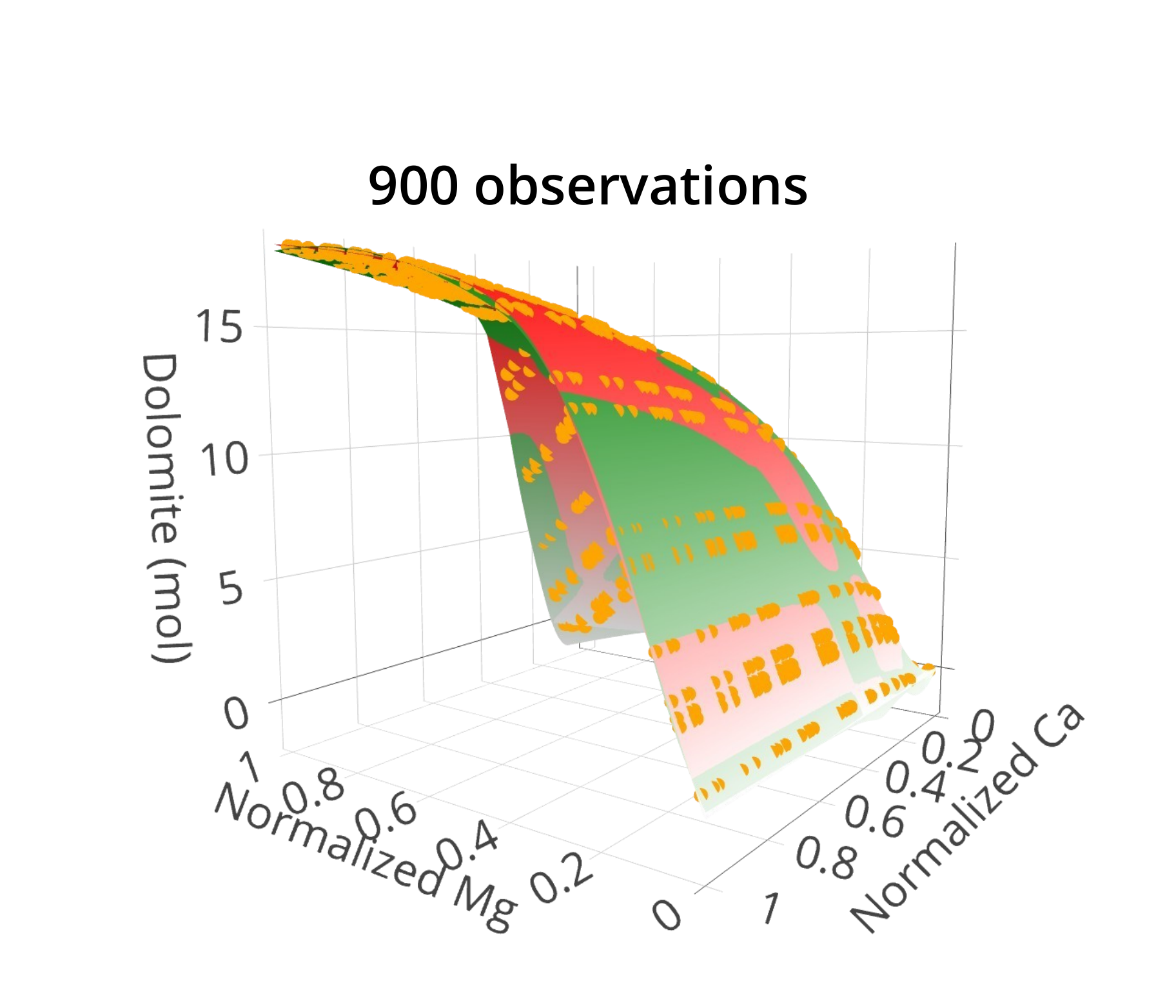}
\includegraphics[width=5.5cm, trim= 0.5cm 1cm 0 1.5cm, clip]{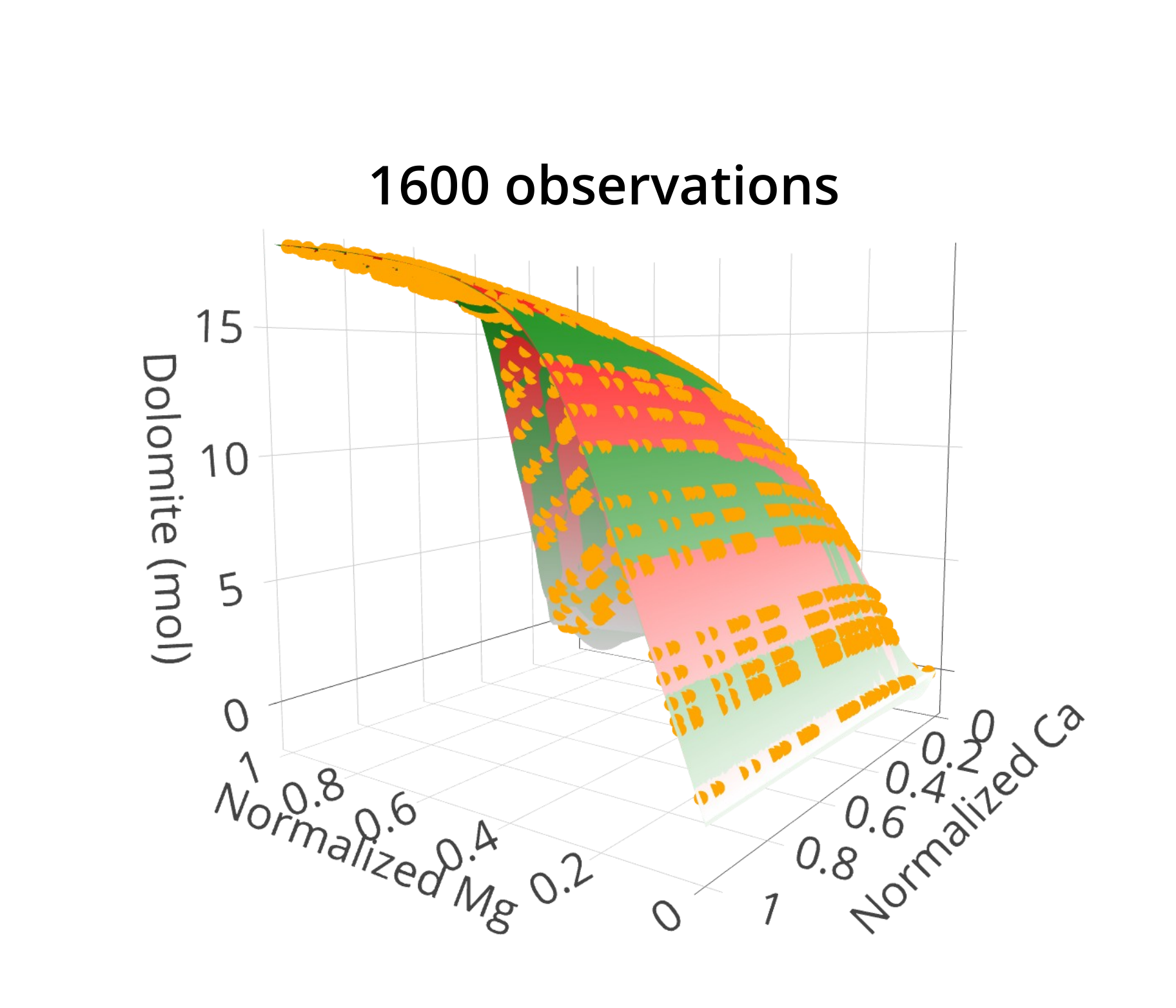}
\caption{Illustration of the estimation of the amount of dolomite depending on the normalized concentration of Ca and Mg for 100 (top left) 625 (top right) 900 (bottom left) and 1600 observations (bottom right).  The red surface describes the true underlying function $f_4$ to estimate, the green surface corresponds to the estimation with GLOBER and the orange bullets are the observation points.}\label{fig:illustration_dolomite2D}

\end{center}
\end{figure}

\end{document}